\documentclass[acmsmall]{acmart}
\usepackage[english]{babel}
\usepackage{subfigure,hyphenat,enumitem}
\usepackage{algorithm}
\usepackage[noEnd]{algpseudocodex}
\usepackage{amsmath,hyperref}
\usepackage{tabularx,tablefootnote}
\usepackage[labelformat=simple,skip=0pt]{subcaption}
\usepackage[normalem]{ulem}
\usepackage{listings}
\usepackage{xcolor}
\usepackage[utf8]{inputenc}

\definecolor{codegreen}{rgb}{0,0.6,0}
\definecolor{codegray}{rgb}{0.5,0.5,0.5}
\definecolor{codepurple}{rgb}{0.58,0,0.82}
\definecolor{backcolour}{rgb}{0.95,0.95,0.92}

\lstdefinestyle{mystyle}{
    backgroundcolor=\color{backcolour},   
    commentstyle=\color{codegreen},
    keywordstyle=\color{magenta},
    numberstyle=\tiny\color{codegray},
    stringstyle=\color{codepurple},
    basicstyle=\ttfamily\tiny,
    breakatwhitespace=false,         
    breaklines=true,                 
    captionpos=b,                    
    keepspaces=true,                 
    numbers=left,                    
    numbersep=5pt,                  
    showspaces=false,                
    showstringspaces=false,
    showtabs=false,                  
    tabsize=2
}

\usepackage{tabularx,wrapfig}
\usepackage[capitalize]{cleveref}

\newcommand{\parabreak}{\vspace*{1.00ex minus 0.25ex}\noindent}

\newcommand{\added}[1]{\textcolor{black}{ #1}}

\newcommand{\sysname}{\textsf{L4Span}}
\newcommand{\sysnames}{\textsf{L4Span's}}

\newcommand{\revise}[1]{{\color{black}#1}}

\setcopyright{cc}
\setcctype{by}
\acmJournal{PACMNET}
\acmYear{2025} \acmVolume{3}
\acmNumber{CoNEXT4}
\acmArticle{25}
\acmMonth{12} \acmPrice{}
\acmDOI{10.1145/3768972}

\crefdefaultlabelformat{\bfseries#2#1#3}

\DeclareCaptionLabelSeparator{emdash}{--- }

\newcommand{\parahead}[1]{\vspace*{1ex plus 0ex minus 0.25ex}\noindent{}{\bfseries #1}}

\setitemize{itemsep=0pt,topsep=3pt,parsep=0pt,partopsep=0pt,leftmargin=1.5em}
\setenumerate{itemsep=2pt,topsep=3pt,parsep=2pt,partopsep=0pt,leftmargin=1.5em}
\setlist{itemsep=2pt,parsep=2pt}

\title[]{\sysname{}: Spanning Congestion Signaling over NextG Networks for Interactive Applications}

\begin{document}

\settopmatter{authorsperrow=4}

\author{Haoran Wan}
\orcid{0000-0001-9726-195X}
\affiliation{%
  \institution{Princeton University}
  \streetaddress{35 Olden Street}
  \city{Princeton} 
  \state{NJ} 
  \postcode{08544}
  \country{USA}}
\email{haoran.w@princeton.edu}

\author{Kyle Jamieson}
\orcid{0000-0002-7940-2867}
\affiliation{%
  \institution{Princeton University}
  \streetaddress{35 Olden Street}
  \city{Princeton} 
  \state{NJ} 
  \postcode{08544}
  \country{USA}}
\email{kylej@princeton.edu}

\begin{abstract}
Design for low latency networking is essential for tomorrow's interactive applications, but it is essential to deploy incrementally and universally at the network's last mile. While wired broadband ISPs are rolling out the leading queue occupancy signaling mechanisms, the cellular Radio Access Network (RAN), another important last mile to many users, lags behind these efforts. This paper proposes a new RAN design, \textbf{L4Span}, that abstracts the complexities of RAN queueing in a simple interface, thus tying the queue state of the RAN to end-to-end low-latency signaling all the way back to the content server. At millisecond-level timescales, L4Span predicts the RAN’s queuing occupancy and performs ECN marking for both low-latency and classic flows. L4Span is lightweight, requiring minimal RAN modifications, and remains 3GPP and O-RAN compliant for maximum ease of deployment. We implement a prototype on the srsRAN open-source software in C++. Our evaluation compares the performance of low-latency as well as classic flows with or without the deployment of L4Span in various wireless channel conditions. Results show that L4Span reduces the one-way delay of both low-latency and classic flows by up to 98\%, while simultaneously maintaining near line-rate throughput. 
The code is available at \href{https://github.com/PrincetonUniversity/L4Span}{\textbf{https://github.com/PrincetonUniversity/L4Span}}.

\end{abstract}


\maketitle

\thispagestyle{empty}

\section{Introduction}
\label{s:intro}

Today and tomorrow's interactive applications (\textit{e.g.}, 
videoconferencing, AR/VR, cloud gaming) all depend
on controlling content senders' rates to 
strike a balance between utilizing the network
and avoiding latency-inducing queues.  Indeed, recent congestion
control algorithms have been evolving to design for this balance
\cite{cardwell_bbr_2019, 
meng_achieving_2022, arun_copa_2018, fouladi_salsify_2018}.
To realize this higher performance bar, 
congestion control algorithms have explored
feedback information from the receiver or network that is 
more timely and richer
than simply packet loss (\textit{e.g.}, ABC \cite{goyal_abc_2020},
XCP \cite{katabi_congestion_2002}). The key to realizing these
protocols on the Internet, however, is end-to-end deployability,
and for this, one front-running approach is the use
of Explicit Congestion Notification (ECN) bits coupled with 
\textit{Low Latency Low Loss Scalable} (L4S) 
congestion signaling \cite{briscoe_low_2023,briscoe_resolving_2019,schepper_dual-queue_2023}.
L4S is gaining traction: it 
is currently rolling out in broadband cable service provider 
networks as \textit{Low Latency DOCSIS} \cite{livingood_comcast_2023,comcast_comcast_2025},
and the mobile standards body, 3GPP, has stated
its intention to standardize L4S in the cellular 
Radio Access Network (RAN)
\cite{3gpp_5g_2023}.

\begin{figure}
\includegraphics[width=\linewidth]{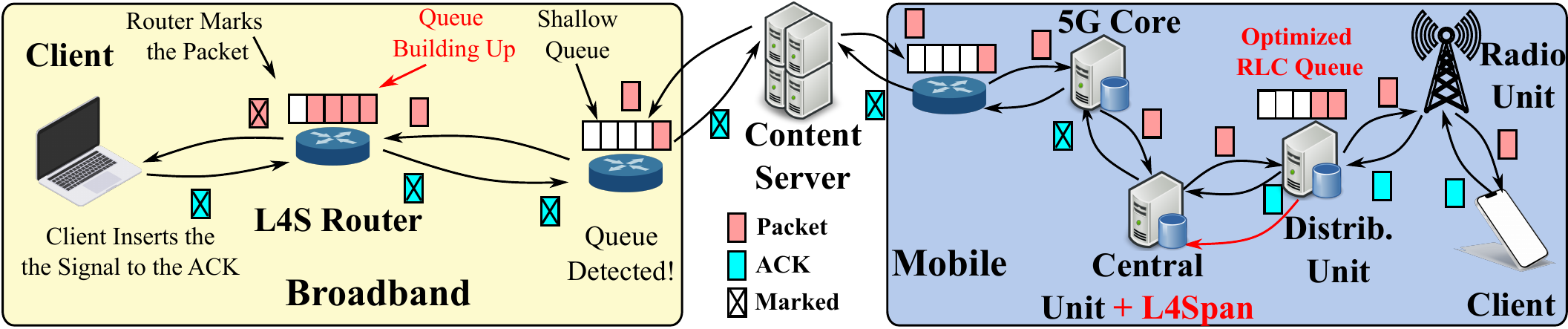}
\caption{\textbf{L4S status quo:} Some backbone and broadband 
ISPs deploy L4S routers (left), but for mobile users, the 5G 
network is key to 
performance, yet lacks L4S functionality (right).}
\label{fig:high_level_path}
\end{figure}
Despite being a very widely-used last-mile network,
the cellular RAN faces unique challenges in its 
ability to signal congestion end-to-end, to
applications and transport protocols.
This is broadly because of its complexity, but 
specifically because the many queues of 5G networks are
hidden deep inside the highly-layered 5G RAN protocol 
suite, whose structure exists alongside, but is mostly
opaque to, Internet protocols at all layers of the Internet's 
protocol stack \cite{khan_experimental_2024,liu_close_2023,liu_handling_2025,grinnemo_optimizing_2025}.
The result is that a very large, multi-hop network (the 5G
Core sub-network and RAN sub-network) 
is hidden from L4S, unintentionally subverting the efficiencies
of the L4S signaling architecture 
(see \S\ref{s:motivation} next).

Furthermore, realizing L4S congestion signaling 
in the RAN faces additional 
challenges stemming from the 
fundamental differences between wired and wireless networks. 
First, the vagaries of wireless propagation, complexities in
the handoff process between cell towers, and 
non-uniform traffic scheduling results in a
highly variable capacity. While
existing wired L4S routers set a very low queuing delay
target (\textit{e.g.}, one millisecond) 
and mark the ECN bit of all packets 
if the sojourn time of the queue
exceeds it, the cellular RAN thus demands
a different approach. Second, compounding the challenge,
the RAN is notably flexible in its configuration 
as a network, adapting to varying traffic demand 
from even a single client with the
capability to dynamically add and remove cell towers from 
the set of those simultaneously serving that client (also known as
carrier aggregation).
Third, as we explain in more detail in \S\ref{s:design},
heterogeneous client capabilities interact with the 
limited quality-of-service queuing capabilities the 5G RAN currently 
offers, interacting with the end-to-end congestion control
loop in subtle ways.

These challenges motivate a fresh look
at the internal data flow signaling mechanisms of the 5G RAN
to ask whether it is possible to bring the benefits of 
L4S in wired networks to 5G.
We seek minimal changes to the 5G network that will allow us
to monitor all relevant queue occupancies, and predict into
the short-term, future queuing delay.  Also required
is an architectural design that spans the two networks' 
(5G and Internet) currently-disjoint signaling mechanisms 
to provide timely congestion signals to L4S-capable as well as 
legacy content senders.

\parabreak{}This paper presents 
\textbf{\sysname{}}, a new design for the 5G RAN 
that passively estimates the queuing delay of each mobile user 
and signals the congestion to the content server sender through 
markings on the packets' header. 
With ECN feedback reflecting the 5G network condition, the L4S sender 
can accurately adjust its sending rate 
to minimize queuing delay in the 5G Radio Link Control (RLC)
layer (see \cref{fig:high_level_path}), while maintaining 
high throughput.  Since wireless throughput is so volatile,
\sysname{} needs to make real-time queuing delay estimates and predictions, for each 
queue in the 5G network.
Our design abstracts the complex queuing and scheduling functions
of the Distributed Unit (DU) into a single, agile rate estimate
that the DU communicates to the Central
Unit (CU) of the RAN, as shown in \cref{fig:high_level_path}.
Classic TCP and UDP traffic flows are still a major 
component of user traffic on the Internet
\cite{sundberg_measuring_2024}, and \sysname{} also 
maintains, and
improves performance for these flows.

Our design makes the following contributions:
\textbf{1)}~To cope with the wireless RAN's high jitter, 
\sysname{} permits slightly larger bounded queues,
but also designs a novel packet marking 
algorithm (\S\ref{s:design:mark_strat}) that takes the telemetry of 
the RAN scheduler and both 
L4S and classic end-to-end congestion control algorithms' 
behavior into account.
\textbf{2)}~To minimize queuing delay, \sysname{} innovates 
a queuing sojourn time prediction algorithm (\S\ref{s:design:occupancy})
that integrates
with the 5G RLC layer. We further analyze its 
performance (\S\ref{s:eval:micro_benchmark}),
shedding light on this important fundamental problem.
\textbf{3)}~\sysname{} provides
a reference implementation
for the cellular RAN that is currently
missing in the L4S ecosystem, achieving 
low latency and high throughput for both L4S and classic
flows. 

\sysname{} is a clean and practical-to-deploy design in the 5G RAN, 
making minimal changes and modifications to
the current 3GPP standardized layer structure and
the 5G network design as a whole.
\revise{\sysname{}} requires only mandatory control messages in 3GPP for its egress rate estimation, 
and conforms to both 3GPP and O-RAN standards.
\sysname{} performs only ECN bit marking in the TCP and IP headers, 
making its existence transparent to other components in the RAN.
Further, our design only reads and reuses currently-existing messages and state
in the 5G RAN to achieve its goal on queuing delay prediction, 
necessitating no intrusive modifications on the bulk of the RAN.
Finally, \sysname{} considers all the corner RAN configuration cases 
that might affect the performance, such as different 5G RLC modes
(\S\ref{s:queue_prediction}) and different 5G 
DRB configurations (\S\ref{s:mark_strategy}).


\parabreak{}\Cref{s:motivation} of this paper goes into
further technical detail of \sysnames{} motivation. 
\Cref{s:related} surveys related work, after which
we present a detailed design in \cref{s:design}
and our implementation in \cref{s:impl}.
Our evaluation
follows in \cref{s:eval}, measuring \sysname{}
operating with many different senders, 
including TCP Prague,
CUBIC, and BBRv2.  Top-line results show that
\sysname{} reduces one way delay 
by up to 98\% while maintaining
near line-rate throughput in a busy 5G 
network with 64 concurrent L4S clients.
Further results show that \sysname{} also optimizes the 
performance of classic flows, achieving up to
97\% RTT reduction in a 16-client RAN
and reducing the finish time of short-lived TCP flows
competing with long-lived background traffic
by a factor of $4\times$.
With regards to video conferencing, our results show that
\sysname{} improves the performance of 
SCReAM interactive video congestion control 
\cite{ericssonresearch_ericssonresearchscream_2025},
reducing RTT time by a factor of $3\times$
while maintaining throughput.
 \Cref{s:concl}
concludes the work. 
The authors attest that this work raises \textbf{no} ethical issues.

\begin{figure*}
        \centering
        \subfigure[\added{CUBIC and L4S in a \textbf{wired network.}}]{
        \includegraphics[width=0.325\linewidth]{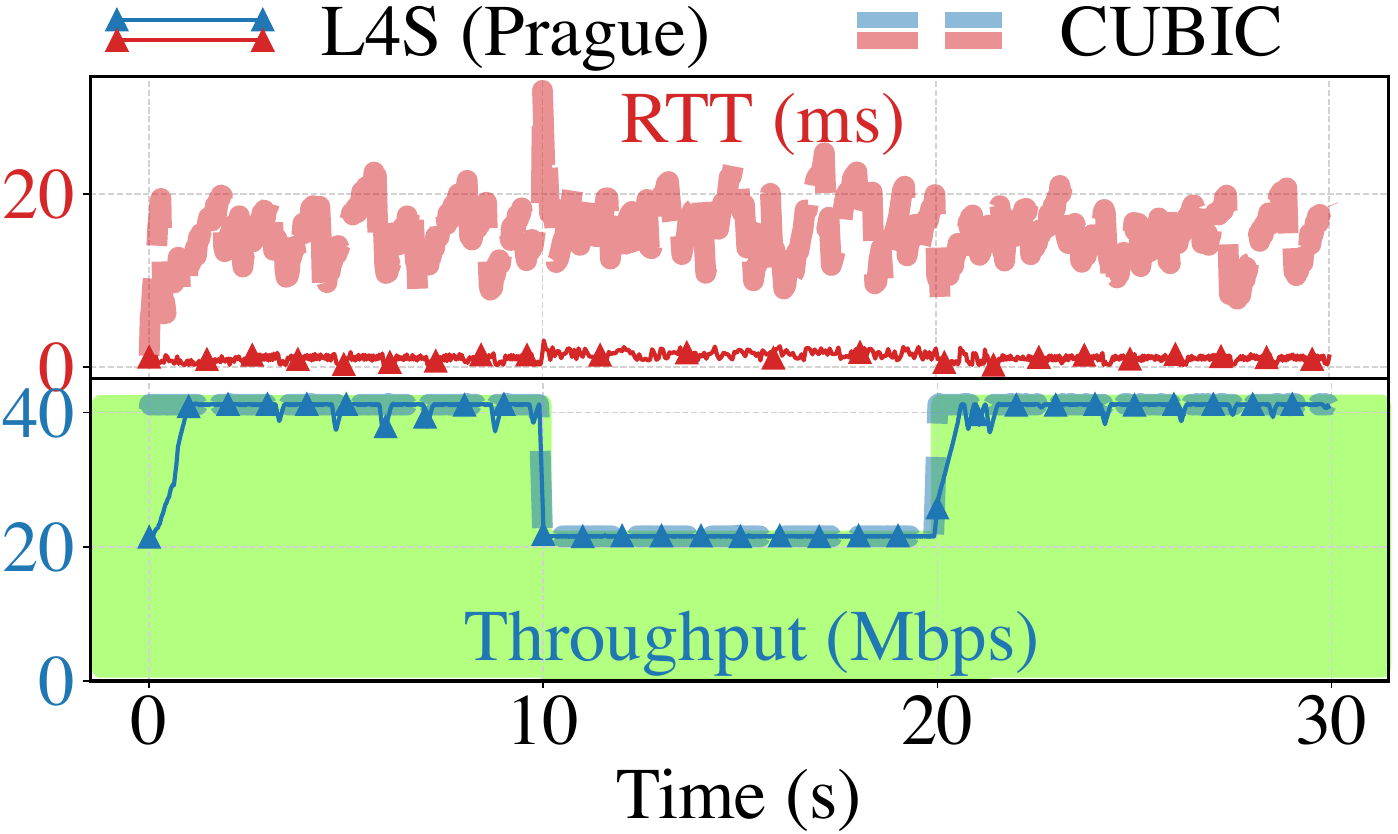}\label{fig:l4s_cubic_wired}}
        \hfill
        \subfigure[\added{CUBIC and L4S in a \textbf{5G network.}}]
        {\includegraphics[width=0.325\linewidth]{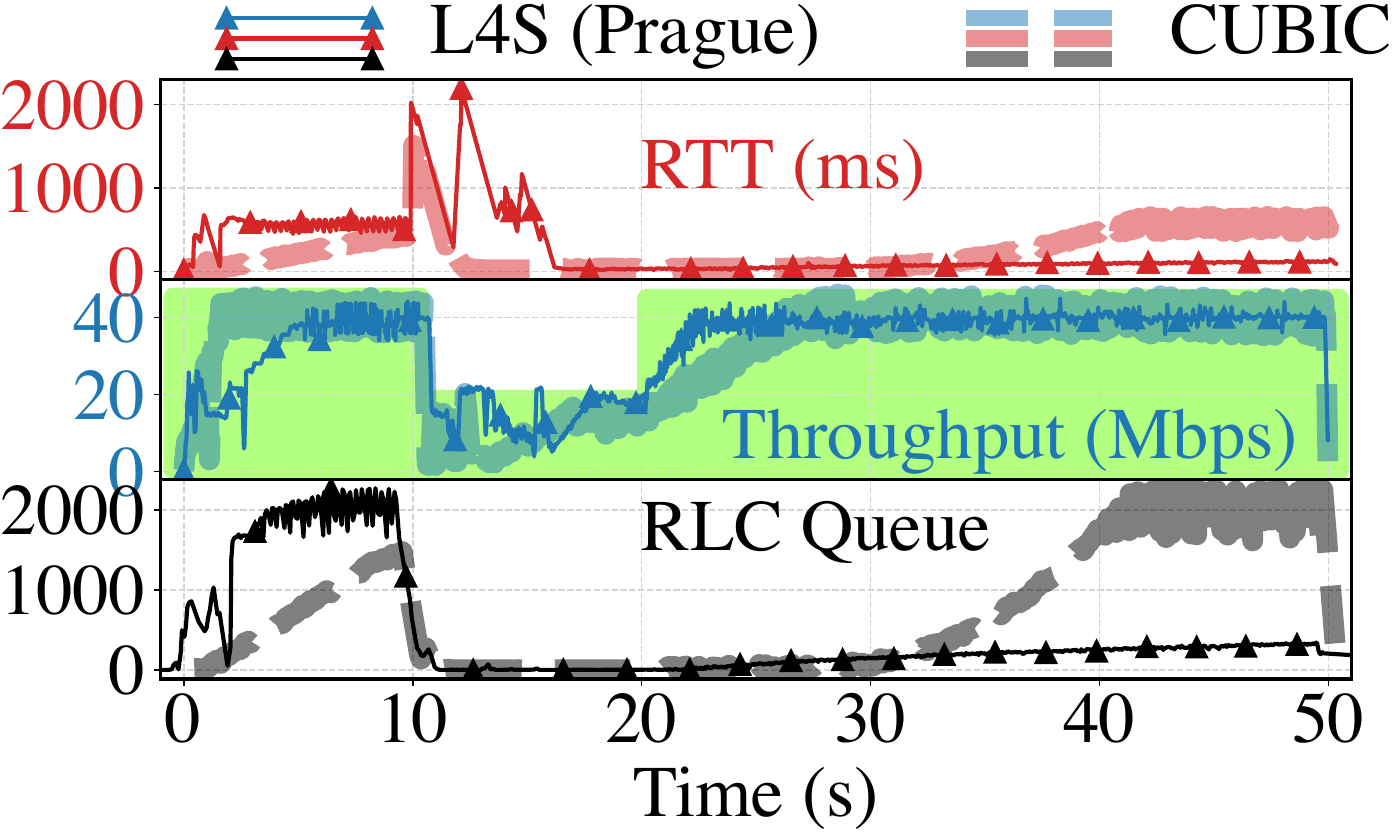}\label{fig:l4s_cubic_ran}}
        \hfil        \subfigure[\added{CUBIC and L4S in 5G + \textbf{\sysname{}}.}]
        {\includegraphics[width=0.325\linewidth]{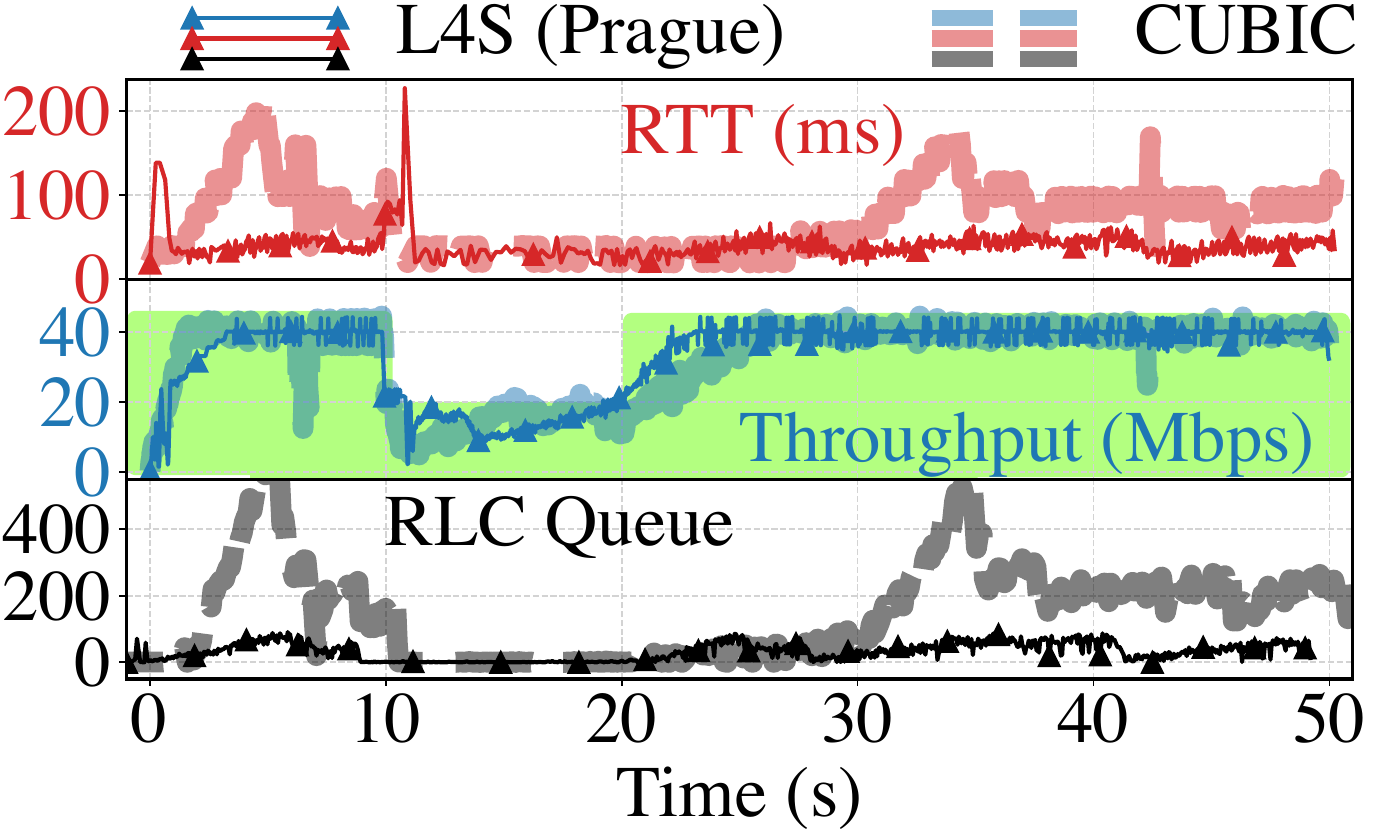}\label{fig:l4s_cubic_l4span}}
        \caption{\textbf{Performance of L4S and CUBIC in different networks:} 
        L4S routers in wired networks achieve line rate and extremely low latency, but 
        are much less effective in 5G networks that don't expose their queues. 
        Both L4S and CUBIC's latency is reduced with \sysname{} in 
        5G, and both maintain line rate (green area). 
        In the latter two figures, the bottleneck shifts from the RAN to wired middleboxes at 10s and shifts back at 20s.}
        \label{fig:l4s_cubic_5g} 
\end{figure*}

%

\section{Background: L4S Signaling Meets the 5G RAN} 
\label{s:motivation}\label{s:background}
\label{s:ran_tx_background}
\label{s:l4s_rlc_queue}

When an internet content server sends a data segment
to a client on an L4S-enabled network, 
any congested L4S-enabled router 
on the forward path 
marks the appropriate ECN bit
when its queuing delay exceeds a 
\emph{sojourn time} threshold $\tau_s$
(defaulting to one millisecond) 
\cite{briscoe_low_2023,schepper_dual-queue_2023}. 
An L4S receiver
then sets the \emph{congestion experienced} (CE) 
bit in the TCP header
of the ACK,
and the information reaches the sender within a round-trip
time, as shown in \cref{fig:high_level_path}
(\textit{upper}).
An L4S sender treats the CE feedback
as a ``lightly-pressed brake,'' meaning the sender 
updates the slow start threshold:
$$\mathrm{ssthresh} \gets \left(1-\alpha/2\right)\cdot\mathrm{cwnd},$$
where $\alpha$ is an exponentially-weighted moving average 
of the fraction of acknowledged bytes with the 
CE bit marked out of
the total bytes receiver over the previous RTT 
(similar to DCTCP~\cite{alizadeh_data_2010}).
However, classic TCP flows treat 
CE feedback as a packet loss, so potentially 
slow dramatically \cite{floyd_addition_2001},
and thus are prone to starvation relative to L4S flows,
and so L4S routers separate L4S and classic flows into 
two different queues,  
ensuring the fair bandwidth share between 
them through a \textit{dual 
queue coupled} (DualPi2) structure \cite{schepper_dual-queue_2023}.
In a toy topology with 
one server, one router, and one client, running 
iperf3 with TCP Prague and CUBIC,
\cref{fig:l4s_cubic_wired} shows TCP Prague (L4S)'s 
low RTT and line-rate throughput performance
in this wired network \revise{thus the RLC buffer is not shown}.
The CUBIC has an RTT around 20 ms, which is the default target for classic flows in the L4S router.

Contrast this with the same content server sending data 
to a mobile client on a 5G network.  The 5G Core (5GC) passes it
to the RAN.
In the RAN's CU, the \textit{Service Data Adaptation 
Protocol} (SDAP) layer maps the packet by its quality of service 
identifier to a \textit{Data Radio Bearer} (DRB), a logical 
channel that spans the 5G architecture from the 5GC all the way
to the UE.
The intervening \textit{Packet Data
Convergence Protocol} (PDCP) and 
\textit{Radio Link Control} (RLC) entities are instantiated
once per DRB.
The PDCP assigns
and records a sequence number for the packet, which is 
then known as a \textit{Service Data Unit} (SDU), 
so that the RLC in the 
DU can run an ARQ
protocol to ensure (more) reliable delivery over the wireless
link. The SDU then moves to the DU
where the RLC queues and retransmits it as necessary. 
It then passes down to lower
layers of the Distributed and Radio 
Units as a \textit{Protocol Data Unit} (PDU)
where it is further (independently) reframed and 
retransmitted as necessary.
\revise{One or several DRBs and the RLC queues are maintained for each UE, resulting in traffic isolation among and within UEs.}
\revise{The MAC layer in the base station schedules data from RLC queues of each UE, and then the traffic shares the over-the-air physical channel.}
Key here is the reframing and encapsulation
of the datagram at the PDCP and lower layers, resulting in
the relevant RLC queue in the DU being
hidden to Layers~3 and above.
\revise{Furthermore, the RLC buffer is designed to be deep for reliable delivery, while, on the contrary, it worsens the sojourn time.}
Hence, the same CUBIC and L4S flows 
both experience large RLC 
queuing delays in a 5G network (\cref{fig:l4s_cubic_ran}, 
\textit{lower}), leading to high end-to-end RTT
(\cref{fig:l4s_cubic_ran}, \textit{upper}), thus impacting 
the performance of interactive applications.
\revise{Also, the latency breakdown in \cref{fig:delay_breakdown} shows that the sojourn time in the RLC buffer makes up the majority portion of the one-way delay without \sysname{}.}
In contrast, \sysname{} 
minimizes queuing delay (\cref{fig:delay_breakdown}) while
simultaneously maintaining throughput 
(\cref{fig:l4s_cubic_l4span}).


\section{Related Work}
\label{s:related}

\textbf{L4S and 5G}
Many existing works discuss enabling L4S feedback mechanism in the 5G network through simulation~\cite{brunello_l4s_2020,brunello_low_2021, son_l4s_2023, pan_optimizing_2024}, which can't capture the resource allocation dynamics between UEs' transmissions\revise{, while \sysname{} observes the allocation outcome and predicts the sojourn time passively}.
Brunello \cite{brunello_l4s_2020,brunello_low_2021} proposes to include the ECN feedback in the PDCP layer, but doesn't have a solution for the dominant classic flows~\cite{sundberg_measuring_2024}, \revise{and \sysname{} proposes solutions for both types of flow}. 
Pan~\cite{pan_optimizing_2024} and Son~\cite{son_l4s_2023} conduct trace-driven evaluations, without real RAN dynamics, \revise{unlike \sysname{}'s over-the-air implementation}.
DChannel \cite{sentosa_dchannel_2023,sentosa_accelerating_2021} guides flows to different physical channels for performance gain, neglecting the impact of the internal queues, \revise{and \sysname{} aims to minimize the internal queue occupancy and achieve full throughput utilization for each UE}.
From the UE perspective, base station selection is explored to achieve performance gain \cite{deng_icellspeed_2020} and energy saving \cite{grinnemo_optimizing_2025}.
XRC \cite{kheirkhah_xrc_2022} is a rate controller in the RAN, requiring customized end-points, while \sysname{} optimizes for existing schemes. 
\revise{OutRAN\cite{kim_outran_2022} proposes to prioritize the short-lived flows over the long-lived flows in the RLC queue, while \sysname{} keeps RLC queue occupancy low and both types of flow benefit, as shown in our evaluation in \cref{fig:long_short}.  
}
\revise{RAPID\cite{diarra_rapid_2022} designs a proxy between the 5G core and the content server to prevent RLC buffer overshooting, but RAPID is too far away from the RAN and thus can't adapt as the RAN's changing throughput as \sysname{} does.}  
\revise{TC-RAN\cite{irazabal_tc-ran_2024,irazabal_preventing_2021} implements the Linux queuing discipline, such as CoDel and ECN-CoDel, inside the RAN between SDAP and PDCP layer, and uses a fixed threshold to drop or mark the packets. Evaluation results show that TC-RAN underutilizes the RAN's capacity while \sysname{}, adapting with the dequeue rate, better utilizes the capacity and achieves low latency. }
Non-queue-building (NQB)~\cite{white_non-queue-building_2025} proposes a low-latency
architecture similar to L4S, but NQB doesn't aim for 
full bandwidth utilization, and \sysname{} can serve similarly.
\sysname{} is the first design that enables congestion feedback for both L4S and classic flows, and is evaluated with live RAN dynamics and various channel conditions.

\parahead{Congestion Control.}
Sprout~\cite{winstein_stochastic_2013} employs a stochastic model to forecast the cellular queue occupancy.
Verus~\cite{zaki_adaptive_2015} uses a packet delay profile for congestion avoidance, and Copa~\cite{arun_copa_2018} uses the RTT measures to minimize the queue occupancy.
CopaD \cite{haile_copa-d_2023} adapts to cellular networks with parameter tuning.  
PBE-CC~\cite{xie_pbe-cc_2020} utilizes the RAN telemetry information to regulate the sender, and a similar design is adopted in piStream~\cite{xie_pistream_2015} and CLAW~\cite{xie_accelerating_2017}.
BBR~\cite{cardwell_bbr_2016} probes the bandwidth and delay periodically, and BBRv2 and 3~\cite{cardwell_bbr_2019} improve the procedure and take the ECN as feedback.
Venkat \cite{arun_starvation_2022} analyzes the starvation in end-to-end congestion controls, while the RAN's behavior is one source of the non-congestive delay.
Ferreira \cite{ferreira_reverse-engineering_2024} proposes a reverse-engineering tool for congestion control algorithms, which better enables middleboxes to assist end-to-end transport. 
Separated in different RAN's queues, flows have contentions for resource allocation partially based on the backlogged data, not fully complying with the observations in the wide area network \cite{brown_how_2023,zeynali_bbrv3_2024}.
\revise{Unlike a new congestion control algorithm that customizes the logic of both sender and receiver, \sysname{} in the last-mile hop (5G network) improves the latency performance for existing algorithms.}

\parahead{In-network Design.}
XCP~\cite{katabi_congestion_2002} uses a control theory framework to achieve high utilization and low queuing delay.
ABC~\cite{goyal_abc_2020} repurposes the ECN field in the IP header as "accelerate" or "brake" command to adjust the sending rate of the sender, making it hard to fit into the L4S architecture.
TACK \cite{li_tack_2020} reduces the frequency of ACK in wireless networks and improves throughput and RTT, \revise{while \sysname{} relies on the ACKs for congestion notification}.
Zhuge~\cite{meng_achieving_2022} delays and/or modifies the ACK in the Wi-Fi for a more timely feedback to the sender, which inspires our RAN short-circuiting design.
Octopus~\cite{chen_octopus_2023} drops the packets 
actively for delay reduction, but requires modifications on the client, server, and RAN, \revise{incurring a heavier burden than \sysname{}}.
SMUFF~\cite{wang_smuff_2024} aims to achieve the line rate in Wi-Fi direct by filling the router's buffer with the optimal amount of data, \revise{and \sysname{} aims for low latency on top of it}.
Sidekick \cite{yuan_sidekick_2024} provides in-wireless-network feedback for QUIC flow, incurring new out-of-band feedback for the sender, \revise{while \sysname{} leverages only in-band feedback in the existing ACK packets, thus it is more bandwidth efficient}.
\section{Design}
\label{s:design}
\label{s:design:overview}

We first summarize \sysnames{} architecture 
(\S\ref{s:design:marking}),
then drill down into its
packet marking strategy
(\S\ref{s:design:mark_strat}),
sojourn time prediction
(\S\ref{s:design:occupancy}), 
and feedback short-circuiting mechanism
(\S\ref{s:design:shortcircuit}).
We develop our design in the context of the O-RAN
7.2x split~\cite{polese_understanding_2023} as it is the dominant design
direction in 5G currently, and our design 
generalizes to other O-RAN splits.

\subsection{\sysname{} Functionality}
\label{s:design:marking}

\begin{figure*}
    \centering
    \includegraphics[width=0.95\linewidth]{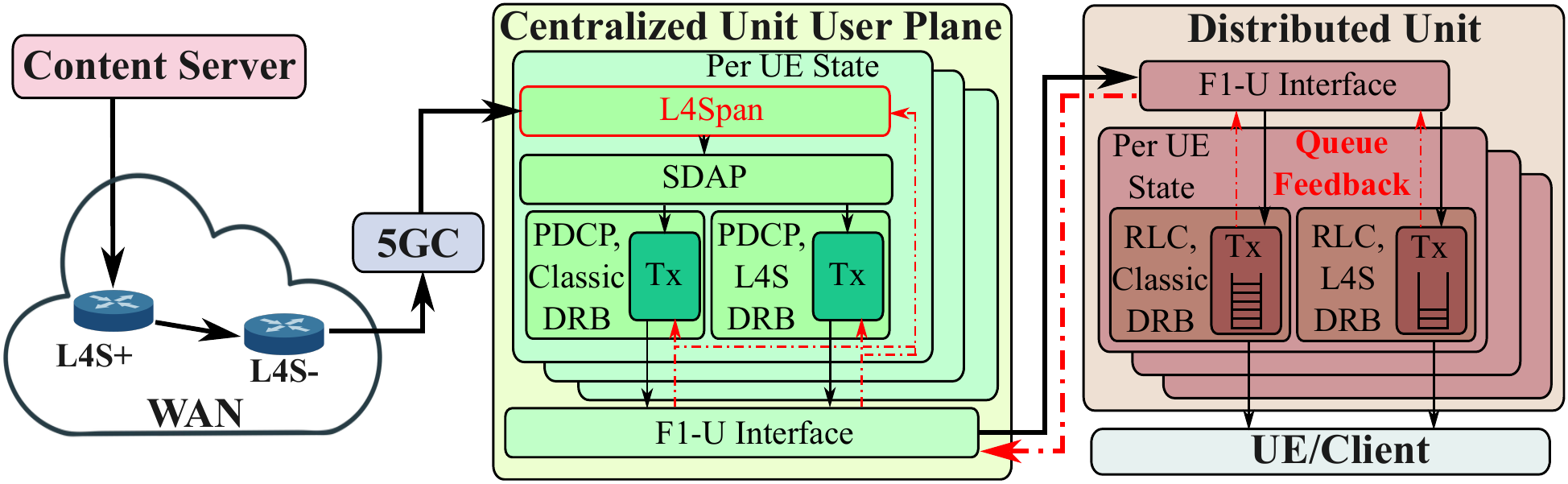}
    \caption{\sysname{} in the end-to-end data path from the content server to the client,
    where the
    {\color{red}{red}} text and arrows inside 
    the CU-UP and DU illustrate our system design, and the 
    black elements inside the 
    RLC entities mark where queues build up inside 
    the 5G network. \sysname{} reuses the existing feedback in the RAN. 
    \textsf{\textbf{L4S+\fshyp{}L4S-}} denotes a router with/without L4S functionality; 
    RU, MAC and PHY layers are not shown.} 
    \label{fig:system-overview}
\end{figure*}

As shown in \cref{fig:system-overview},
we situate most of \sysnames{} functionality in 
modules inside the RAN CU,
above the SDAP and PDCP layers that constitute 
the CU's
per-UE state.
A new queue feedback path directs 
downlink data delivery status messages from 
the RLC in the DU.
Three classes of events trigger \sysnames{}
functions: \textbf{1)}~receiving a downlink
datagram from the 5GC, \textbf{2)}~receiving 
RAN feedback, and \textbf{3)}~receiving 
an uplink ACK---pseudocode can be found
in \cref{appd:pseudo_code}.

\parahead{Receiving a downlink datagram:} when 
it receives a datagram from 
the 5GC, \sysname{} creates a mapping between the 
five-tuple \cite{matthews_stateful_2011}, and a 
UE and Data Radio Bearer (DRB) tuple---this separates
classic and L4S flows \revise{by their ECN field (01 for L4S ECN flows, 10 for classic ECN flows)}, as introduced above
in \cref{s:motivation}.  
\revise{The five-tuple contains the source and destination IP addresses, ports, and transport protocol, and is unique for each flow and UE.}
The (UE, DRB) tuple
indexes \sysnames{} \textit{packet profile table} 
(\S\ref{s:f1u_feedback}) for egress rate prediction (\S\ref{s:design:queue_delay_prediction}) and 
marking decisions.

\parahead{Receiving RAN queue feedback:}
as shown in \cref{fig:system-overview}, \sysname{} receives 
downlink RLC data delivery events over the F1-U 
interface (\S\ref{s:f1u_feedback}), a mandatory 3GPP API 
from the RLC to the PDCP entity.
Upon receiving the feedback, the \sysname{} layer uses this 
data to update the corresponding packets' status in the 
packet profile table and make a marking decision (\S\ref{s:mark_strategy}).

\parahead{Receiving an uplink ACK:}
\sysname{} only operates on the uplink packet if the feedback short-circuiting is possible (TCP, see \S\ref{s:design:shortcircuit}).
\sysname{} first reverse-maps the ACK to the correct DRB 
, and then updates the packet's 
relevant fields based on the marking decision.

\subsection{Marking Strategy}
\label{s:mark_strategy}
\label{s:design:mark_strat}

ECN marks inform L4S and classic senders 
of congestion, so they can change their behavior. 
But 
as observed above (\S\ref{s:background}),
different senders react to ECN markings differently, and so we mark
packets with different strategies, in analogy to 
DualPi2 \textit{Active Queue Management} (AQM) 
\cite{schepper_dual-queue_2023}. 
In an \sysname{} 5G network, a
DRB may serve L4S flows only (\S\ref{s:design:marking:l4s}), 
classic flows only (\S\ref{s:design:marking:classic}), or 
a mix of both (\S\ref{s:design:marking:shared_drb}).
\revise{\sysname{} classifies the above scenarios by checking the ECN fields, and work with each.}

\subsubsection{L4S-Only DRB}
\label{s:design:marking:l4s}


\begin{figure}
    \centering
    \includegraphics[width=0.9\linewidth]{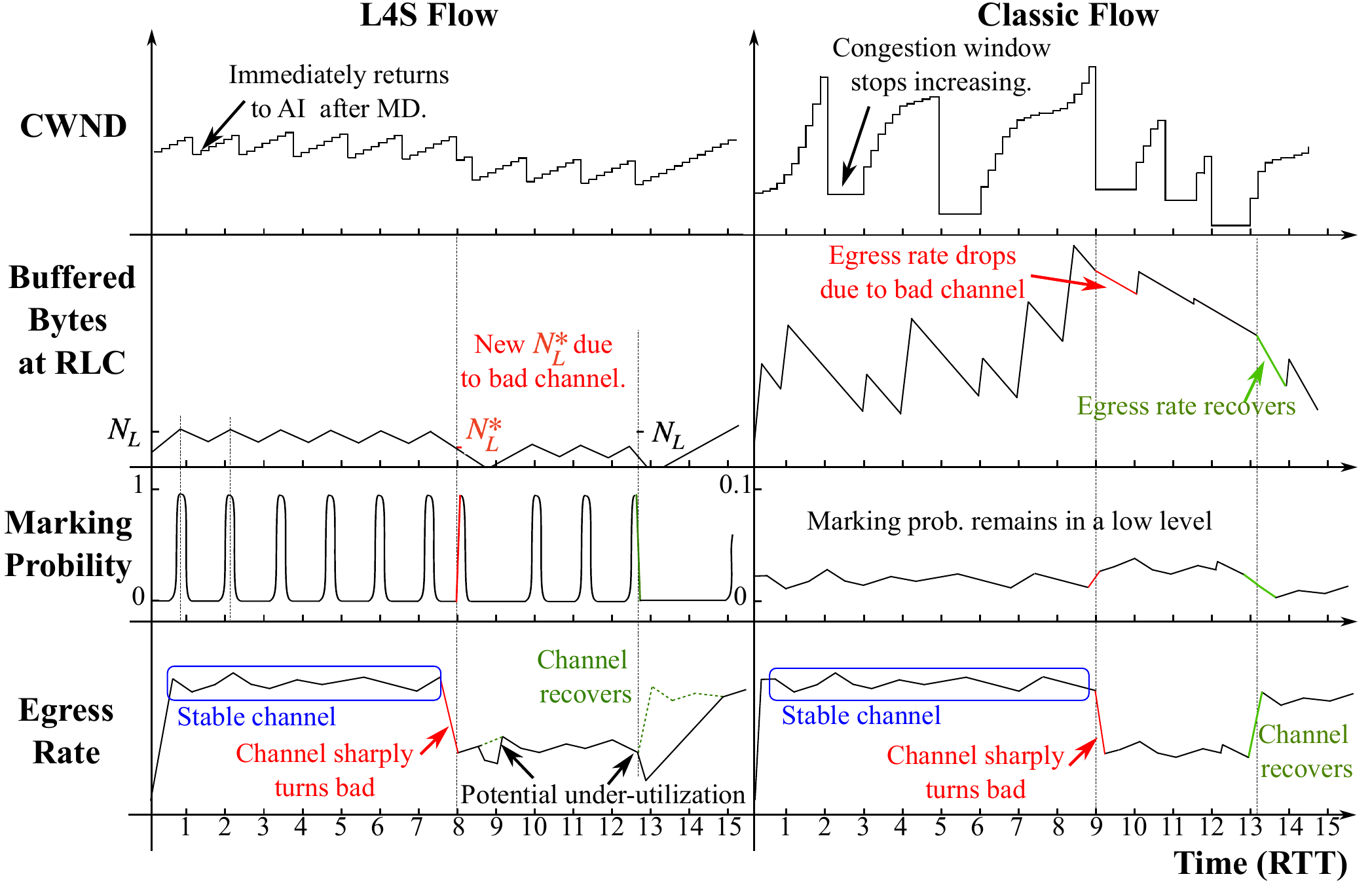}
    \caption{\sysname{}'s behavior on L4S and classic flows with wireless channel variations.}
    \label{fig:l4span-behavior}
\end{figure}

Here \sysname{} drives the queuing delay low by marking 
aggressively, because the L4S sender treats the CE ECN feedback as a 
``slightly-pressed brake,'' resuming additive increase
immediately upon receiving non-CE ACKs after the
congestion window (\textsf{cwnd}) is cut (\S\ref{s:motivation}).
Hence an L4S sender's \textsf{cwnd} changes
frequently and converges to a small saw-tooth around the optimal 
operation point \cite{briscoe_implementing_2018}.
\textbf{Marking algorithm:} Given a predicted \revise{egress rate} $\hat{r}_e$ and its error distribution $e_{r_e}$
(\S\ref{s:queue_prediction} describes how 
to make the estimates), 
\sysname{} marks the packet with probability $p_{\mathrm{L4S}}$,
the likelihood the actual egress rate satisfies the sojourn time 
threshold $\tau_s$, given $N_{\text{queue}}$ 
queued bytes: 
\begin{figure}[H]
    \begin{minipage}[b]{0.48\linewidth}
    \begin{eqnarray}
    p_{\text{L4S}} = P\left\{ \left. r_e \ge \frac{N_{\text{queue}}}{\tau_{\text{thr}}} \right | \hat{r}_e, e_{r_e} \right\} =  P \left\{e_{r_e} \le \frac{N_{\text{queue}}}{\tau_{\text{thr}}} - \hat{r}_e \right\},  \label{eq:l4s-marking}
\end{eqnarray}
    \end{minipage}
    \hfill
    \begin{minipage}[H]{0.33\linewidth}
    \includegraphics[width=0.9\linewidth]{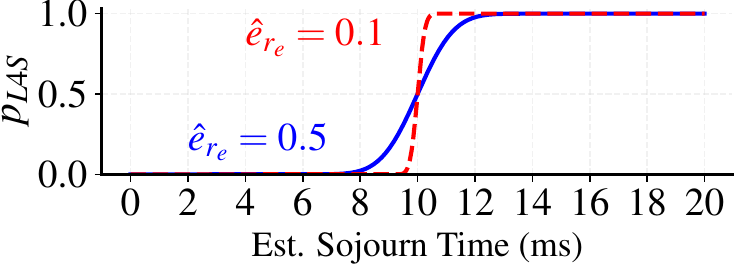}
    \end{minipage}
\end{figure}
\noindent
where $e_{r_e} \sim \mathcal{N}(0, \hat{e}^2_{r_e})$ and $\hat{e}_{r_e}$ 
is the standard deviation of the egress rate over the latest estimation window.
In a wired network, DualPi2 estimates the sojourn time by 
subtracting the ingress timestamps of the queue head packet 
and tail packet, and marks all packets when the sojourn time 
exceeds one millisecond.
However, the 5G network has a volatile egress rate, making such estimation infeasible. 
Instead, \sysname{} uses a varying distribution to calculate mark
probability (Eq.~\ref{eq:l4s-marking}): if the egress 
rate is volatile ($\hat{e}_{r_e} \uparrow$), the distribution has a 
flatter edge at $\tau_{\text{thr}}$ to avoid potential under-utilization. 
If the egress rate is stable ($\hat{e}_{r_e} \downarrow$),
the distribution has a sharper edge to pursue low latency more aggressively.
If the egress rate is invariant ($\hat{e}_{r_e}=0$), \sysnames{} 
marking strategy reduces to the DualPi2 strategy.
We set a sojourn time threshold of 10~milli\-seconds (see \S\ref{s:eval:micro_benchmark} 
for justification), as the 5G MAC layer
requires an adequately filled buffer for resource scheduling.

The behavior of \sysname{}, the L4S sender, and the RAN is
shown in the running example of \cref{fig:l4span-behavior} \emph{(left)}. 
We define a \emph{bytes threshold} $N_L = \hat{r}_e \cdot \tau_s$ in the figure, equivalent to the sojourn time threshold.
At the first RTT, the RLC buffer builds up 
to the threshold, as the L4S sender paces packets in \textit{additive increase} 
(AI) \cite{l4steam_l4steamudp_prague_2025}.
In the second RTT, the buffered bytes trigger the marking in Eq.~\ref{eq:l4s-marking}, then the L4S sender observes 
CE signal and conducts a \textsf{cwnd} 
\textit{multiplicative decrease} (MD). 
Immediately after the MD, the sender returns to AI, until the next CE signal.
In the stable channel, \sysname{} and the sender maintain a small 
sawtooth pattern around the delay threshold.

In the seventh RTT, the wireless channel sharply degrades, 
worsening the egress rate.
\sysname{} detects this and adjusts its mark threshold in the 
subsequent RTT, forcing the sender to once more cut \textsf{cwnd}.
Here, the RLC buffered bytes may drop to zero, and 
the UE would experience a brief throughput under-utilization, but
this is promptly remedied by the sender immediately returning to 
AI and refilling the RLC buffer.
In the 13th RTT, the channel recovers, and 
increased throughput drains buffers 
causing potential under-utilization until 
the L4S sender uses AI to refill the buffer.

\subsubsection{Classic-Only DRB}
\label{s:design:marking:classic}
Unlike L4S, a classic sender (\textit{e.g.} CUBIC, Reno) 
treats the congestion feedback the same way as the packet loss, 
and reacts by cutting its slow start threshold multiplicatively. 
With this in mind, we should not aim for 
a shallow queue for the classic flows, as the 
TCP endpoint would frequently receive the CE feedback, 
cut its slow start threshold and suffers from severe starvation \cite{briscoe_low_2023}.
The design goal is to prevent the well-documented buffer bloat \cite{jiang_understanding_2012, guo_understanding_2016,gettys_bufferbloat_2011} and 
in the meantime, maintain a suitable amount of bytes in the buffer to avoid underutilization.


To prevent buffer bloat, \sysname{} marks 
the packets with the probability that  
matches the average ingress rate with the RAN egress 
rate to balance the buffer size.
The TCP throughput is modeled as $r_\mathrm{classic} \approx {\text{MSS} \cdot K} / (\text{RTT}\sqrt{p_{\text{classic}}})$, where $K = \frac{1+\beta}{2}\sqrt{\frac{2}{1-\beta^2}}$ and 
$\beta$ is the MD parameter \cite{padhye_modeling_1998} (0.5 for Reno \cite{paxson_tcp_1999}).
\sysname{} predicts the RAN egress rate $\hat{r}_e$ from its packet profile table (see \S\ref{s:design:queue_delay_prediction}).
To complete the equation, \sysname{} estimates the initial $\widehat{RTT}^*$ using the interval between the first two forward TCP packets (SYN and the subsequent ACK) of each flow.
In the further operation, we add $\widehat{RTT}^*$
with the predicted sojourn time $\hat{\tau}^{r}_{\text{s}}$ over the last coherence time window (\S\ref{s:design:queue_delay_prediction}) as the RTT estimates $\widehat{RTT} = \widehat{RTT}^* + \hat{\tau}_s^r$.
The marking probability is calculated as:
\begin{equation}
    \hat{r}_e = \frac{\mathrm{MSS} \cdot K}{\widehat{RTT} \sqrt{p_\mathrm{classic}}}\ \rightarrow\  p_\mathrm{classic} = \left (\frac{\mathrm{MSS} \cdot K}{\widehat{RTT} \hat{r}_e}\right )^2. 
    \label{eq:p_classic}
\end{equation}
$\widehat{RTT}$ is an
overestimation of the RTT, as the UE's DRB may have other backlogged data when the handshake happens.
Instead of harming the performance, the slightly lower $p_\mathrm{classic}$, resulting in a higher ingress rate, helps to build an adequate RLC buffer size and prevent under-utilization (see the evaluation in \S\ref{s:eval:dualpi2_and_rlc_queue}).
If $\widehat{RTT}$ is not available, \textit{e.g.} UDP flows, we use $2\hat{\tau}^{r}_{\text{s}}$ as the RTT estimates.


Our running example continues in 
\cref{fig:l4span-behavior} \emph{(right)} with 
the behavior of \sysname{}, a classic sender, and the RAN.
Initially, the classic sender sends the packet burst to the RAN, and increases 
its \textsf{cwnd} with the CUBIC function. 
The marking probability, calculated from \cref{eq:p_classic}, remains low.
When receiving a congestion signal marked by \sysname{}, the classic sender cuts 
its \textsf{cwnd} and pauses the AI for one RTT, during which the RAN 
dequeues the packets.
As the channel turns bad, the dequeue rate decreases, resulting in a higher marking probability to regulate the ingress rate. 
In this period, the sender receives relatively more congestion signals, thus the 
RAN can drain the queue.
As the channel recovers, and the \sysnames{}'s
marking probability returns to a lower value, and the sender returns to AI.
Compared with the L4S flow's short and more volatile marking 
behavior, the marking behavior for a classic flow spans a longer time and operates 
at relatively small values, where the differences are due to the distinct 
behaviors of the classic and L4S senders.


 
\subsubsection{L4S-Classic Shared DRB}
\label{s:design:marking:shared_drb}

To achieve optimal performance, 
best practice is to keep L4S and classic flows in separate DRBs of each UE.
However, some lower-end UEs do not support multi-DRB 
configuration, and a new marking scheme is needed, as both types of flows experience performance drop (in RTT or throughput) in such a scenario if unattended (see \S\ref{s:eval:marking}).
To achieve good resource utilization,
we keep the classic packet marking probability ($p_{\text{classic}}$).
The L4S flow's throughput (MBytes/s)
is inversely proportional to the 
packet mark rate ($r_{\text{L4S}}$) \cite{schepper_understanding_nodate, briscoe_implementing_2018}:
$r_{\text{L4S}} \approx 2 \text{MSS}/(\text{RTT}\cdot p_{\text{L4S}})$, and the $r_\mathrm{classic}$ is discussed above.
To balance two types of flows' throughputs, we mark 
the L4S flow with a coupled probability 
$p_{\text{L4S}} = \alpha\sqrt{p_{\text{classic}}}$, 
where \revise{$\alpha$ is the solution of $r_{L4S} = r_\mathrm{classic}$ equation}, assuming an equal RTT.
Here, L4S and
classic flows compete fairly in
one UE's DRB, while the fairness among different 
UEs is achieved by the MAC scheduler, working orthogonally with \sysname{}.

\subsection{RAN Queue Occupancy Prediction}
\label{s:queue_prediction}
\label{s:design:occupancy}

\revise{Here we introduce the design details of \sysname{}. To begin with, \sysname{} re-purposes the existing F1-U messages in the base station.}
\sysname{} utilizes the F1-U 
feedback messages from the RLC to the PDCP entity (\S\ref{s:f1u_feedback})
to estimate the dequeue rate, as well as the estimation errors based on a \emph{shadow table}
described next (\S\ref{s:packet_table}).
\revise{Then, \sysname{} uses the egress rate estimations to predict the sojourn time (\S\ref{s:design:queue_delay_prediction}), and based on the estimations, makes the marking decisions. }

\begin{figure}
\begin{minipage}[b]{0.52\linewidth}          
    \centering
    \includegraphics[width=\linewidth]{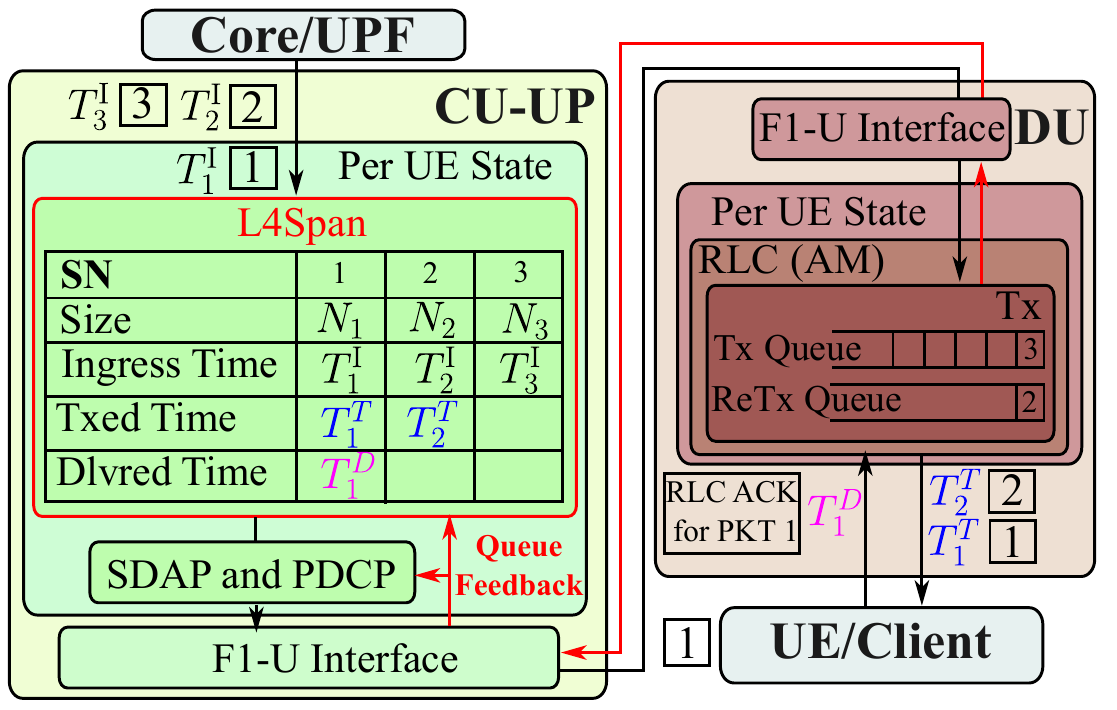}
    \caption{\sysnames{} packet profile table based on the RAN feedback information. 
    $T^{\{I, T,D\}}_{i}$ denotes the $i^{\mathrm{th}}$ packet's 
    \textbf{I}ngress, \textbf{T}ransmission, and \textbf{D}elivery timestamps.
    RU, MAC and PHY components are not shown.}
    \label{fig:pkt_profile_queue}
\end{minipage}
\hfill
\begin{minipage}[b]{0.44\linewidth}            
\centering
    \includegraphics[width=\linewidth]{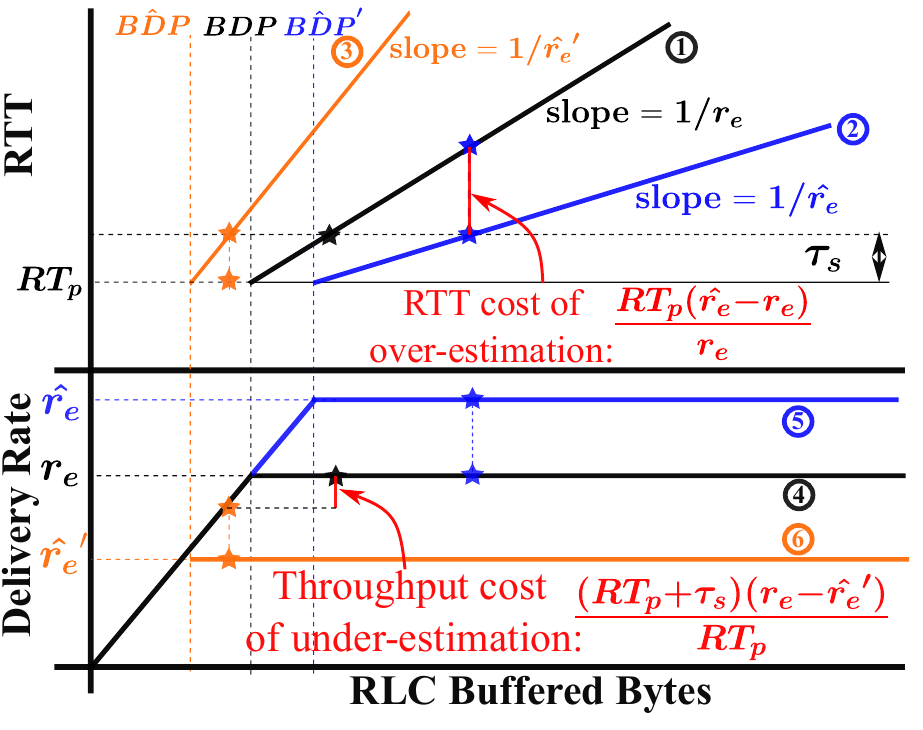}
    \caption{Cost of errors. \sysname{} marks the packet using the estimated egress rate and a queuing delay threshold $\tau_{s}$. Stars ($\star$) mark the ideal operation points with the estimated egress rate.} 
    \label{fig:error_analysis}
\end{minipage}
\end{figure}

\subsubsection{F1-U Feedback Loop}
\label{s:f1u_feedback}
\revise{The RAN's behavior and the feedback information vary with the
RLC mode configured in each DRB: there are two RLC modes 
for user data delivery: RLC \textit{Acknowledged Mode} AM and 
RLC \emph{Unacknowledged Mode} (UM), where 
the latter omits retransmissions and the 
retransmission queue, and doesn't provide delivery time feedback.}
The purpose of the F1-U feedback loop is to expose the RAN's congestion 
queue to the upper \sysname{} layer. 
There are three types of 3GPP-standard~\cite{3gpp_ts38425_2025} F1-U messages:
\textsf{downlink user data}, \textsf{downlink data delivery status},
and \textsf{assistance information}.
To make our design as general as possible, we 
use only the mandatory fields of the \textsf{downlink data delivery} message:
the highest transmitted PDCP \textit{sequence number} (SN) and 
the highest delivered PDCP SN.
The RLC triggers a timestamped feedback message when it 
transmits the PDCP SDU down to the MAC/PHY, and when it
receives an RLC ACK that indicates SDU delivery from the UE in 
RLC AM. As we
explain next, \sysname{} handles both AM and UM.

\subsubsection{Packet Profile Table: Timekeeping.} 
\label{s:packet_table}

The \sysname{} packet profile table
tracks packets' progress through the RLC in order to 
predict queue occupancy.  \Cref{fig:pkt_profile_queue} provides a
running example, packets with sequence numbers 
1, 2 and 3 go through different procedures with a timestamp on each: 

\noindent
\textbf{1)} Entry to CU-UP \sysname{} layer, recording of 
\emph{ingress timestamps} $T^{\text{I}}_1, T^{\text{I}}_2$, and $T^{\text{I}}_3$. 

\noindent
\textbf{2)} MAC transmits Packets~1 and~2,
RLC layer reports the status to PDCP and \sysname{}, with recorded \emph{transmission timestamps} ($T^{\text{T}}_1, T^{\text{T}}_2$).

\noindent
\textbf{3)} RAN delivers Packet~1 
to UE, which sends an RLC acknowledgment. 
RLC sends the highest delivered sequence number
and \emph{delivery timestamp} 
$T^{\text{D}}_1$ to the PDCP and \sysname{}.


\parabreak{}From the packet profile table, we 
calculate the actual queuing delay of each packet by 
subtracting the transmitted time and the ingress time 
(\textit{i.e.}, $T^{\text{T}}_1 - T^{\text{I}}_1$ 
and $T^{\text{T}}_2 - T^{\text{I}}_2$), 
and retransmission delay 
(\textit{i.e.}, $T^{\text{D}}_1 - T^{\text{T}}_1$), 
if the DRB is configured with RLC AM.
To ensure \sysname{} works in both AM and UM, we 
estimate queue status with the 
intersection of the feedback information for 
both RLC modes---the packet transmit times in the feedback information.
However, the calculated queuing delay reflects the
status of the previously transmitted packets (Packets~1 and 2 in 
\cref{fig:pkt_profile_queue}), while ideally we want to estimate 
the queuing delay of the current standing queue (Packets~3 and later 
in \cref{fig:pkt_profile_queue}), and set the feedback accordingly.
To do that, we need to predict the RAN egress rate.



\subsubsection{Sojourn Time Prediction and Error Estimation}
\label{s:design:queue_delay_prediction}
To predict the RLC queue's packet sojourn time,
\sysname{} monitors and 
predicts the queue egress rate within a short time window.
Upon receiving RAN feedback, 
\sysname{} remembers the highest transmitted packet sequence number, 
say $k$, then calculates the current egress rate of the DRB using 
a window with time duration $\tau_c$:
\begin{equation}
    r^{T}_{k} = \left(\sum\nolimits_{i \in \{i | T^T_{k} - \tau_c < T^T_{i} \le T^T_{k} \} } N_i \right)/\tau_c,
\end{equation}
where $N_i$ is the packet size and $T^{\text{T}}_i$ is the transmission 
time of the $i^{\text{th}}$ packet.
We choose $\tau_c$ to be half a pre-set \textit{channel coherence time} \revise{(the time that the channel response is the same)}, \revise{where the $\tau_c$ is measured by \cite{wang_doppler_2018} in a driving scenario, and can cover most of the daily usage scenario}. 
Then we use another $\tau_c$-long window to calculate the average egress rate, as the smoothed egress rate:
\begin{equation}
    \hat{r}_{e} = \left(\sum\nolimits_{i \in \{i | T^T_{k} - \tau_c < T^T_{i} \le T^T_{k} \}} r^{T}_{i} \right)/ \left|\{i | T^T_{k} - \tau_c < T^T_{i} \le T^T_{k} \}\right|,
\end{equation}
where $\left|\cdot\right|$ denotes the number of elements.
In this way, all the packets used for egress rate estimation are transmitted within $2\tau_c$, the channel coherence time, during which the wireless channel is considered stable.
With $N_{queue} = \sum\nolimits_{i \in \{ i | T^I_{i} > T^I_k \} }N_i $, the estimated sojourn time thus becomes:
\begin{eqnarray}
    \hat{\tau}^{r}_{\text{s}} &=&\frac{ N_{queue}}{\hat{r}_e}.
\end{eqnarray} 
\revise{\sysname{} uses the estimated sojourn time for RTT estimation and marking decision making (\S\ref{s:mark_strategy}). }

Here we analyze the cost of egress rate estimation errors if the generic DualPi2 strategy is adopted.
\cref{fig:error_analysis} shows a snapshot of the queue, where the $r_e$ is the dequeue rate and $RT_p$ is the round-trip propagation time.
Ideally, DualPi2 operates on the line \textcircled{1} and \textcircled{4} in the figure, achieving sojourn time and throughput targets.
However, when it over-estimates the egress rate ($\hat{r}_e > r_e$, line \textcircled{2} and \textcircled{5} in the figure), causing an under-estimated sojourn time and further more queued bytes, the RTT would increase by $\frac{RT_{p}(\hat{r}_e - r_e)}{r_e}$, marked by the blue star on the line \textcircled{1}. 
While if the egress rate is under-estimated ($\hat{r}_e' > r_e$, line \textcircled{3} and \textcircled{6} in the figure), resulting in overaggressive markings, the throughput would decrease by $\frac{(RT_{p} + \tau_s)(r_e-\hat{r}_e')}{RT_{p}}$, marked by the orange star on the line \textcircled{4}.

To adapt to the volatile wireless throughput, \sysname{} considers the egress rate estimation errors for the marking strategy. 
Based on our evaluation in \S\ref{s:eval:micro_benchmark}, the egress rate estimation error ($e_{r_e} = \hat{r}_e - r_e$) follows the Gaussian distribution with a near-zero mean. 
Given $\hat{r}_e \approx \mathbb{E}(r_e)$ over the latest estimation window, $e_{r_e}$ has the similar variation as $r_e$.
We estimate $e_r$'s standard deviation using the groundtruth dequeue rate's standard deviation within the last $\tau_c$-long window ($\hat{e}_{r_e}$), getting $e_{r_e} \sim N(0, \hat{e}_{r_e}^2)$. We take the sojourn time and egress rate error estimations for the marking decision  (\S\ref{s:mark_strategy}).
ML based forecasters \cite{perez_aichronolens_2025,gholian_deexp_2025,basit_predictability_2024} could potentially improve the prediction accuracy, but are unsuitable for microsecond-level packet and RAN feedback processing in \sysname{} (see \S\ref{s:micro:proc_time}).

\subsection{Feedback Short-circuiting}
\label{s:ack_marking}
\label{s:design:shortcircuit}

\begin{figure}
\begin{minipage}[b]{0.43\linewidth}          
    \centering
    \includegraphics[width=\linewidth]{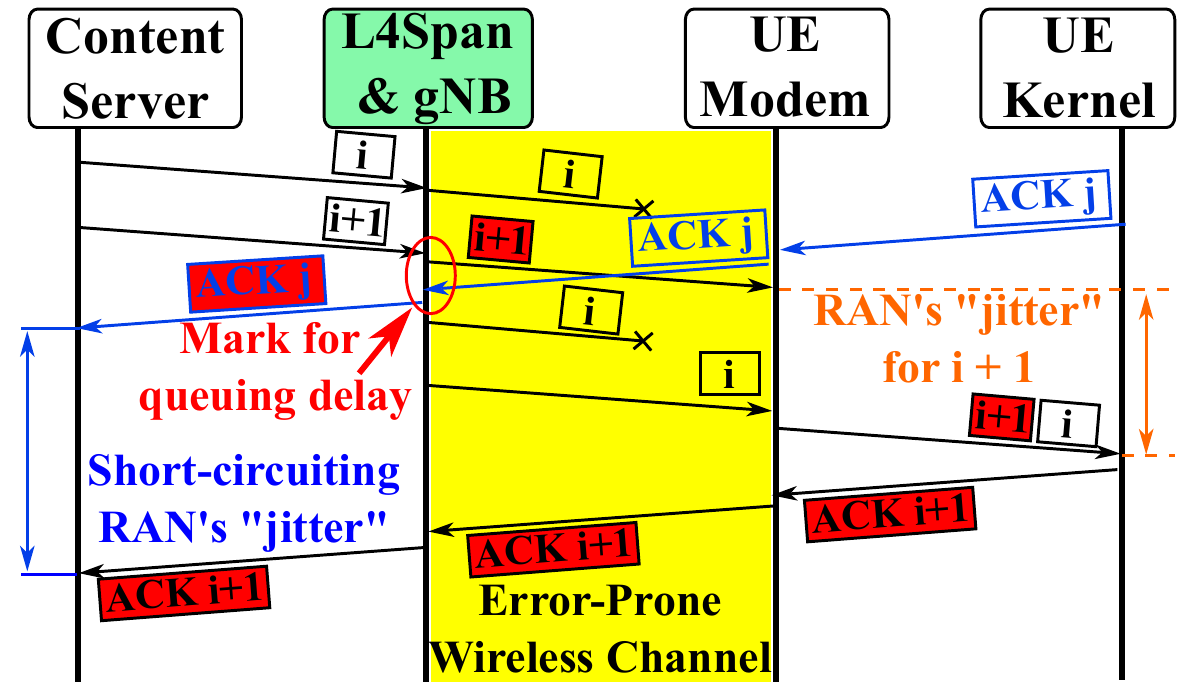}
    \caption{\sysnames{} can short-circuit the feedback by modifying the ACK instead of the IP packet for timely adjustments.}
    \label{fig:feedback_short-circuiting}
\end{minipage}
\hfill
\begin{minipage}[b]{0.53\linewidth}            
\centering
    \centering
    \includegraphics[width=\linewidth]{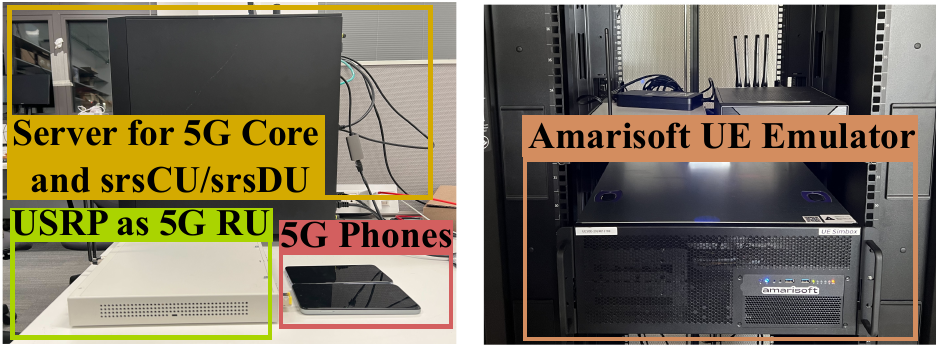}
    \caption{Hardware used in \sysname{} implementation. One desktop server runs the 5G Core, srsCU and srsDU, and the USRP is served as the RU. We use both real phones and Amarisoft UE emulator as the UEs for the evaluation.}
    \label{fig:hardware}
\end{minipage}
\end{figure}

In the L4S architecture~\cite{briscoe_rfc_2023}, the sender relies on 
feedback from the client, experiencing the entire RTT, but
the 5G network delays packet delivery further, 
as shown in \cref{fig:feedback_short-circuiting}.\footnote{The MAC/PHY and
RLC ARQ delay the transport block by eight
ms~\cite{xie_pbe-cc_2020,wan_nr-scope_2024,yi_athena_2024}, and up to 100~ms 
\cite{yi_automated_2025,3gpp_specification_2025}, respectively, and the 
scheduling delay approaches tens of milliseconds as UE numbers increase (\S\ref{s:eval:transport}).}
To prevent RAN jitter and delays from delaying the feedback, we propose to short-circuit the RAN by modifying the uplink feedback (\cref{fig:feedback_short-circuiting}), inspired by prior work \cite{meng_achieving_2022}.
L4S adopts several feedback format variations;
\sysname{} supports all of them: \textbf{\textit{1)}} AccECN~\cite{briscoe_more_2022} 
uses ECN-Echo and option field in the ACK TCP header (Prague~\cite{briscoe_implementing_2018} or BBRv2 \cite{cardwell_bbr_2019}), \textbf{\textit{2)}} classic ECN uses~\cite{floyd_addition_2001} 
ECN-Echo in the ACK's TCP header as feedback \cite{baker_ietf_2015}, and 
\textbf{\textit{3)}} data inside the data payload 
(SCReAM~\cite{ericssonresearch_ericssonresearchscream_2025} or QUIC~\cite{google_quichequichequiccorecongestion_controlbbr2_sendercc_nodate}).

\parahead{Short-circuiting the RAN.}
\sysname{} first classifies between AccECN and classic ECN, by checking the TCP header's option field, used by AccECN.
If AccECN protocol is adopted, the feedback bits in the TCP header contain the number of packets marked as CE and bytes marked as CE, ECT-1, and 
ECT-0~\cite{briscoe_more_2022}.
Upon marking, \sysname{} tentatively marks a downlink packet by storing the number of packets and bytes of all three types.
Instead of changing the packet's ECN field, \sysname{} uses the latest ratio to split the ACKed bytes, and updates the ACK's TCP header accordingly~\cite{briscoe_more_2022}.
For classic ECN, \sysname{} keeps marking the ECN-Echo field in the ACK header upon decision, until the CWR flag is set in the downlink packets \cite{matthews_stateful_2011}. 
\sysname{} serves as a "bookkeeper" for the client, and short-circuits the RAN's complex jitters, resulting in a better performance in RTT (\S\ref{s:eval:short-circuiting}).

\parahead{Fallback to Mark Downlink Packet.}
For the UDP flow, the feedback could be encrypted (QUIC \cite{langley_quic_2017}) or presented in customized format (SCReAM \cite{johansson_self-clocked_2014,ericssonresearch_ericssonresearchscream_2025} and UDP Prague \cite{l4steam_l4steamudp_prague_2025}), \sysname{} fallbacks to mark downlink packets' IP ECN field. 
Furthermore, \sysname{} can also be configured to drop packets selectively instead of ECN marking to provide feedback to non-ECN flow senders/

\section{Implementation}
\label{s:impl}

We implement \sysname{} on top of srsRAN~\cite{system_srsran_2023} with approximately 2,000 lines of C++ code (excluding reused code).
The \sysname{} layer is in the CU-UP and inside the UE context, meaning that
during the PDU session creation for each UE upon its initial connection, the RAN creates one
\sysname{} entity and connects \sysname{} to the GTP-U interfance and lower SDAP layer.
\revise{When a packet is sent from the 5G core to the CU, \sysname{} checks the packet's ECN field to identify its type. 
Then, \sysname{} makes the marking decision with the information from the lower layers and the estimated egress rate and sojourn time.
}
In the meantime, \sysname{} keeps a copy of the QoS flow and DRB mapping table, 
which will be used for packet profile table creation and feedback short-circuiting.
For the feedback information, we reuse the RAN's message by spawning a dedicated thread in the F1-U interface to call \sysname{}'s feedback handler function upon receiving messages from the RLC/DU, which we use 
to predict the queue occupancy and make marking decisions.
For the coherence time $\tau_c$, we use a measurement in \cite{wang_doppler_2018} 
with a 3.5~GHz base station and a moving UE with 70~km/h speed ($\tau_c = 24.9$~ms), 
covering most of the sub-6 GHz scenarios  (evaluated in \S\ref{s:eval:micro_benchmark}), 
as the higher the center frequency and the faster the UE, the shorter the coherence time.
As for the downlink packet processing, \sysname{} \revise{marks the packets on its ECN field if using downlink marking (for UDP or QUIC flows), then it recalculates the CRC checksum on its IP header.}
\revise{For the uplink feedback short-circuiting, \sysname{} updates the TCP header's Nonce, CWR, and ECN-Echo and TCP options for the AccECN feedback. 
\sysname{} then calculates and updates the TCP checksum.}
%

\section{Evaluation}
\label{s:eval}
We proceed top-down, first introducing our overall methodology, then evaluating it with end-to-end congestion control performance improvements, and we evaluate \sysname{}'s in micro-benchmarks.

\subsection{Methodology}
\label{s:eval:methodology}
We evaluate \sysname{} in the open-source srsRAN 5G network and Open5GS~\cite{open5gs_httpsopen5gsorg_nodate} as the 5G Core, as shown in~\cref{fig:hardware}.
We run srsCU and srsDU on a desktop machine, where \sysname{} is integrated into the CU.
The cell is on a TDD band n78 with 30kHz subcarrier spacing, and the cell's center frequency is 3750~MHz with 20~MHz bandwidth, yielding a 40~Mbits/s capacity. 
We use two types of UEs to generate traffic in the RAN -- \textbf{\textit{1)}} commercial 5G phones, and \textbf{\textit{2)}} a production grade test equipment -- Amarisoft UE emulator, which can emulate up to 64 UE over-the-air and emulate different channels. 
\revise{Senders are two Microsoft Azure instances with ping times of 38ms and 106ms.}

To begin with, we evaluate the following congestion control schemes' performance in \sysname{}:
\begin{itemize}
    \item Prague~\cite{briscoe_implementing_2018}. TCP Prague is implemented for the L4S architecture, where the client sends the feedback using TCP header fields with AccECN~\cite{briscoe_more_2022}.
    \item CUBIC~\cite{ha_cubic_2008}. CUBIC sender cuts its congestion window to a fixed ratio upon loss or CE.
    In steady state, it increases its congestion window following the cubic function.
    \item BBRv2~\cite{cardwell_bbr_2019}. BBRv2 includes the DCTCP (or L4S)-like congestion window adjustments upon receiving the AccECN signal. It proactively probes the bandwidth and RTT.
\end{itemize}
We also evaluate BBR~\cite{cardwell_bbr_2016} and Reno~\cite{paxson_tcp_1999} with \sysname{}. Please check the Appendix \ref{appd:bbr_and_reno} for their result.

Furthermore, We evaluate two application-level algorithms to analyze \sysname{}'s performance on interactive applications:
\begin{itemize}
    \item SCReAM~\cite{johansson_self-clocked_2014,ericssonresearch_ericssonresearchscream_2025}. SCReAM is a congestion control algorithm designed for webRTC~\cite{blum_webrtc_2021} over UDP and supports L4S congestion feedback.
    The receiver reads the number of CE bytes and feedback through RTP feedback.
    \item UDP Prague~\cite{l4steam_l4steamudp_prague_2025}. UDP Prague is designed for interactive application in the L4S architecture, where the receiver sends the feedback to the sender in the UDP payload.
\end{itemize}


\begin{figure*}
        \centering
        \subfigure[16 UEs, default RLC queue length, \revise{38 ms RTT}.]
        {\includegraphics[width=0.48\linewidth]{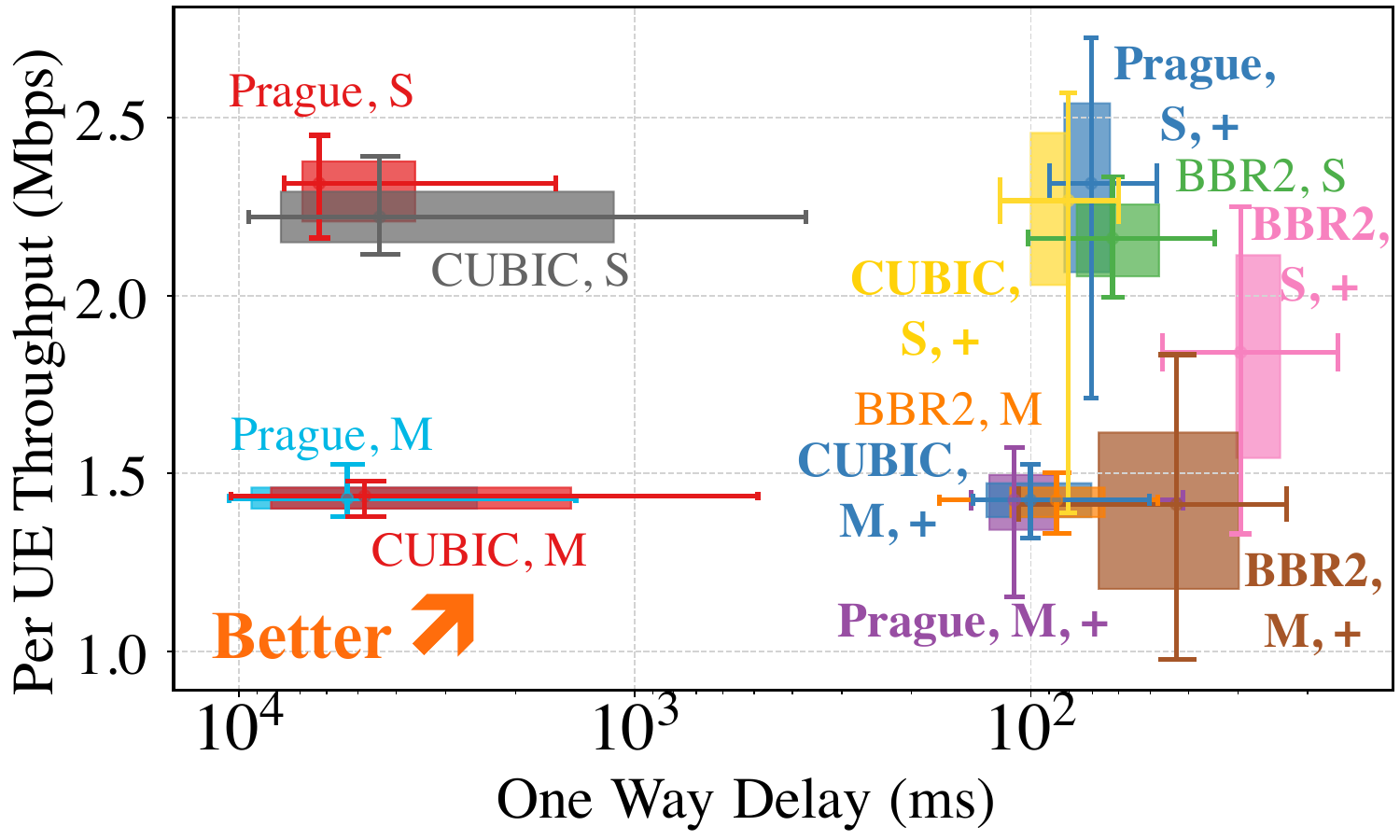}\label{fig:16ue_1000}}
        \hfill
        \subfigure[64 UEs, default RLC queue length, \revise{38 ms RTT}.]
        {\includegraphics[width=0.48\linewidth]{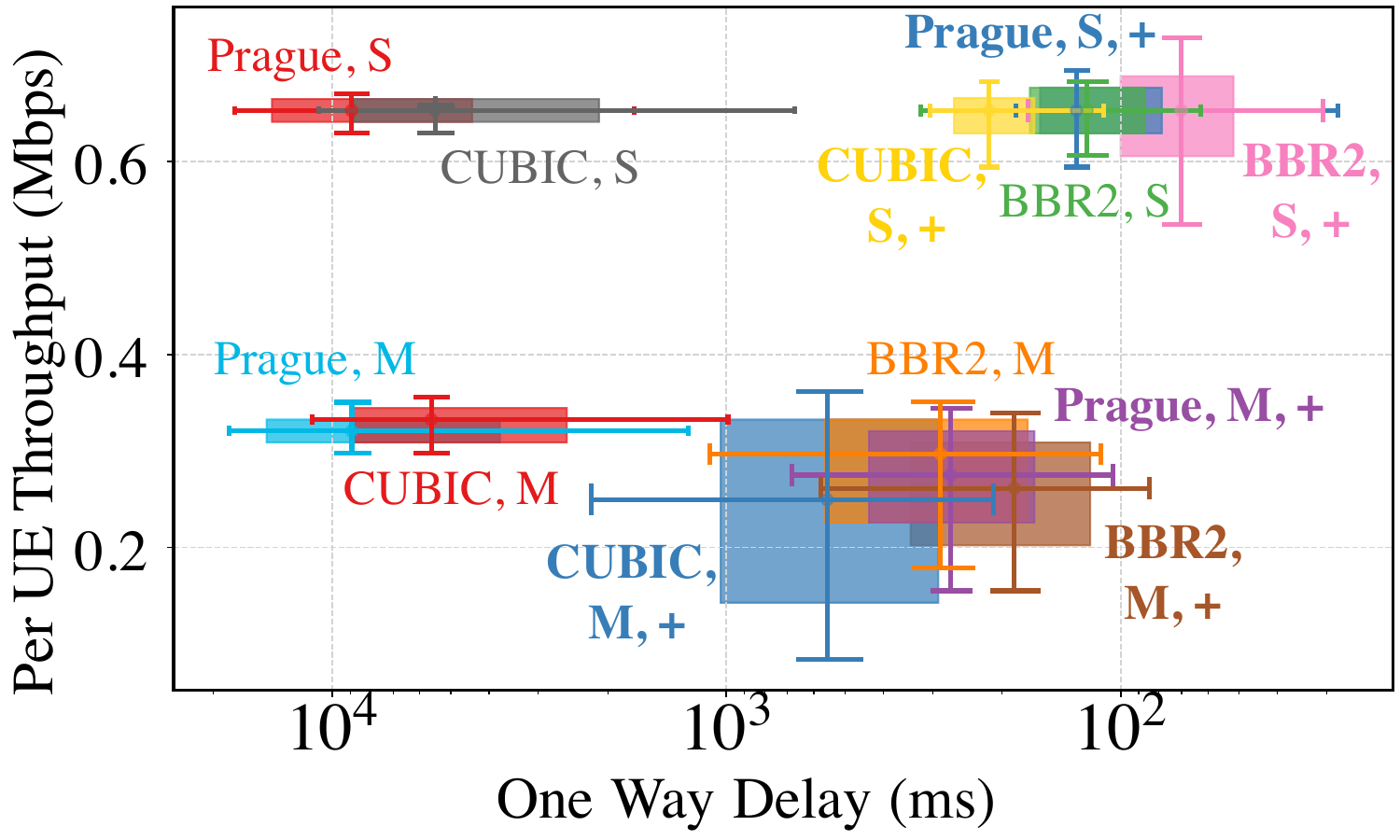}\label{fig:64ue_1000}}

        \subfigure[16 UEs, RLC queue length 256, \revise{38 ms RTT}.]
        {\includegraphics[width=0.48\linewidth]{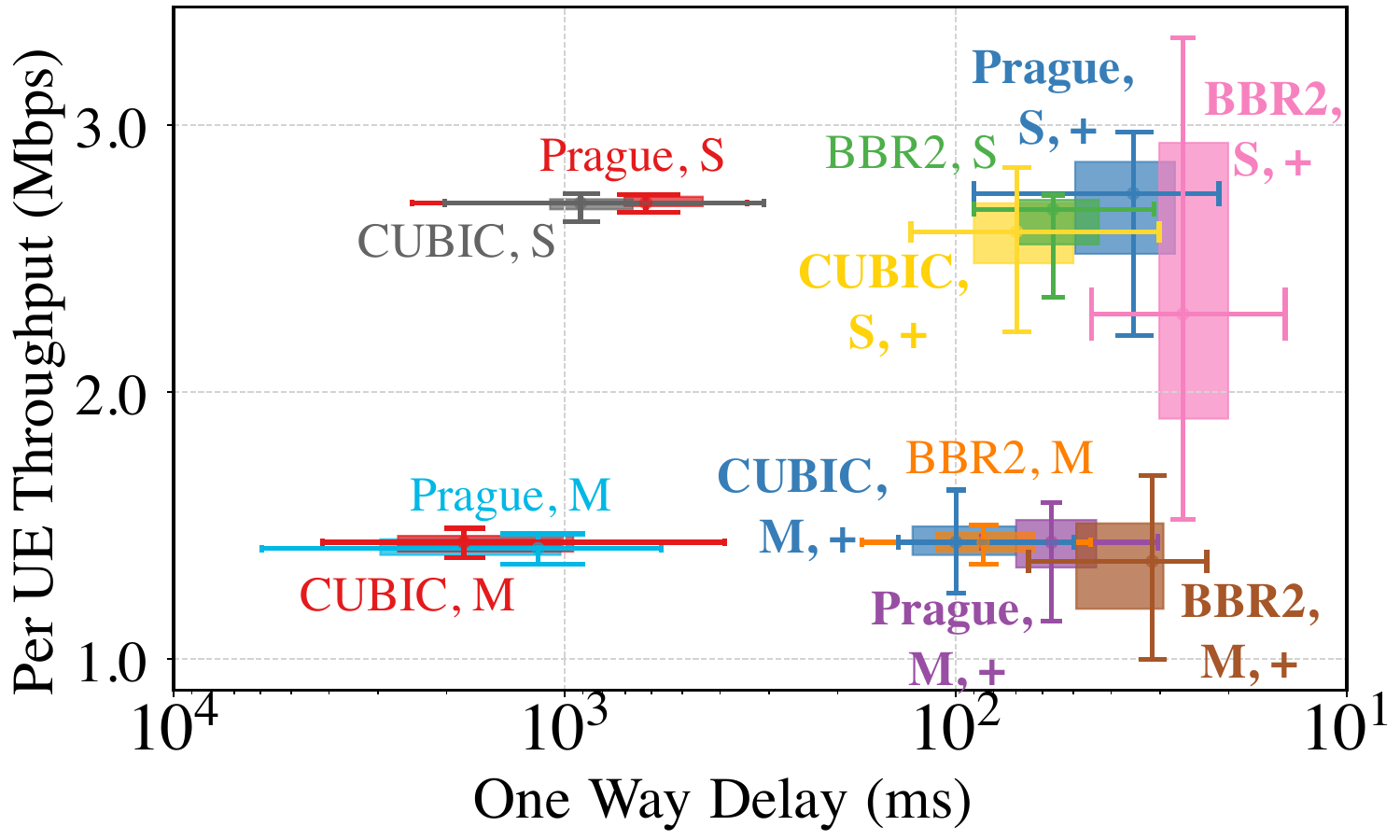}\label{fig:16ue_256}}
        \hfill
        \subfigure[64 UEs, RLC queue length 256, \revise{38 ms RTT}.]
        {\includegraphics[width=0.48\linewidth]{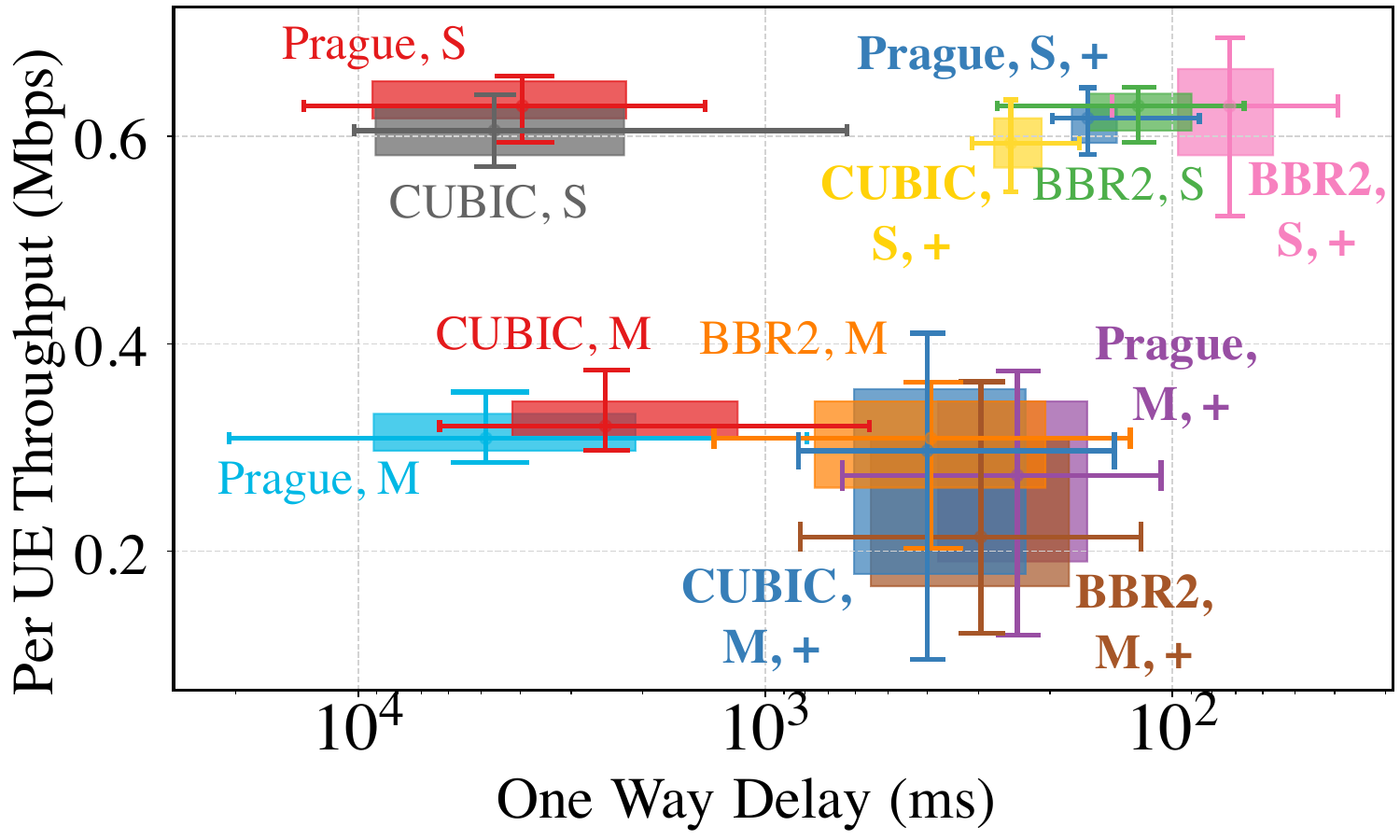}\label{fig:64ue_256}}

        \subfigure[\revise{16 UEs, default RLC queue length, 106 ms RTT}.]
        {\includegraphics[width=0.48\linewidth]{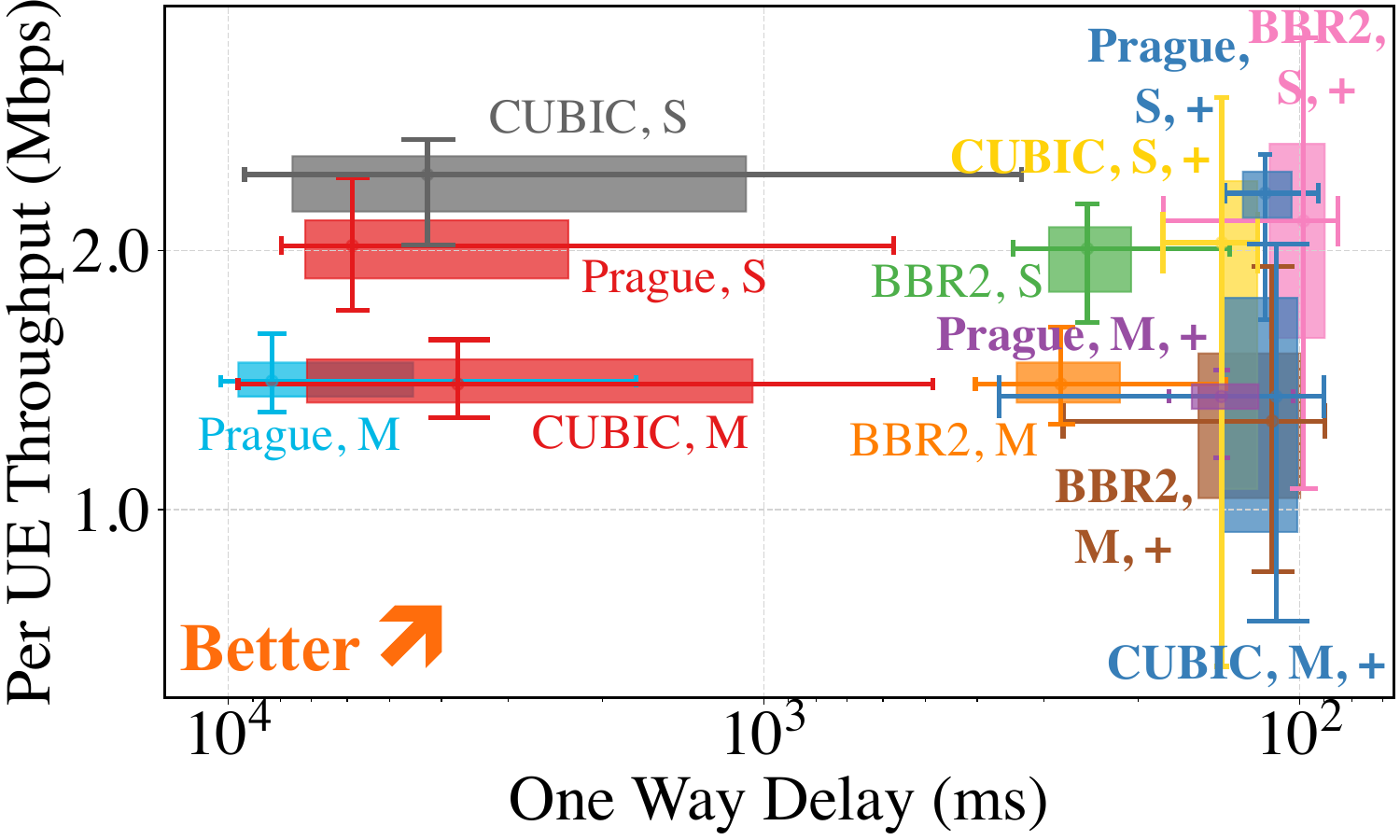}\label{fig:16ue_0916}}
        \hfill
        \subfigure[\revise{64 UEs, default RLC queue length, 106 ms RTT}.]
        {\includegraphics[width=0.48\linewidth]{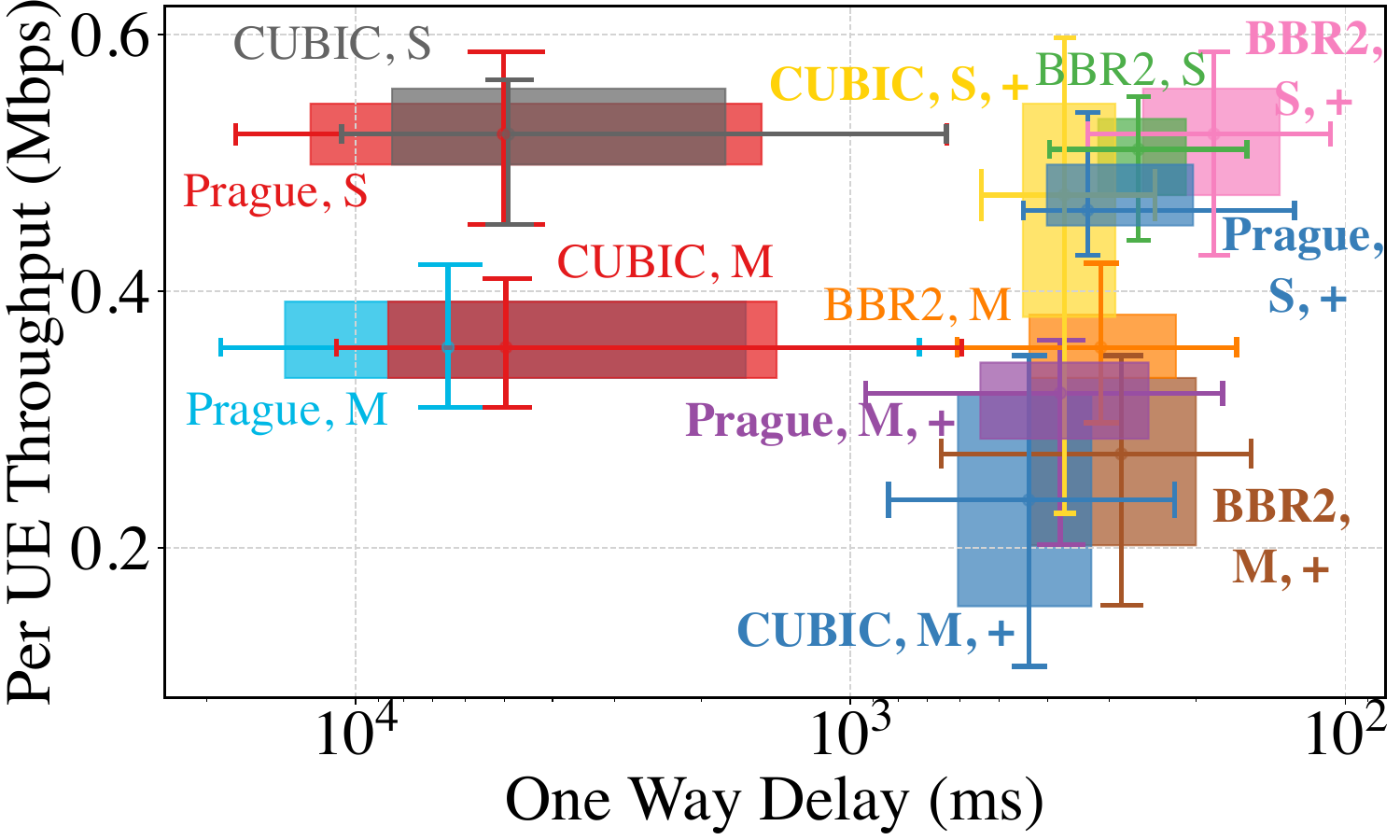}\label{fig:64ue_0916}}
        
        \subfigure[\revise{16 UEs, RLC queue length 256, 106 ms RTT}.]
        {\includegraphics[width=0.48\linewidth]{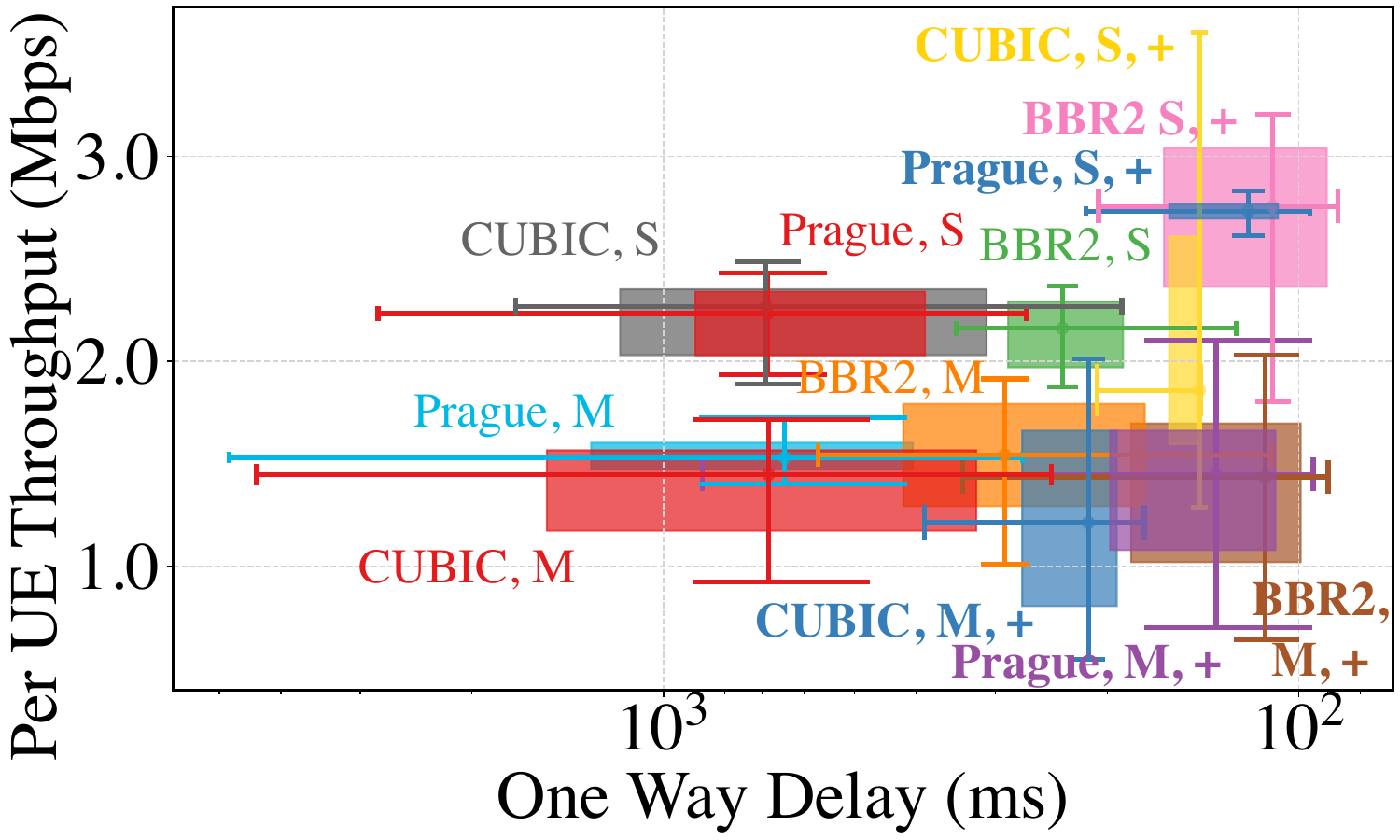}\label{fig:16ue_0916_256rlc}}
        \hfill
        \subfigure[\revise{64 UEs, RLC queue length 256, 106 ms RTT} .]
        {\includegraphics[width=0.48\linewidth]{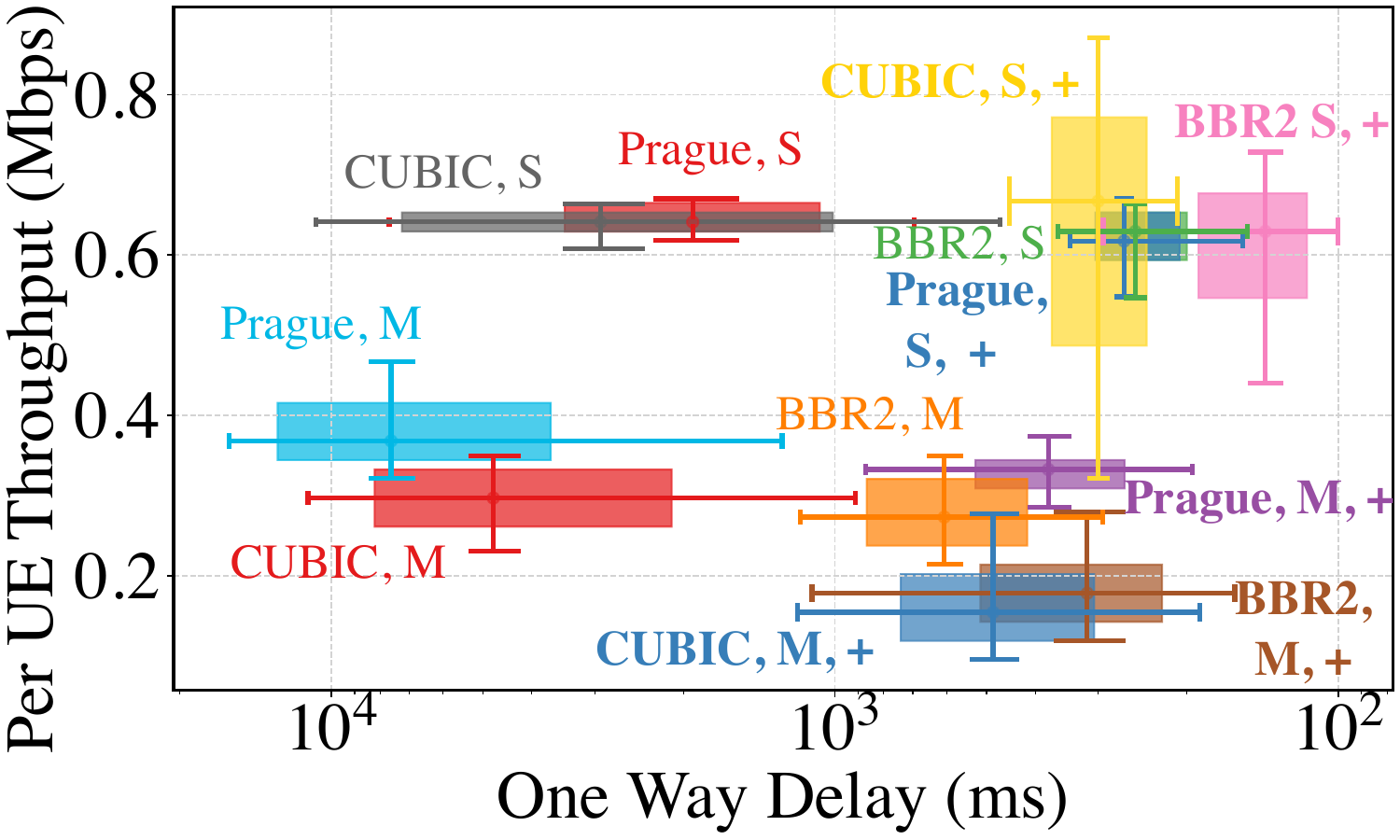}\label{fig:64ue_0916_256rlc}}
        \caption{\added{\sysname{} improves the performance of various congestion control algorithms in reducing RTT while maintaining throughput, under severely congested RAN and different channel conditions (S: Static, M: Mobile). $+$ indicates \sysname{} is deployed. \revise{Senders are Azure instances with uncongested RAN ping time marked in the caption.} Center point: median, box edges: 25th- and 75th-percentile, and whiskers: 10th- and 90th-percentile.}}
        \label{fig:l4s_tcp_end_to_end}   
\end{figure*}

\subsection{Transport Layer Performance}
\label{s:eval:transport}
In this evaluation, we evaluate: \sysname{}'s performance impact on \textbf{\textit{1)}} widely deployed TCP and \textbf{\textit{2)}} interactive video congestion control algorithms;  
\textbf{\textit{3)}} \sysname{}'s impact on the fairness of the RAN, \textbf{\textit{4)}} the short-circuiting design's effectiveness, and \textbf{\textit{5)}} when L4S and classic flows share the same DRB.

\subsubsection{Performance Impacts on TCP Congestion Control} 
Here we evaluate \sysname{}'s performance impact on congestion control algorithms, including Prague (L4S), BBRv2 (L4S), and CUBIC (classic).
%
%
We connect 16 and 64 UEs into the RAN through 
the Amarisoft UE emulator through the over-the-air channel, and all the UEs are 
doing concurrent TCP downloading through \texttt{iperf3}, making the RAN extremely congested.
We emulate different channels with the UE emulator, including static, pedestrian- and vehicular-speed channels, and combine the latter two as mobile.
We compare different RLC layer queue size settings, including the default srsRAN RLC queue length of 16384 and 256 SDUs. 
\cref{fig:l4s_tcp_end_to_end} shows the evaluation result.
Across all scenarios, \sysname{} can massively reduce the one-way delay and maintain a high throughput level.
In the default RLC queue setting \revise{and with the server of 38ms RTT}, \sysname{} reduces the median one-way delay of Prague by 98.87\% and 97.93\% in static and mobile channels for 16 UEs with a median throughput drop of less than 1\%.
As for BBRv2, \sysname{} reduces 
the RTT by 52.48\% and 52.27\% in static and mobile channels, with the cost of 
9.8\% and 0.08\% throughput drop. 
For CUBIC flow, \sysname{} can also reduce its one-way delay by 98.85\% and 97.11\% in static and mobile channels with little median throughput drop.
Similar trends can be found in 64 UEs and different RLC queue settings, \revise{as well as the 106ms RTT server, shown in \cref{fig:16ue_0916} to \cref{fig:64ue_0916_256rlc}}.
One thing to note is that 256 RLC queue reduces the one-way delay but is less effective than \sysname{}, as packet drops and retransmissions happen more frequently.

\begin{figure}
    \begin{minipage}[b]{0.48\linewidth}
    \includegraphics[width=0.98\linewidth]{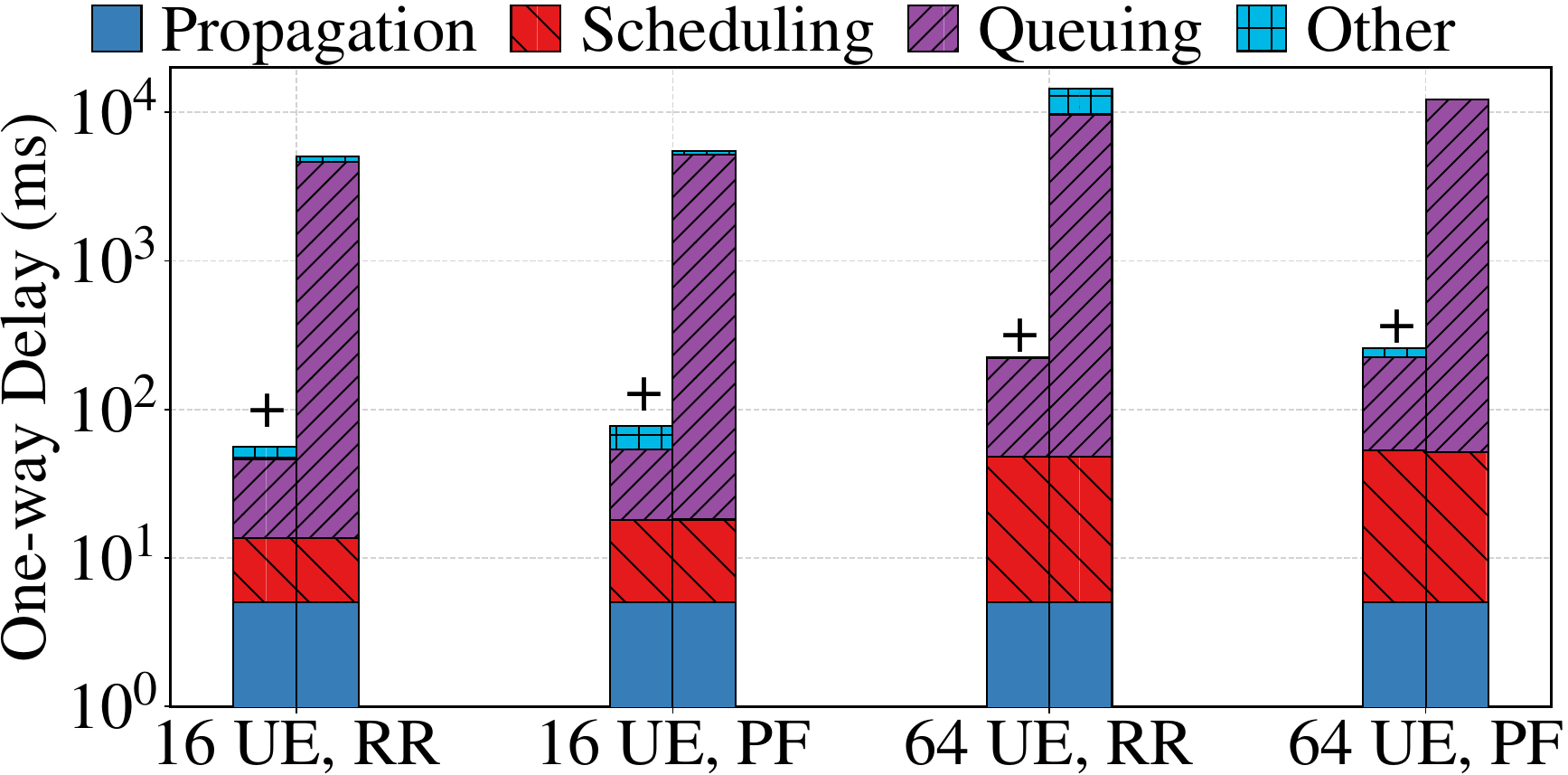}
    \caption{Average delay break-down for different scheduling algorithms in the srsRAN (RR: round-robin, PF: proportional fair) with different numbers of UEs in the RAN. + marks the \sysname{} is deployed.}
    \label{fig:delay_breakdown}
    \end{minipage}
    \hfill
    \begin{minipage}[b]{0.48\linewidth}
    \includegraphics[width=0.97\linewidth]{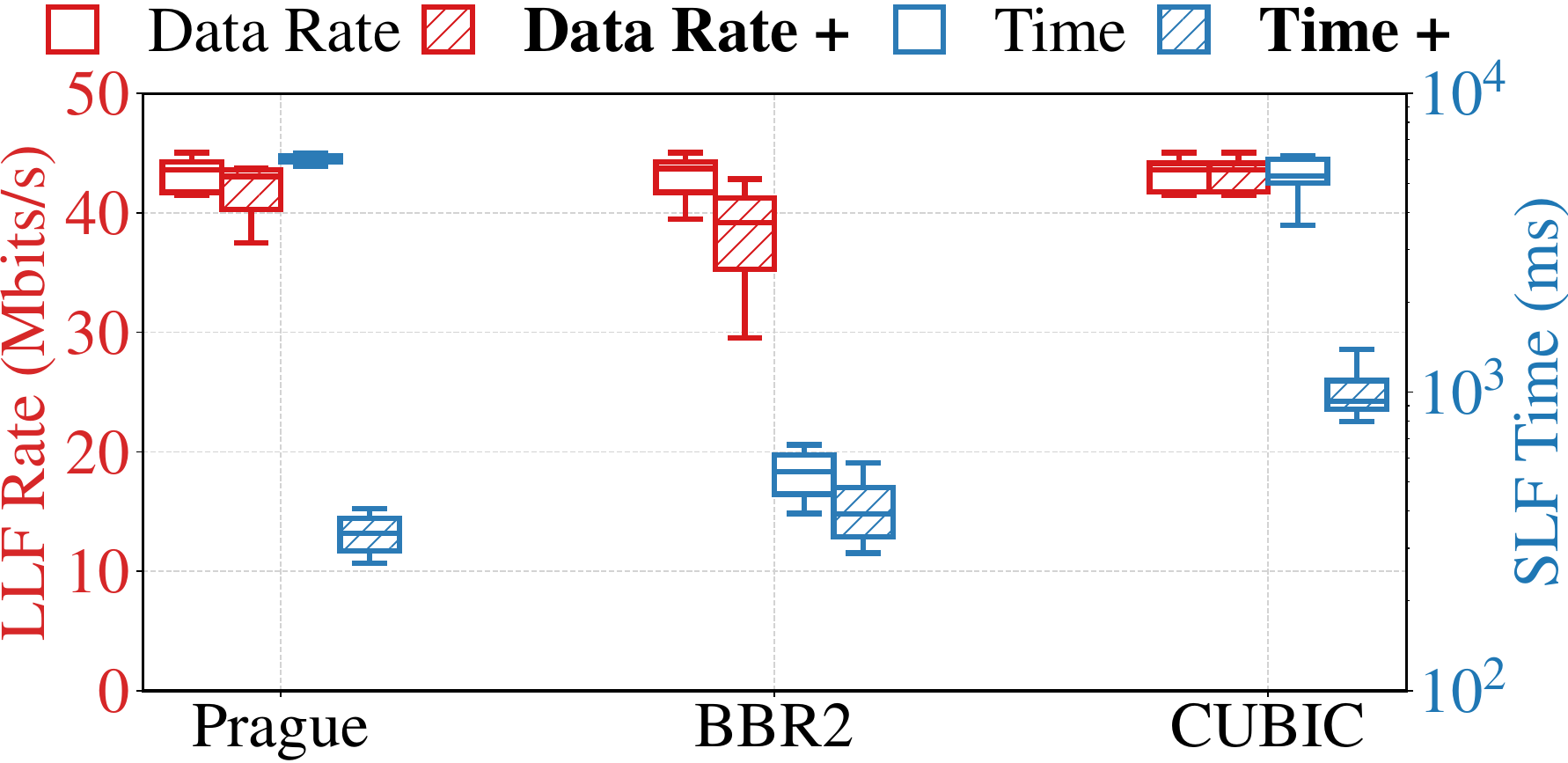}
    \caption{\sysname{} improves the finish time of the SLF (14 kilo-bytes) while keeping the LLF' throughput usage (center: median, box edges: 25th- and 75 th-percentile, whiskers: 10th- and 90th-percentile).}
    \label{fig:long_short}
    \end{minipage}
\end{figure}

Here we break down Prague's one-way delay, including propagation, scheduling, and queuing delays.
The propagation delay is calculated from the 
NTP-synchronized server and the 5G core's packet timestamp.
We breaks the sojourn time into queuing and scheduling delay.
The scheduling delay is the wait time for a packet in the queue head for the next transmission opportunity, collected from the MAC layer log.
Queuing delay is the time between when a packet enters the queue and it reaches the head of the queue, calculated by subtracting the scheduling delay from the sojourn time.
We compare two scheduling methods -- round robin and proportional fair, with 16 and 64 UEs in the RAN.
\cref{fig:delay_breakdown} shows the result of this evaluation, \sysname{} works with both scheduling schemes.

%
%

Applications host both \textit{long-lived flows} (LLF) and \textit{short-lived flows} (SLF) are prevalent, such as video on-demand, and web browsing, where the LLF delivers the content and SLF delivers the interactions.
Here we evaluate the impact of \sysname{} on such applications, by running two TCP flows within one commercial UE -- one LLF and one SLF, where the SLF's size is 14 kilo bytes.
\cref{fig:long_short} shows the result; \sysname{} can reduce the finish time of the SLF, while keeping the LLF's throughput.
For TCP Prague, \sysname{} reduces the SLF's finish time by \added{94.59\%}, with \added{10\%} of throughput drop for the LLF.
Similar performance improvements can be found in BBRv2 and CUBIC flows.
%
\begin{figure*}
        \centering
        \subfigure[\revise{Performance of Prague under \sysname{} and TC-RAN}]
        {\includegraphics[width=0.475\linewidth]{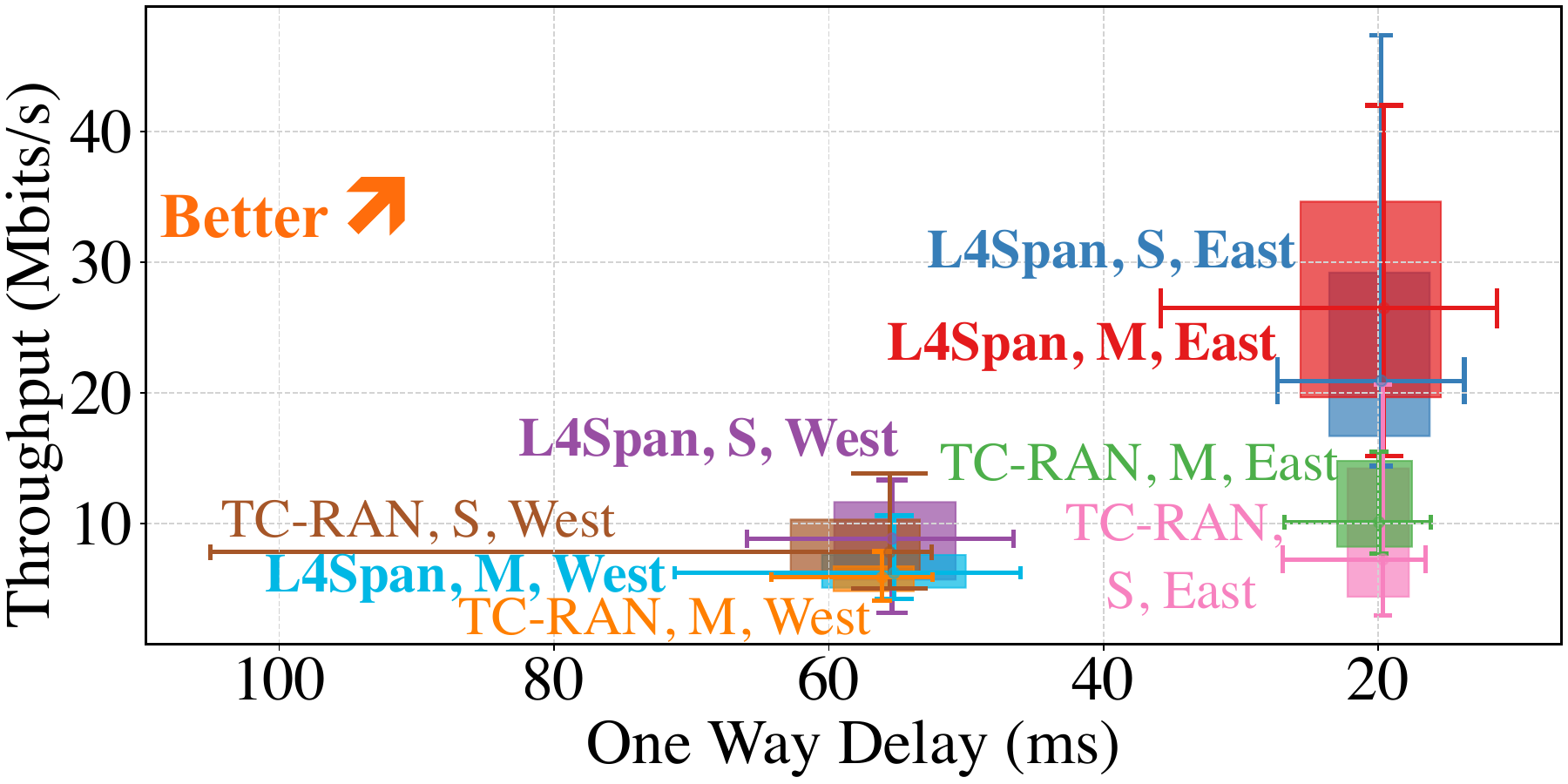}\label{fig:tcran_prague}}
        \hfill
        \subfigure[\revise{Performance of CUBIC under \sysname{} and TC-RAN}]
        {\includegraphics[width=0.475\linewidth]{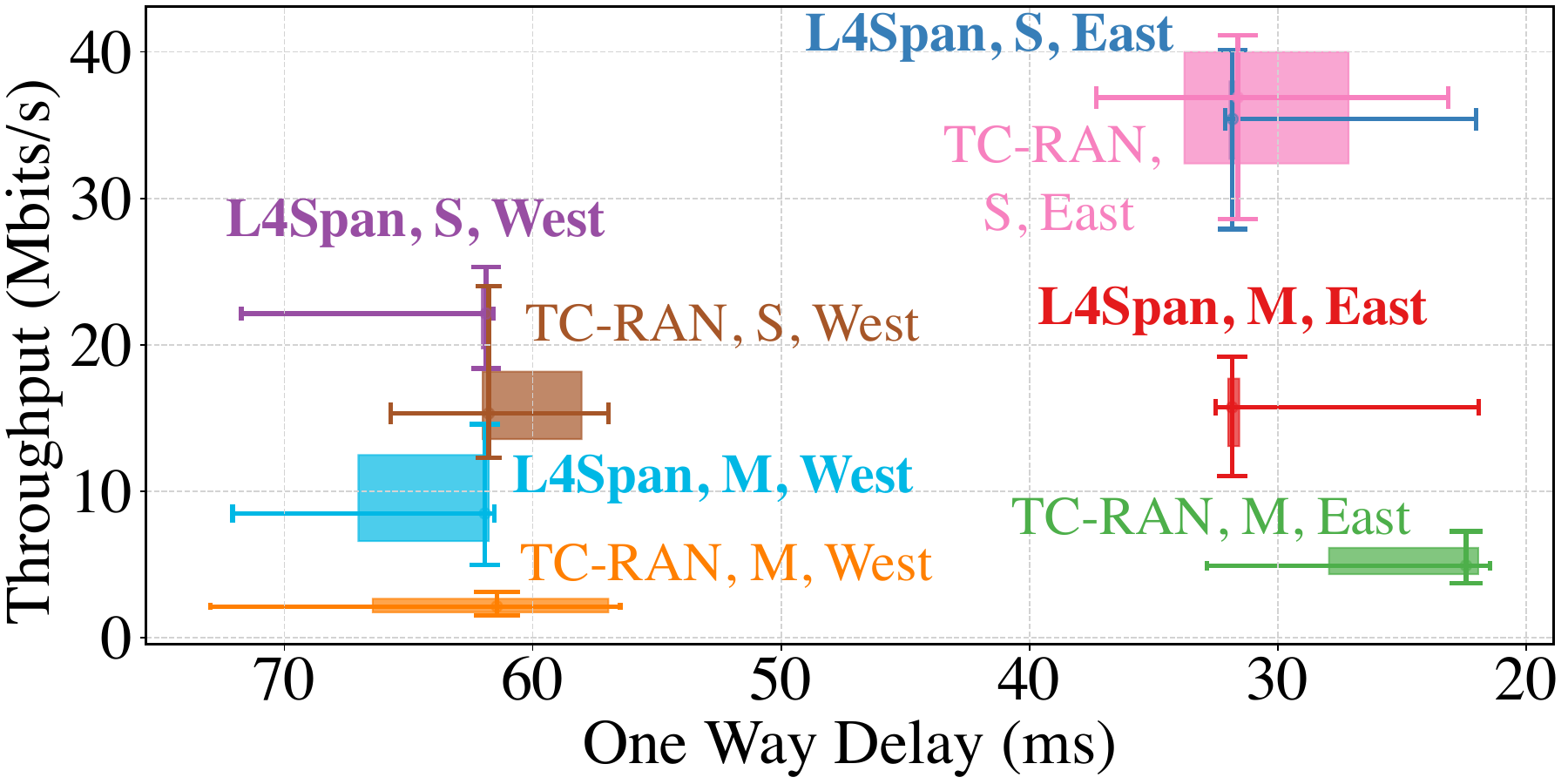}\label{fig:tcran_cubic}}
        \caption{\revise{Comparison between \sysname{} and TC-RAN(S: Static, M: Mobile, East: 38ms RTT, West: 106ms RTT).}}
        \label{fig:l4span_tcran}   
\end{figure*}

\subsubsection{\revise{Comparison with Baseline}}
\revise{Here we compare \sysname{} with the baseline method TC-RAN\cite{irazabal_tc-ran_2024}, by connecting one UE into both RANs.
We use the same RAN configurations (cell bandwidth, MCS table, etc.) in both \sysname{} and TC-RAN for a fair comparison. 
We evaluate the performance of TCP Prague and CUBIC, with default ECN-CoDel and CoDel configuration in the TC-RAN, respectively, and deploy senders in the two Azure instances mentioned above (west with 106ms RTT and east with 38ms RTT).
\cref{fig:l4span_tcran} shows the result of this evaluation.
\sysname{} achieves a similar delay performance with TCP Prague as TC-RAN, but achieves better throughput utilization with improvements of 148\% for the static channel and 6\% for the mobile channel, as shown in \cref{fig:tcran_prague}. 
Note that there is a latency spike in the TC-RAN, which happens at the beginning of the session due to the longer propagation delay, while \sysname{} mitigates this with the RAN short-circuiting design. 
For CUBIC, \sysname{} achieves a better channel utilization as it tries to align the throughput of the sender and RAN, while TC-RAN uses a fixed threshold in CoDel, leading to under-utilization.
}

\begin{figure*}
        \centering
        \subfigure[Static channel condition.]{
        \includegraphics[width=0.3\linewidth]{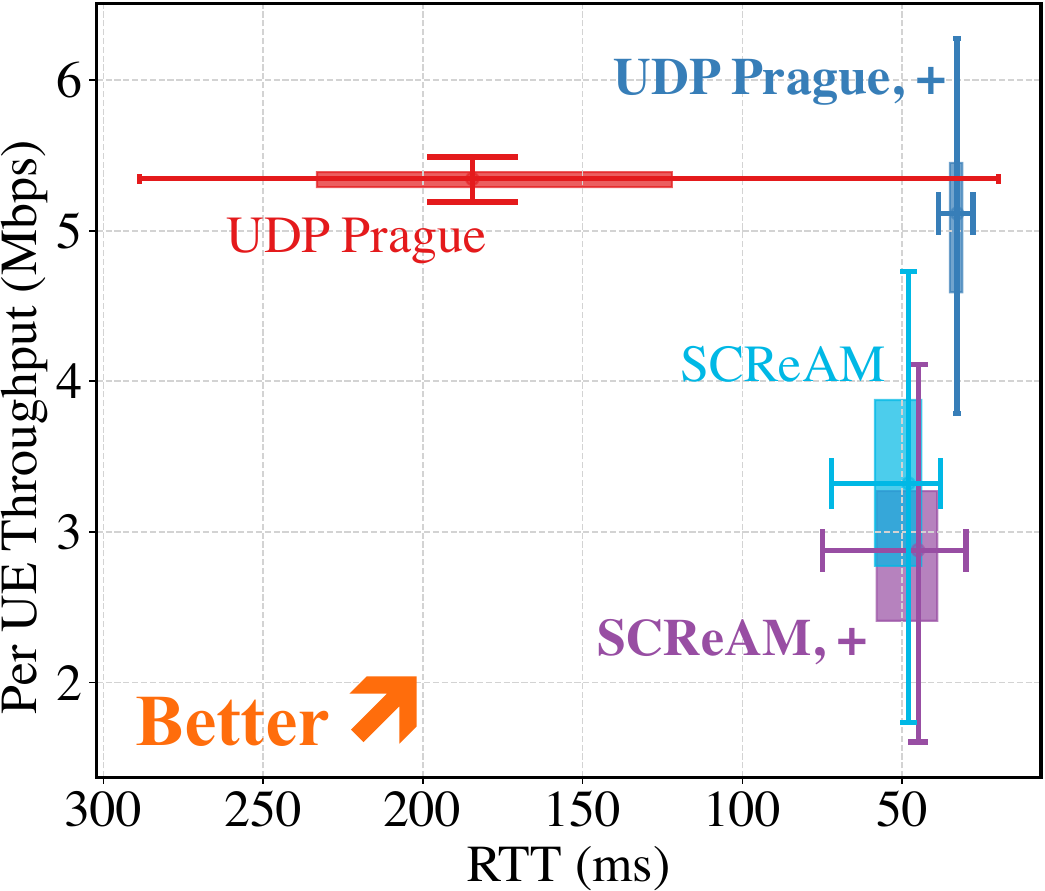}\label{fig:static_udp}}
        \hfill
        \subfigure[Pedestrian channel condition. ]
        {\includegraphics[width=0.3\linewidth]{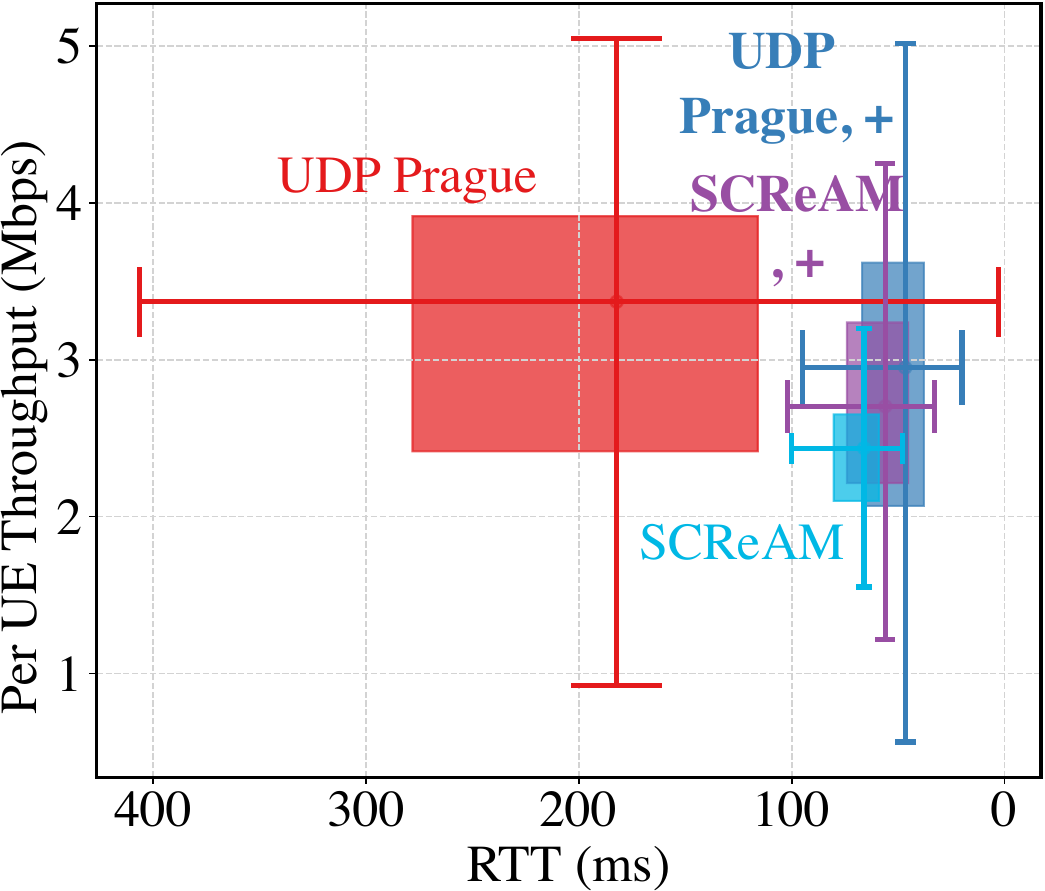}\label{fig:epa_udp}}
        \hfill
        \subfigure[Vehicular channel conditiopn.]
        {\includegraphics[width=0.3\linewidth]{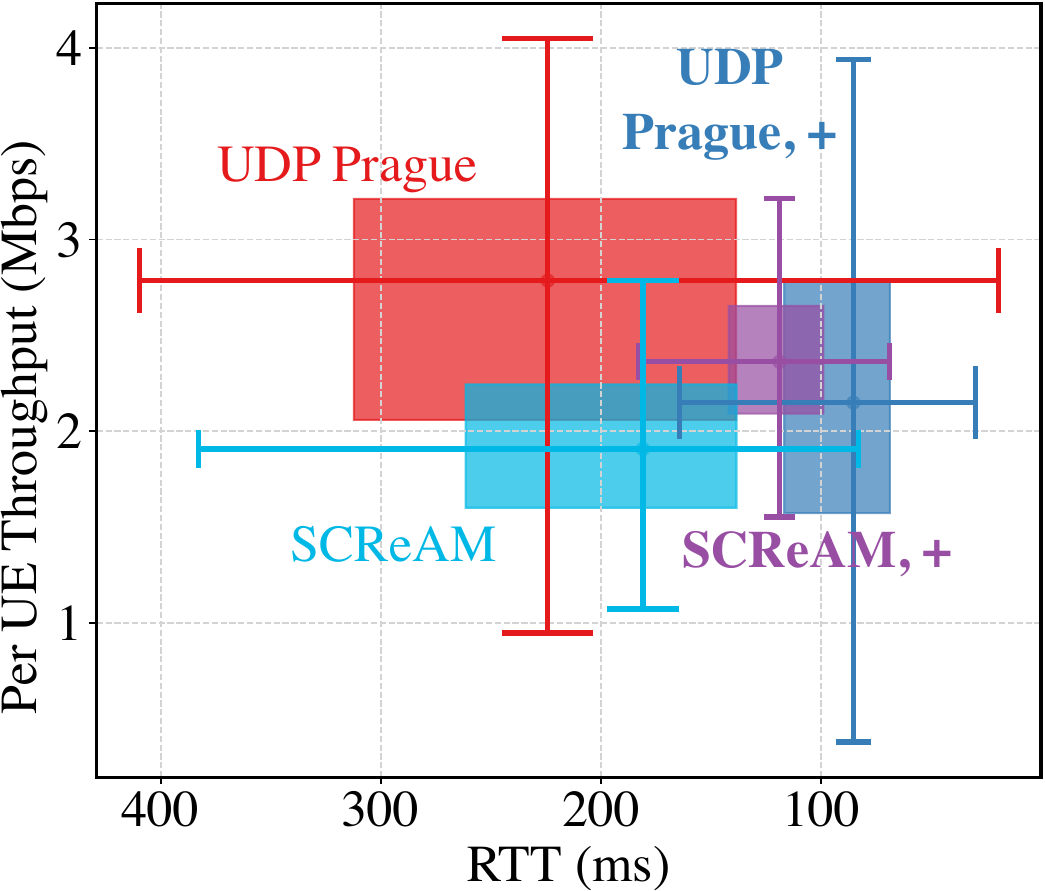}\label{fig:eva_udp}}
        \caption{SCREAM and UDP Prague's performance with \sysname{} under different channel conditions.}
        \label{fig:l4s_udp}   
\end{figure*}

\subsubsection{Interactive Application's Congestion Control} 
\label{s:eval:method:app_cong}
Here we evaluate the performance SCReAM and UDP Prague, designed for interactive video applications.
Both algorithms support L4S over UDP, thus \sysname{} disables
the RAN short-circuiting.
We connect 8 Amarisoft UEs into the network to perform concurrent downlink transmission from a local server, under different channel conditions.
\cref{fig:l4s_udp} shows the results, where \sysname{} improves the the RTT performance of both schemes under all channel conditions. 
Specifically, \sysname{} reduces the RTT of UPD Prague in static, pedestrian and vehicular channel by 76.33\%, 38.04\%, and 44.83\%, while slightly reduces its throughput by 5.64\%, 8.25\%, and 17.74\%.
For SCReAM, \sysname{} reduces its RTT by 12.60\%, 11.20\%, and 38.44\% in static, pedestrian and vehicular channel conditions, with a little bit higher throughput variations.


\subsubsection{Fairness}
\begin{figure*}
        \centering
        \subfigure[Three L4S Prague flows with similar RTT in \sysname{}.]{
        \includegraphics[width=0.24\linewidth]{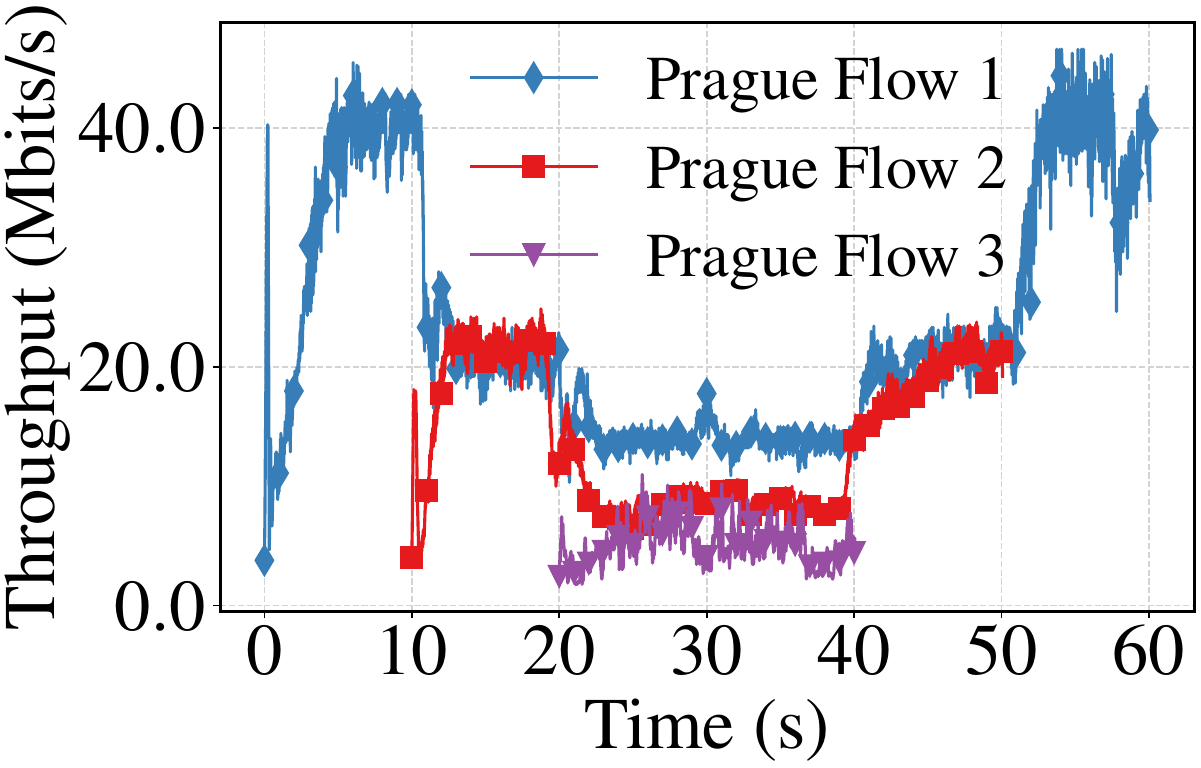}\label{fig:fairness_3l4s_similarrtt}}
        \hfill
        \subfigure[Three L4S Prague flows with distinct RTT in \sysname{}.]
        {\includegraphics[width=0.24\linewidth]{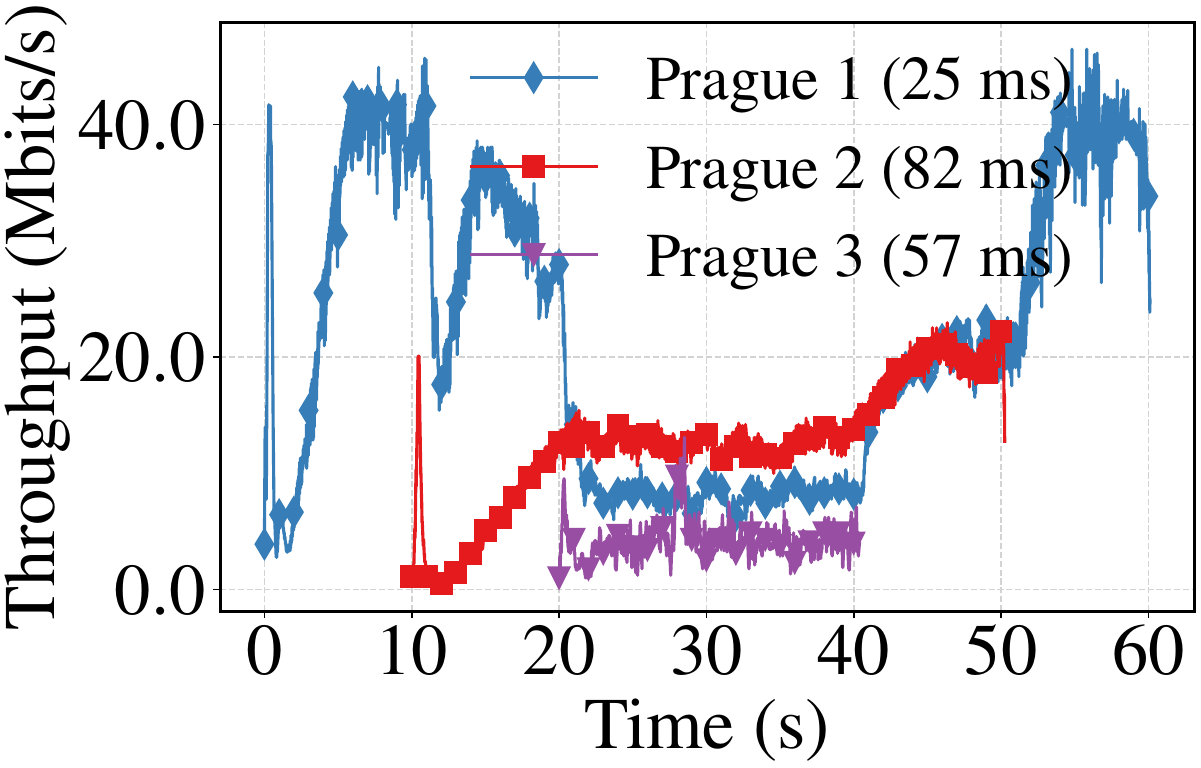}\label{fig:fairness_3l4s_diffrtt}}
        \hfill
        \subfigure[Two L4S Prague flows and one Cubic flow in \sysname{}.]{
        \includegraphics[width=0.24\linewidth]{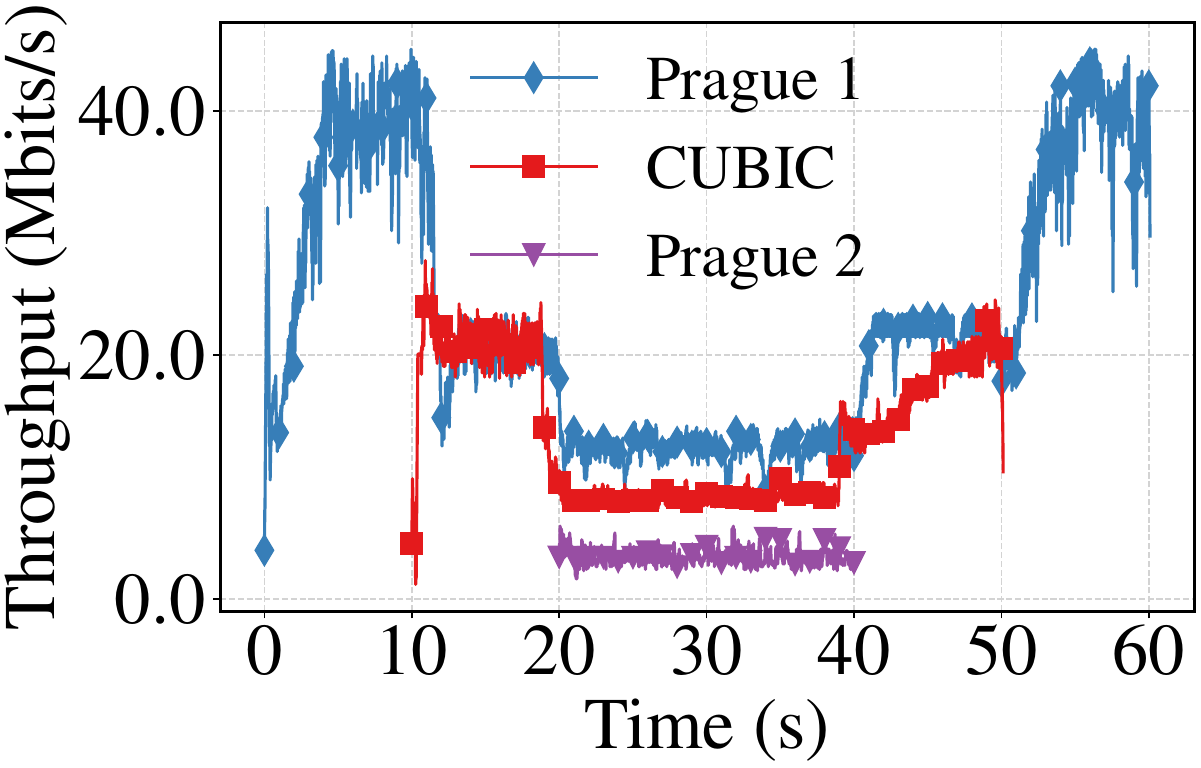}\label{fig:fairness_2l4s_1cubic}}
        \hfill
        \subfigure[Two L4S Prague flows and one BBRv2 flow in \sysname{}.]{
        \includegraphics[width=0.24\linewidth]{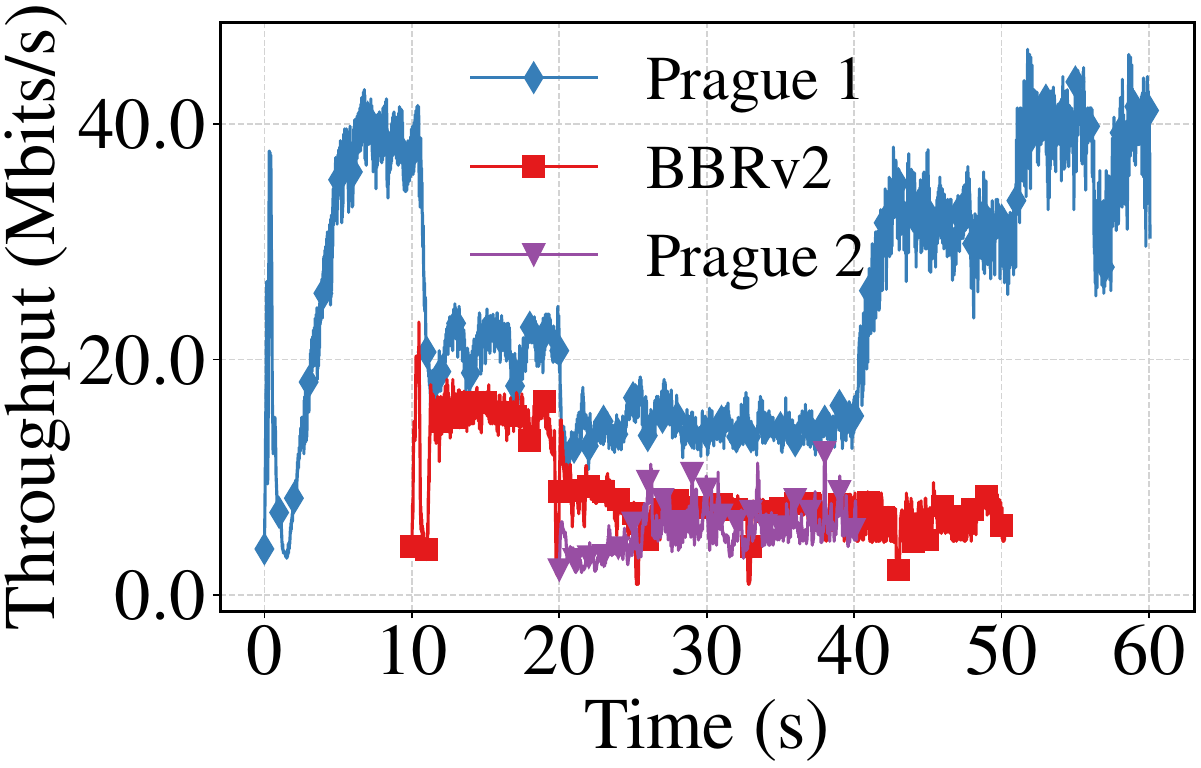}\label{fig:fairness_2l4s_1bbrv2}}
        \caption{\sysname{}'s impact on the TCP throughput fairness among different flows.}
        \label{fig:l4span_fairness}   
\end{figure*}

We evaluate \sysname{}'s impact on the TCP fairness between different UEs in the network.
In this evaluation, we connect three commercial UEs into the 5G network with starting time of 0, 10 and 20 seconds and ending time of 60, 50, 40 seconds, and each UE carries one TCP flow.
As a result, three Prague flows can maintain their fair share rate as shown in \cref{fig:fairness_3l4s_similarrtt}, and the Prague flow with higher RTT would need more time to converge to the fair share throughput as shown in \cref{fig:fairness_3l4s_diffrtt}.
When sharing the RAN with another UE using buffer filling TCP congestion control algorithm (CUBIC), the \sysname{} maintains the balance among three UEs~\cref{fig:fairness_2l4s_1cubic}.
BBRv2, also treated as L4S flow, takes longer to recover, as shown in \cref{fig:fairness_2l4s_1bbrv2}.
\revise{When three flows share the RAN simultaneously between 20 and 40s, the MAC scheduler determines the resource allocation based on each UE's configurations and channel conditions, resulting in slight different throughputs.}

\subsubsection{Feedback Short-circuiting}
\label{s:eval:short-circuiting}
Here we evaluate the effectiveness of \sysname{}'s short-circuiting design.
During this evaluation, one UE is connected to the 5G network with L4S Prague or classic CUBIC flow and communicates with a local server to rule out other delay impacts.
\cref{fig:l4span_mark_ack} shows the results of this evaluation.
\sysname{} can achieve a lower RTT for both Prague (\added{28.52 ms v.s. 33.87 ms}) and CUBIC (75.27 ms v.s. 85.42 ms) flows on average and on 99.9 percentile tail for Prague (\added{52.97 ms} v.s. 179.34 ms) and CUBIC (160.42 ms v.s. 190.53 ms) with short-circuits, as shown in \cref{fig:l4span_mark_ack_rtt}.
Also, the short-circuiting doesn't affect the throughput performance much (\cref{fig:l4span_mark_ack_tput}).

\subsubsection{DRB shared by L4S and classic flows}
\label{s:eval:marking}
\begin{figure}
\begin{minipage}[b]{0.64\linewidth}
    \subfigure[\added{RTT performance.}]{
    \includegraphics[width=0.48\linewidth]{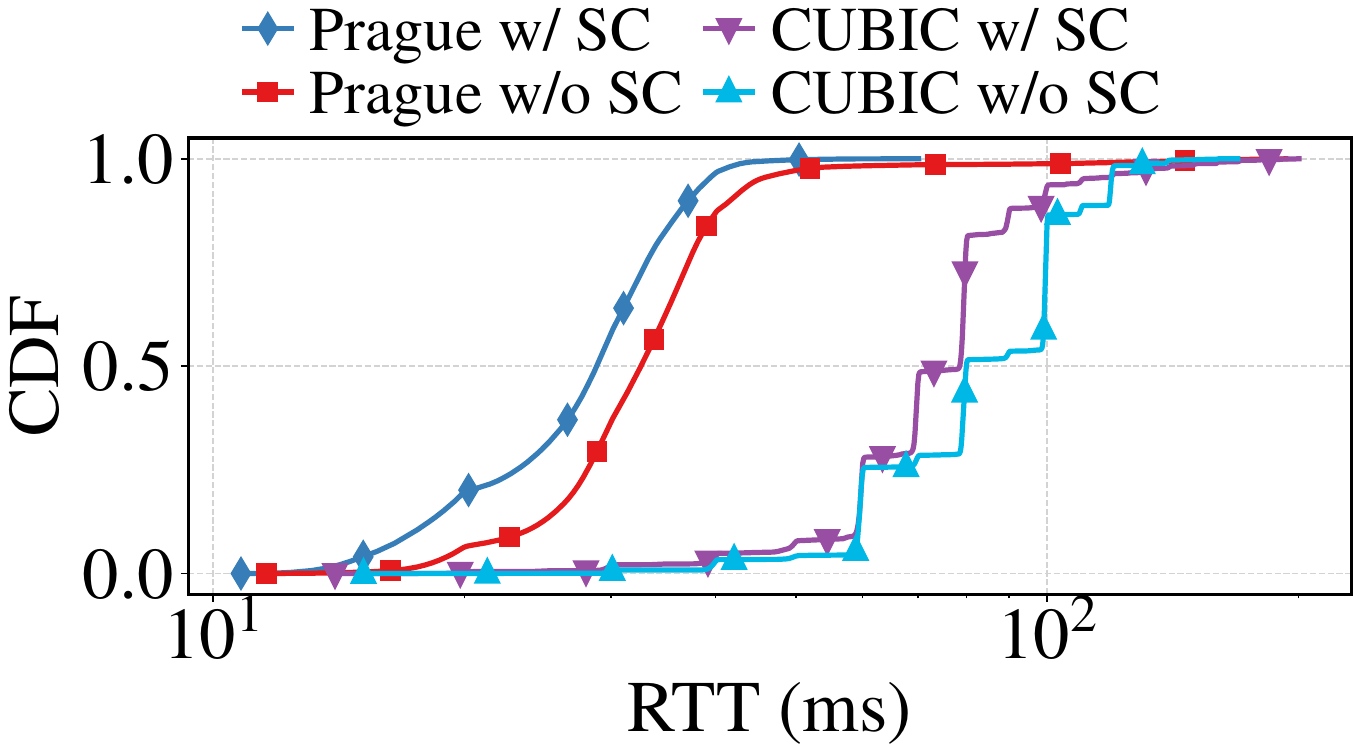}\label{fig:l4span_mark_ack_rtt}}
    \hfill
    \subfigure[\added{Throughput performance.}]
    {\includegraphics[width=0.48\linewidth]{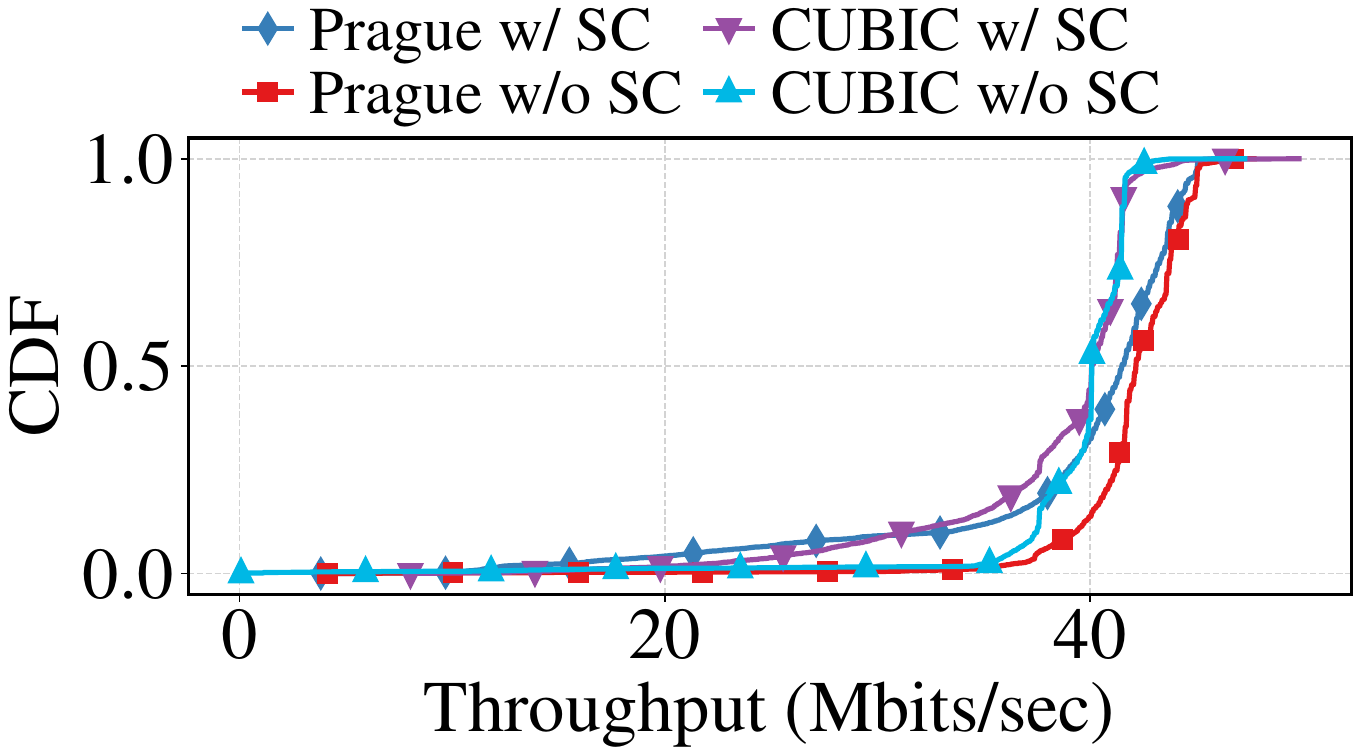}\label{fig:l4span_mark_ack_tput}}
    \caption{\sysname{}'s performance comparison when it uses feedback short-circuiting (SC in the figure) or not for L4S and classic flows.}
    \label{fig:l4span_mark_ack}  
\end{minipage}
\hfill
\begin{minipage}[b]{0.32\linewidth}          
    \centering
    \includegraphics[width=\linewidth]{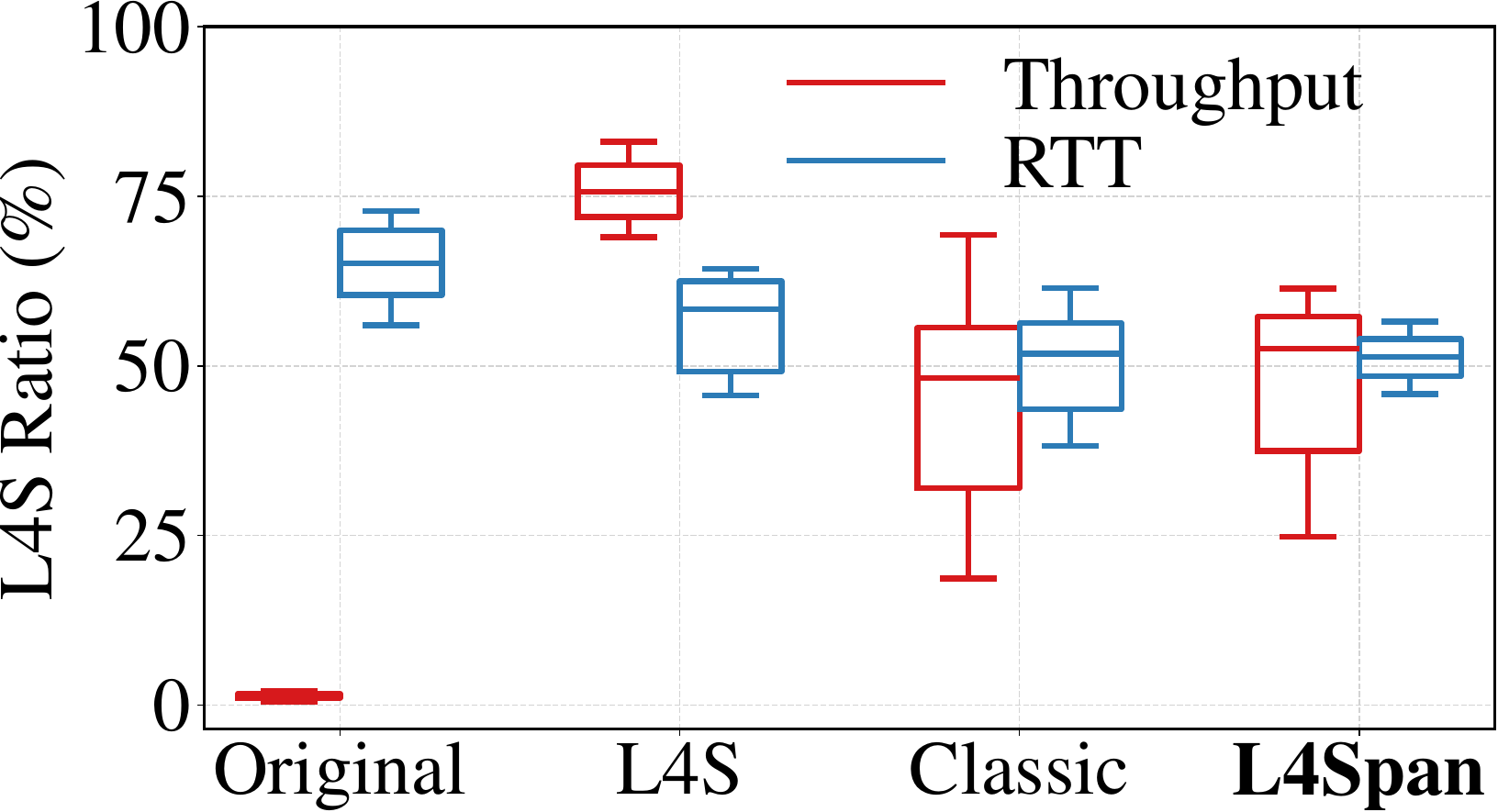}
    \caption{When L4S and classic flows share a DRB, \sysname{} achieves fair throughput and RTT.}
    \label{fig:drb_sharing}
\end{minipage}
\end{figure}

\sysname{} has a separate marking strategy when both flows share the same DRB (\S\ref{s:design:marking:shared_drb}).
During this evaluation, we connect one phone into the 5G network and start two flows -- one for Prague and one for CUBIC.
\cref{fig:drb_sharing} shows the result of this evaluation, where the y-axis is the ratio of L4S in RTT ($RTT_{L4S}/ (RTT_{L4S} + RTT_{Classic})$) and throughput ($r_{L4s}/ (r_{L4s} + r_{Classic})$).
We evaluate four marking methods, as listed in the x-axis labels.
Firstly, we keep the original marking strategy as they don't share the queue ("Original" in \cref{fig:drb_sharing}), where the L4S flow starves.
Secondly, we mark both flows with the L4S strategy in \S\ref{s:design:marking:l4s} ("L4S" in \cref{fig:drb_sharing}), where the classic flow is starved with a throughput ratio of round 25\%, as the classic flow slows it sending more than the L4S flow.
Thirdly, we mark both flows with the classic strategy in \S\ref{s:design:marking:classic} ("Classic" in \cref{fig:drb_sharing}), causing a larger throughput variation.
The marking strategy in \S\ref{s:design:marking:shared_drb} ("\sysname{} in \cref{fig:drb_sharing}") performs the best -- fair share capacity and the smallest variations.


\subsection{Microbenchmarks} 
\label{s:eval:micro_benchmark}
In this evaluation, we \textbf{\textit{1)}} illustrate the Dualpi2's marking strategy is unsuitable for wireless networks, \textbf{\textit{2)}} validate the parameter selections in our design, and \textbf{\textit{3)}} evaluate \sysname{}'s processing time.

\subsubsection{Marking Behavior}
\label{s:eval:dualpi2_and_rlc_queue}
We compare \sysname{} against \textbf{\textit{1)}} 
Dualpi2~\cite{schepper_dual-queue_2023}, and 
\textbf{\textit{2)}} DualPi2 + 10ms threshold to show the marking strategy in the wired network can't be applied directly to the wireless scenario.
We reimplement DualPi2 to replace \sysname{} and evaluate TCP Prague and BBRv2.
DualPi2's sojourn time calculation can't capture the wireless channel variations, resulting in severe under-utilization -- 
73\% and 28\% lower throughput with thresholds of 1 and 10 ms, respectively.

We evaluate the queue length of the L4S flow and classic flow with \sysname{} to demonstrate that the queue occupancy rarely reaches zero, leading to underutilization.
\cref{fig:rlc_queue_cdf} shows the result, where in all the scenarios, the classic flow RLC queue length doesn't fall to zero. 
And the L4S flow maintains a low queue occupancy and achieves ultra-low queuing delay.

\subsubsection{Parameter Selection}
\begin{figure}
\begin{minipage}[b]{0.64\linewidth}
    \centering
    \subfigure[16 UE scenario.]{
    \includegraphics[width=0.485\linewidth]{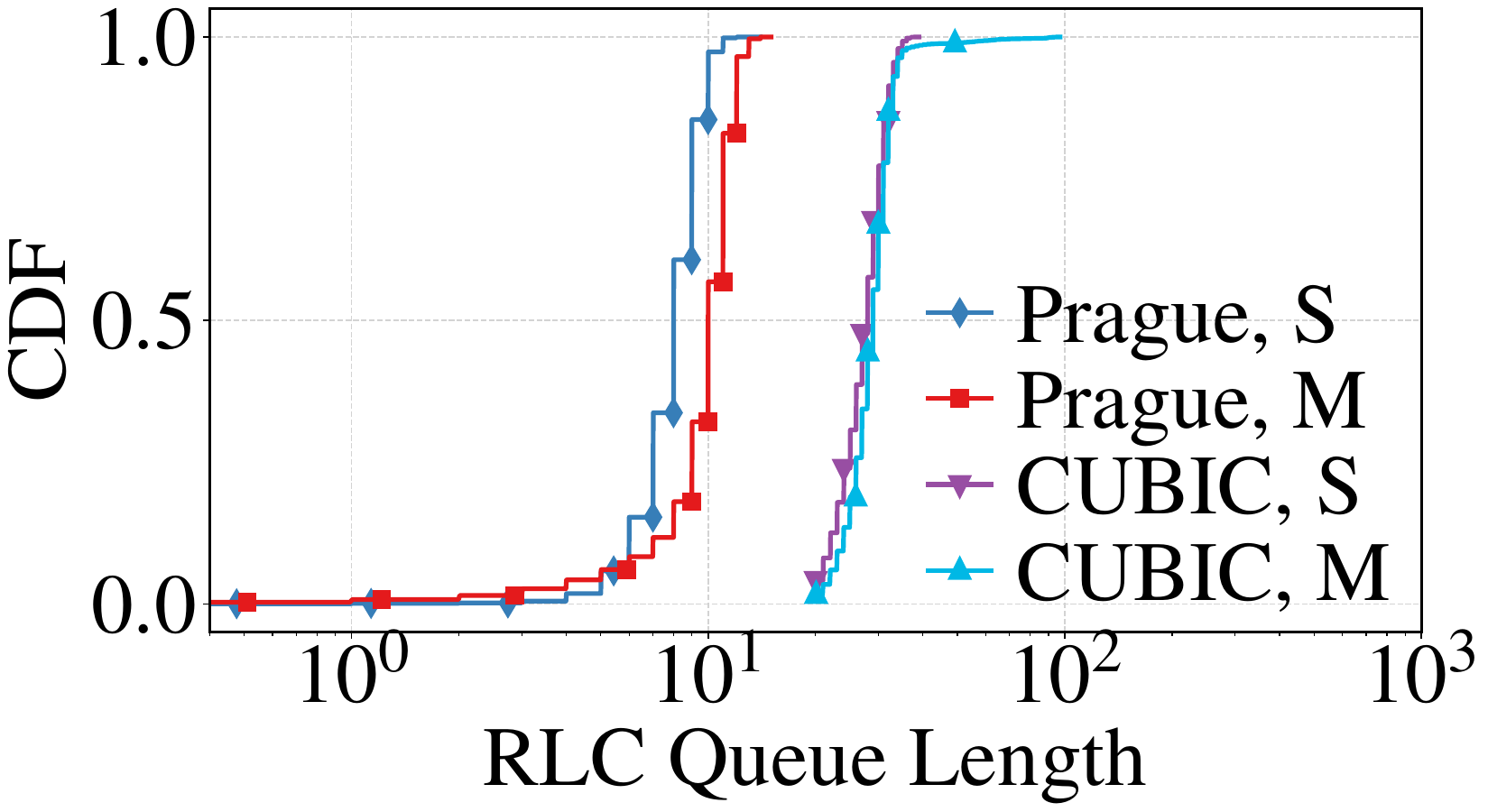}\label{fig:rlc_queue_16}}
    \subfigure[64 UE scenario.]
    {\includegraphics[width=0.485\linewidth]{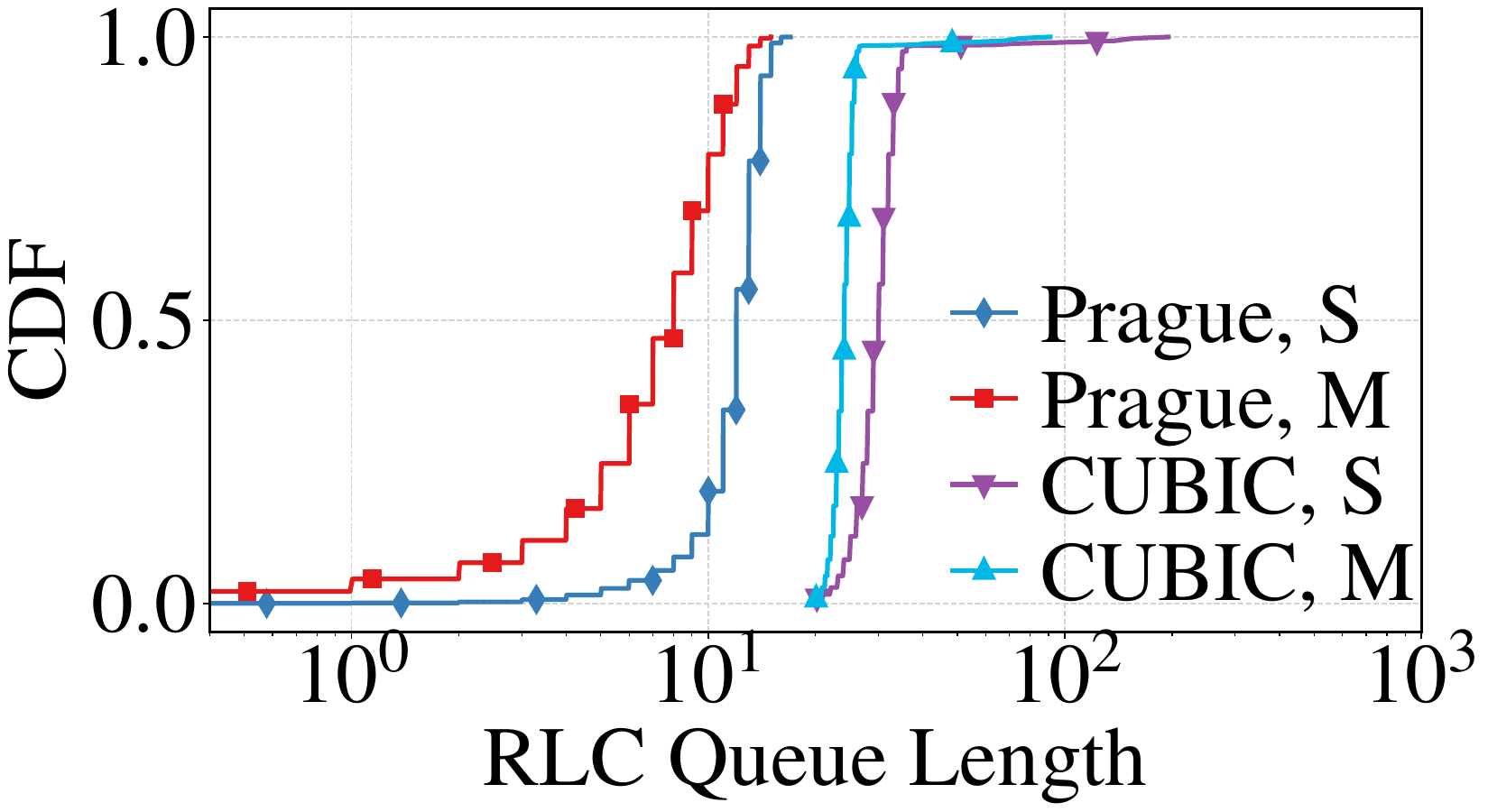}\label{fig:rlc_queue_64}}
    \vskip -0.08in
    \caption{RLC queue CDF (S: Static, M: Mobile).}
    \label{fig:rlc_queue_cdf}   
\end{minipage}
\hfill
\begin{minipage}[b]{0.32\linewidth}          
   \centering
    \includegraphics[width=\linewidth]{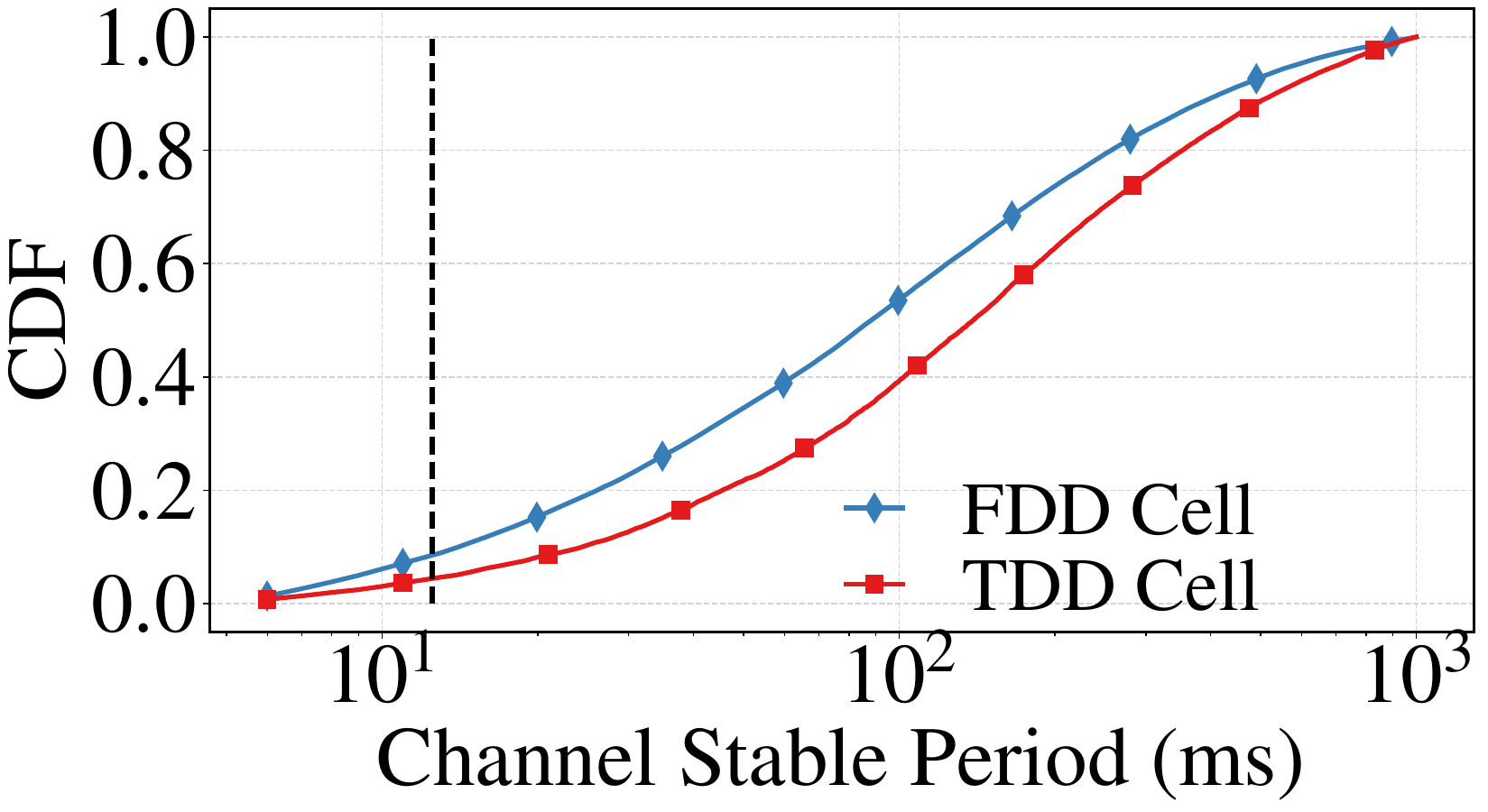}
    \caption{Channel stable period of T-Mobile cells.}
    \label{fig:channel_stable_period}   
\end{minipage}
\end{figure}

To verify our window selection -- half of the pre-set coherence time (24.9 ms), for egress rate estimation, we 
leverage a telemetry tool called NR-Scope \cite{wan_nr-scope_2024} to collect the DCIs in the two 
commercial base stations (a 2.5GHz TDD cell and a 600 MHz FDD 
cell) near our lab.
We count the period during which the MCS index's deviation is within 5 as an estimation of channel stable period, and include periods shorter than 1 s in the statistics.
\cref{fig:channel_stable_period} shows the measurement result, and the dashed vertical line marks our window sizes, where most of the time (>90\%) the channel stable period is higher than our window size\footnote{One thing to note is that MCS depends on both channel condition and buffered bytes, not fully reflecting the channel conditions. For example, if the buffered bytes are low, the RAN would use a low MCS even if the channel condition is great.}.

\begin{figure}
\begin{minipage}[b]{0.48\linewidth}
    \centering
    \subfigure{
    \includegraphics[width=0.8\linewidth]{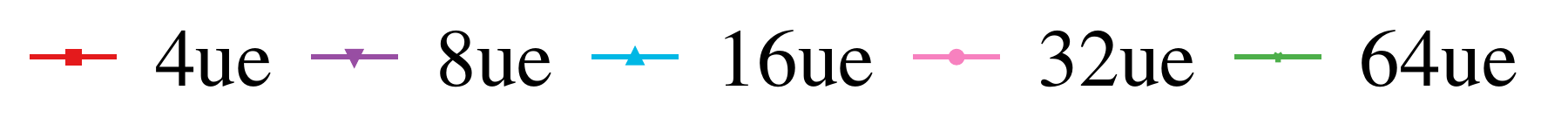}}
    \vskip -0.16in
    \subfigure[\added{$\tau_s$'s impact on RTT.}]{
    \includegraphics[width=0.48\linewidth]{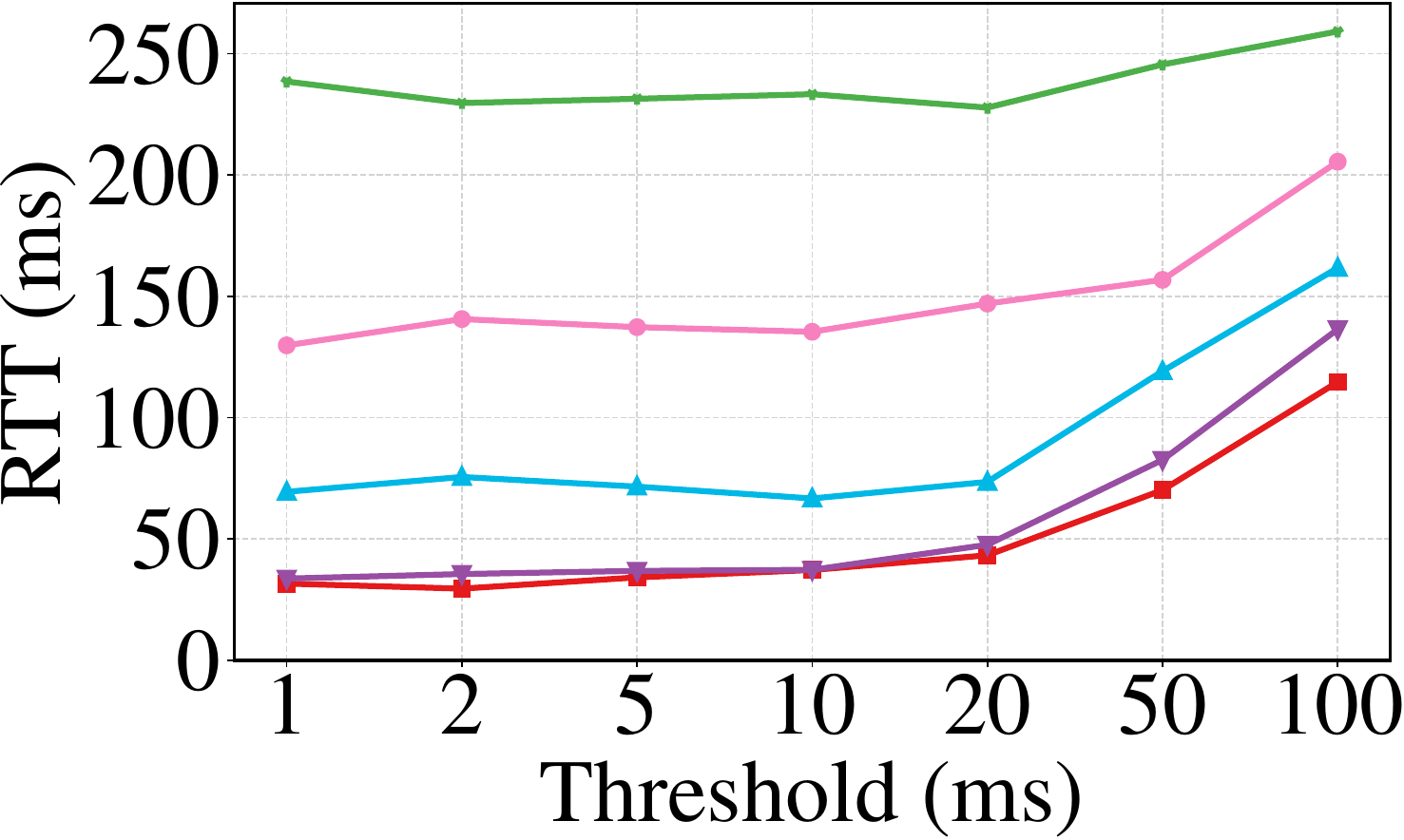}\label{fig:thres_rtt}}
    \subfigure[\added{$\tau_s$'s impact on rate.}]
    {\includegraphics[width=0.48\linewidth]{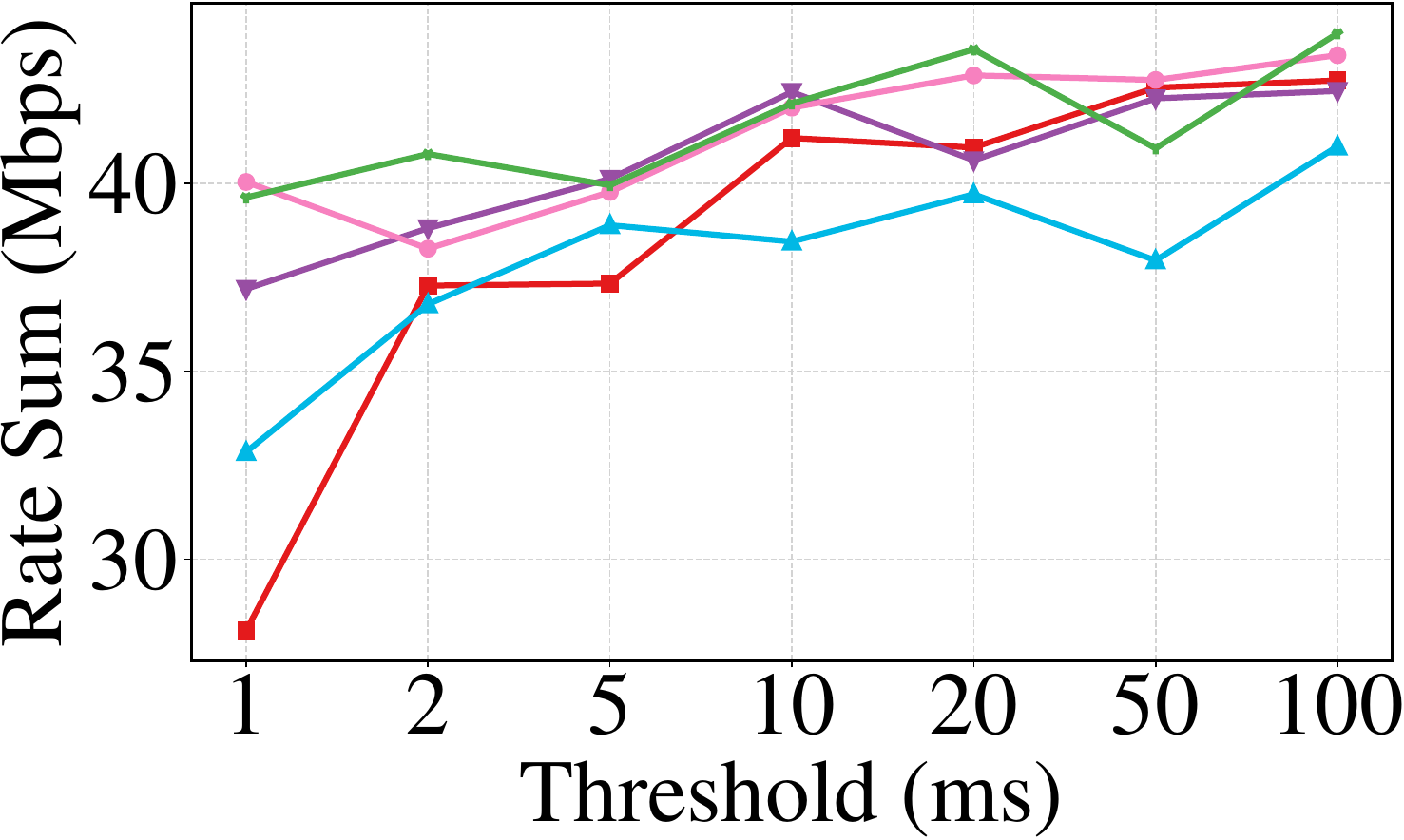}\label{fig:thres_tput}}
    \vskip -0.08in
    \caption{Impact of $\tau_s$ on performance.}
    \label{fig:thres_selection}   
\end{minipage}
\hfill
\begin{minipage}[b]{0.24\linewidth}          
    \centering
    \includegraphics[width=\linewidth]{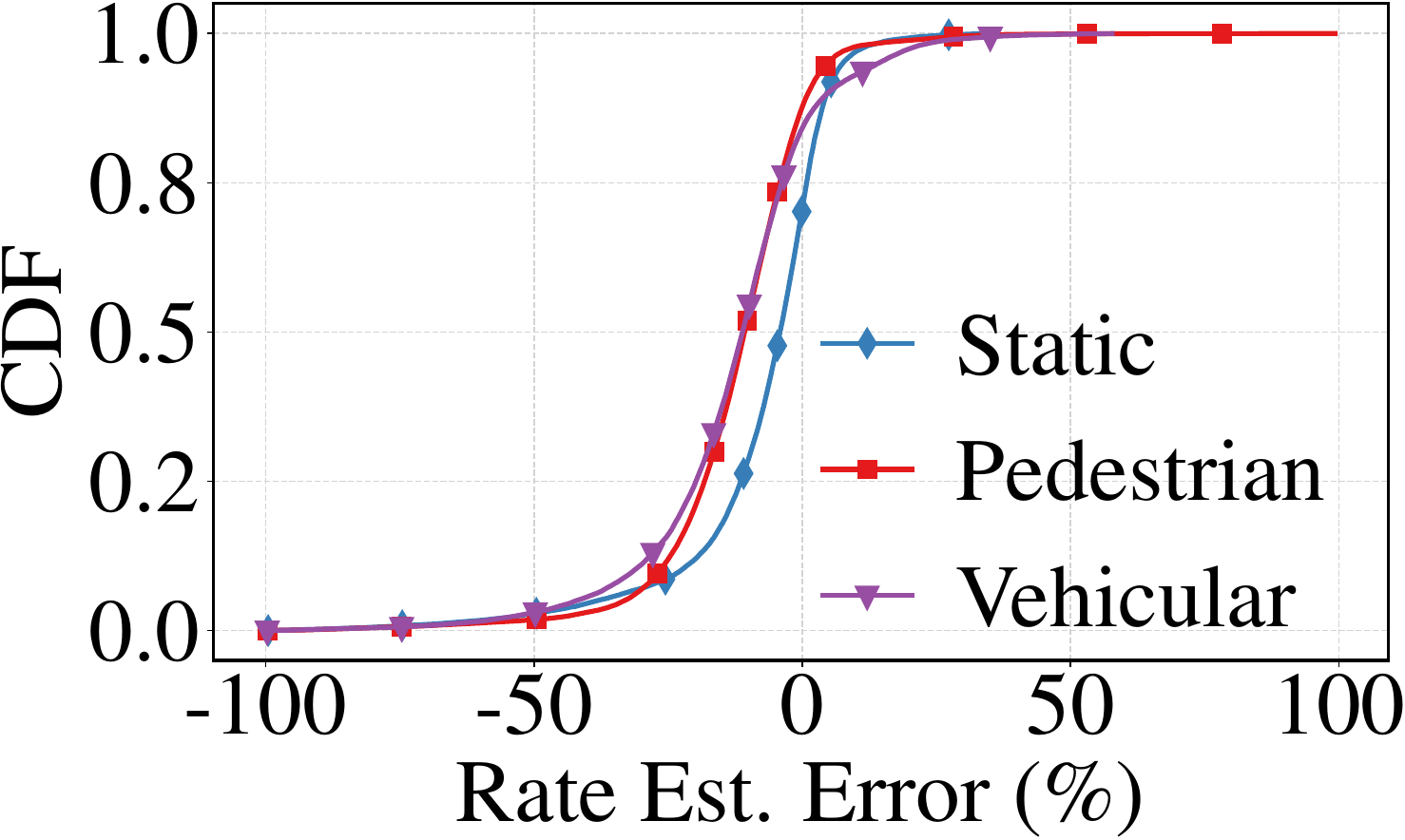}
    \caption{\added{Egress rate estimation errors.}}
    \label{fig:egress_rate}
\end{minipage}
\hfill
\begin{minipage}[b]{0.24\linewidth}            
\centering
    \includegraphics[width=\linewidth]{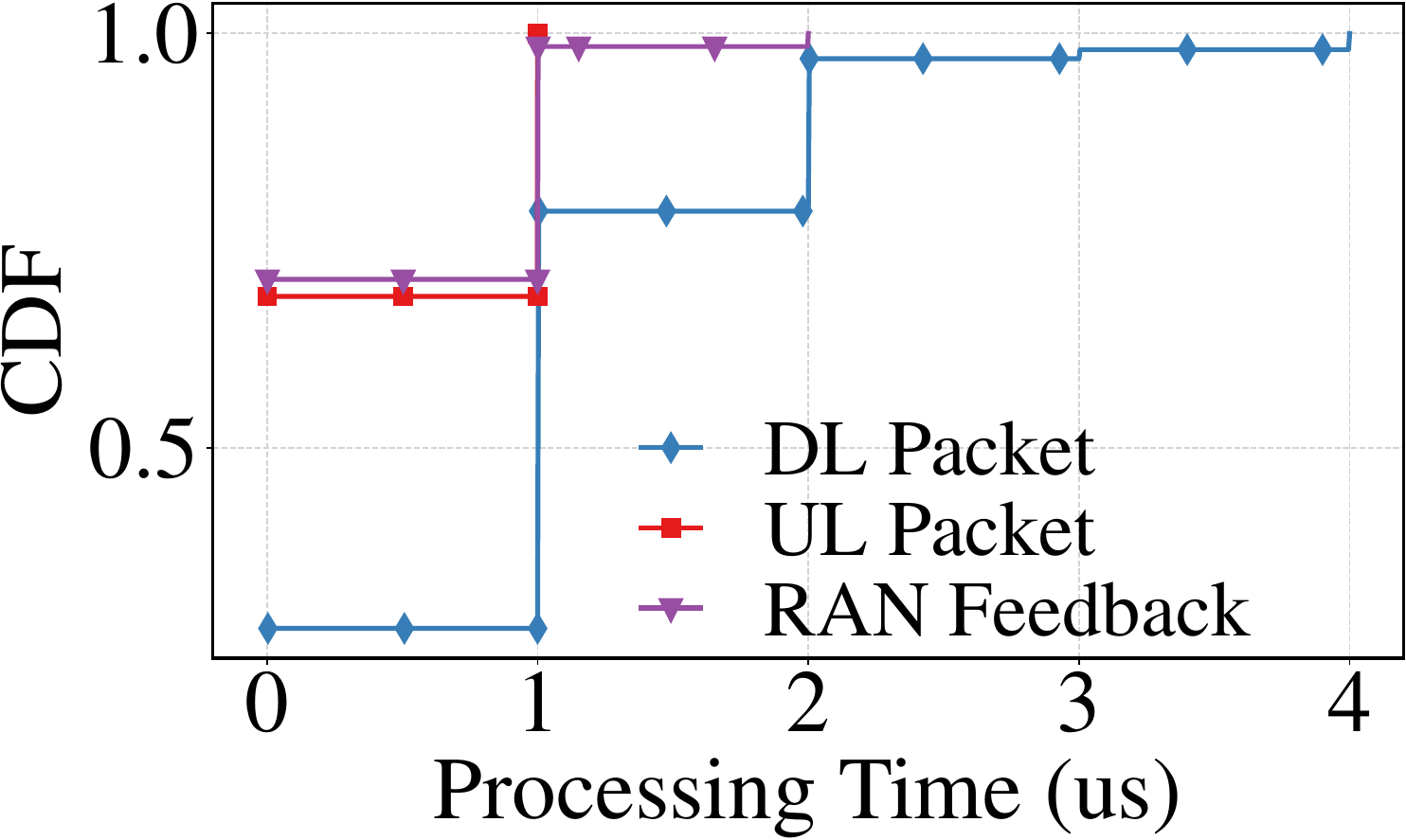}
    \caption{Processing time of \sysname{} on three events.}
    \label{fig:proc_time}
\end{minipage}
\end{figure}

To keep a low queuing delay for the L4S flow, \sysname{} utilizes a sojourn time threshold ($\tau_s$), which is 10 millisecond in the paper.
Here we evaluate how $\tau_s$ affects the performance of Prague.
\cref{fig:thres_selection} shows the result of this evaluation with each point showing the mean value.
As the queuing delay threshold grows higher, the throughput reaches a good level at $10 ms$ threshold with a low RTT. 

\begin{table*}[]
\small
\begin{tabular}{ccccc}
\hline
\textbf{Client Types} & srsRAN (idle) & srsRAN+L4Span (idle) & srsRAN (busy) & srsRAN+L4Span (busy) \\ \hline
\textbf{CPU Usage}    & $2.54\%$      & $4.44\%$      & $13.44\%$     & $14.96\%$     \\
\textbf{Memory Usage} & $6.93\%$      & $6.94\%$      & $7.28\%$      & $7.30\%$      \\ \hline
\end{tabular}
\caption{\revise{\sysname{}'s CPU (13700K) and memory (64GB) usage compared with the original srsRAN.}} \label{tab:cpu-mem-usage}
\end{table*}

To evaluate the RLC layer egress estimation accuracy, we connect 16 UEs into the network with three channel conditions and compare the RLC layer log with the estimated egress rate at \sysname{}.
\cref{fig:egress_rate} shows the measurements, and most of the time the RLC layer egress rate errors at \sysname{} are near $0\%$ across three channel conditions.

\subsubsection{\revise{System Performance}}
\label{s:micro:proc_time}
Here we evaluate the processing time of the three events of \sysname{} (\S\ref{s:design:overview}), on RAN feedback, on downlink and uplink packets.
We run srsRAN with \sysname{} on a 24-Core i7-13700K machine \revise{with 64 GB memory}, connect 64 UEs to the RAN, and collect the processing time of \sysname{} when all the UEs are downloading simultaneously.
\cref{fig:proc_time} shows the results, \sysname{} finishes its job within 2 microseconds for uplink packet and feedback information processing, and above 50\% of the processing is done within 1 microsecond.
For the downlink packet processing, \sysname{} finishes 97\% of its process in less than 2 microseconds, and in rare cases, it takes 4 microseconds. 

\revise{We evaluate the CPU and memory usage of \sysname{} by comparing it with the original srsRAN on the same machine and in two different operation states: 1) idle -- no user, and 2) busy -- 64 UEs downloading concurrently.
\cref{tab:cpu-mem-usage} shows the result of this evaluation, and \sysname{} incurs less than $2\%$ of more CPU and less than $0.02\%$ of more memory usage compared with the original srsRAN.
} 

\section{Discussion}
\label{s:discussion}

\noindent
\textbf{Compatibility with other RAN technologies.} \sysname{} is compatible with other RAN lower or upper layer technologies, such as \textit{carrier aggregation} (CA) \cite{ye_dissecting_2024,li_ca_2023}, MIMO, slicing \cite{chen_channel-aware_2023,an_helix_2024,cheng_oranslice_2024}, \revise{and handover.}
CA and MIMO only change the workflow of MAC and PHY layers, captured by \sysname{} egress rate prediction.
Slicing maps users into different UE context groups
\cite{balasingam_application-level_2024}, while \sysname{} 
works with the physical resources allocated to the client.
\revise{Upon handover, the buffered bytes are sent to a new RAN, and the markings are already done based on the old estimates.
The negative effect won't last long as the buffer is kept low by \sysname{}, we leave the evaluation as future work.}

\noindent
\textbf{QUIC Transport Protocols.} QUIC flow is another 
prevalent Internet traffic type, equipped with end-to-end data encryption. 
To harness the advantage of L4S architecture, Google implements the Prague \cite{google_quichequichequiccorecongestion_controlbbr2_sendercc_nodate} and BBRv2 as the underlying congestion control schemes, where the receiver reads the clear IP ECN field marked by middleboxes and bounces back the feedback.
\sysname{} can improve QUIC flows' performance by marking on their outer IP headers, similar to UDP flows (\S\ref{s:eval:method:app_cong}). 


\noindent
\textbf{5G uplink.} The design in this paper focuses on downlink; we leave 
the complementary uplink design to future work. In the uplink, the buffer is inside each client \cite{khan_how_2024}, and different strategies, such as pacing the packet to 
follow the RAN's transmission patterns, can be adopted.

\noindent
\textbf{Alternate Designs.}
Here we list design possibilities other than the foregoing for the 
placement of \sysnames{} and arguments against their adoption:
\textit{\textbf{1)} 5G Core/UPF:} There is no DRB state
stored inside the 5G Core/UPF, so the gNB has to send that state and the queue status, to the UPF, which incurs extra 
communication delay, lengthening the feedback loop. 
\textit{\textbf{2)} SDAP (CU-UP):} 
While we could implement marking in the SDAP layer, 
\sysname{} operates in the TCP/IP layer, while the SDAP is its own layer (with an associated 
header) in the 5G protocol stack, so layering principles favor this design less.
\textit{\textbf{3)} PDCP (CU-UP) and DU:} While each PDCP entity only has visibility
into its own packet queue, 5G may route an uplink feedback packet via a different DRB than
the downlink packet that elicited it.  Hence, this design choice precludes 
our proposed feedback short-circuiting design (\S\ref{s:ack_marking}). 
IP header compression in 
the PDCP layer \cite{3gpp_ts38321_2025} precludes packet marking in lower layers.
\textit{\textbf{4)} O-RAN RICs:} RICs incur extra networking delay~\cite{polese_understanding_2023}, real-time RICs only work in DU~\cite{ko_edgeric_2024,lacava_dapps_2025}.

\noindent
\textbf{Malicious Behaviors.} \revise{In the 5G scenario, if a sender or a middlebox mingles the ECN field or adversarially ignores the ECN feedback, the damage to the network is minimal. UEs' queues are isolated (\S\ref{s:background}), and filling one queue only affects one specific UE's traffic in one DRB, while traffic in other DRBs and  UEs still get their fair share of resources guaranteed by the MAC scheduler.}

\noindent
\textbf{Incentives for ISPs.} \revise{Deploying L4S in the 5G and wired networks improves the latency performance, and benefits the ISPs' competitiveness. 
Furthermore, L4S functionality only needs to be deployed at the bottleneck of the data path with a controllable cost. 
Thus, many applications (Apple FaceTime\cite{team_testing_2023}, Google QUIC\cite{google_quichequichequiccorecongestion_controlbbr2_sendercc_nodate}) and ISPs (Comcast\cite{comcast_comcast_2025}, T-Mobile\cite{saw_tmobile_2025}) are already rolling out L4S.}

\section{Conclusion}
\label{s:concl}
In this paper, we have proposed \sysname{}, the 
first network architecture proposal that
spans L4S signaling over 5G networks in a 
practical and high performance implementation.
Our experimental evaluation demonstrates that \sysname{} can optimize 
the performance of both L4S and classic flows in different
applications and transport layer technologies.
Furthermore, \sysnames{} design covers the many
possible 5G configurations, making it practical for deployment
today.

\section*{Acknowledgements}

This material is based upon work supported by the National Science Foundation under grants AST-2232457, CNS-2223556, ITE-2326928, and OAC-2429485. We gratefully acknowledge a gift from Princeton NextG industrial affiliate program member Qualcomm Corporation.

\bibliographystyle{concise2}
\begin{raggedright}
\bibliography{paws-zotero}
\end{raggedright}

\appendix
\newpage
\section{\sysname{} Layer Pseudo Code}
\label{appd:pseudo_code}
Here, we list the pseudo code of \sysname{} triggered by three events: on receiving a downlink IP packet (\cref{fig:on_dl_pkt}), on receiving the RAN feedback (\cref{fig:on_tcp_ack}, top) and on receiving an uplink packet (\cref{fig:on_tcp_ack}, down).

\begin{figure}[h]
\begin{minipage}[b]{0.45\linewidth}
\begin{lstlisting}[language=C++]    
void on_dl_pkt(buffer pkt, qos_flow_id qfi) {
  /* Map the QFI to DRB */
  auto drb_id = qfi_to_drb(qfi);
  /* Save the five-tuple mapping to DRB */
  auto five_tuple = extract_five_tuple(pkt);
  five_tuple_to_drb[five_tuple] = drb_id;
  /* Read the ECN bits and update DRB flows */
  auto flow_type = classify_flow(pkt);
  update_drb_flows(drb_id, flow_type);
  /* Save to packet profile queue */
  add_to_profile_queue(drb_id, pkt);
  /* Perform marking if it's a UDP packet */
  if (is_udp(pkt))
    mark_pkt(drb_id, pkt);
  /* Pass the packet to lower SDAP layer */
  to_sdap(pkt, qfi);
}
\end{lstlisting}
\caption{The \sysname{} layer's processing upon receiving a downlink IP packet.}
\label{fig:on_dl_pkt}
\end{minipage}
\hfill
\begin{minipage}[b]{0.48\linewidth}
    \begin{lstlisting}[language=C++]    
void on_ran_feedback(uint32_t txed_sn, uint32_t dlvred_sn, drb_id_t drb_id, timestamp ts) {
  /* Update the pkt profile queue */
  update_pkt_prof(txed_sn, dlvred_sn, drb_id, ts);
  /* Standing packets queuing delay prediction */
  auto stand_pkt_qdelay = qdelay_predict();
  /* Update marking decision*/
  update_drb_mark_state(drb_id, stand_pkt_qdelay);
}
\end{lstlisting}
\begin{lstlisting}[language=C++]    
void on_ul_packet(buffer pkt) {
  if(is_tcp_ack(pkt)) {
    /* Extracts the ACK's downlink five-tuple */
    auto five_tuple = extract_ack_five_tuple(ack);
    /* Reverse map the five-tuple to DRB id */
    auto drb_id = five_tuple_to_drb[five_tuple];
    /* Mark ECN bits based on the mark state */
    mark_pkt(drb_id, pkt);
  }
  to_upf(pkt); /* Pass the packet to the core UPF */
}
\end{lstlisting}
\caption{The \sysnames{} processing on receiving the RAN feedback (top), and the uplink packet (down).}
\label{fig:on_tcp_ack}
\end{minipage}
\end{figure}


\section{BBR and Reno's Evaluation Result with \sysname{}}\label{appd:bbr_and_reno}
Here we present the evaluation result of BBR and Reno:
\begin{itemize}
    \item BBR~\cite{cardwell_bbr_2016}. BBR sender periodically switches between bottleneck bandwidth and propagation delay probing state, and doesn't react much to the packet loss or CE feedback.
    \item Reno~\cite{paxson_tcp_1999}. TCP Reno cuts its congestion window to half upon packet loss and adds one onto its congestion window size per RTT in steady state.
\end{itemize}
\sysname{} can decrease the RTT of Reno by 97.61\% and 97.12\% in static and mobile channels, yielding a higher variation in throughput.
For other channels and different UE numbers, \sysname{} can constantly improve the performance of RTT by more than 90\% and incurs very little performance drop in throughput.
BBR, doesn't react much to the ECN or packet loss, yields higher variations in RTT and throughput with L4S, while the median values stay the same most of the time. 

\begin{figure*}
        \centering
        \subfigure[16 UEs, default RLC queue length, \revise{38 ms RTT}.]
        {\includegraphics[width=0.48\linewidth]{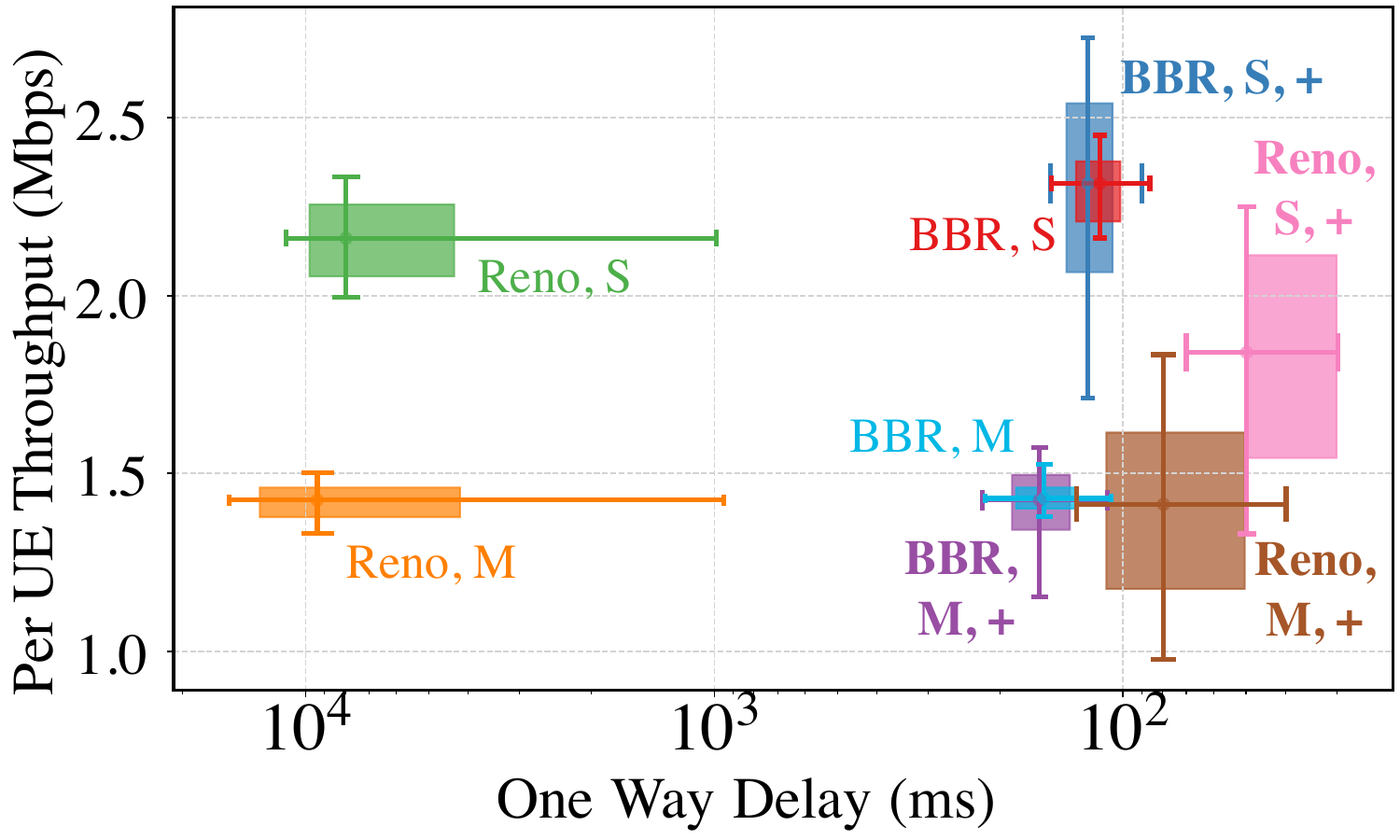}\label{fig:16ue_1000_reno}}
        \hfill
        \subfigure[64 UEs, default RLC queue length, \revise{38 ms RTT}.]
        {\includegraphics[width=0.48\linewidth]{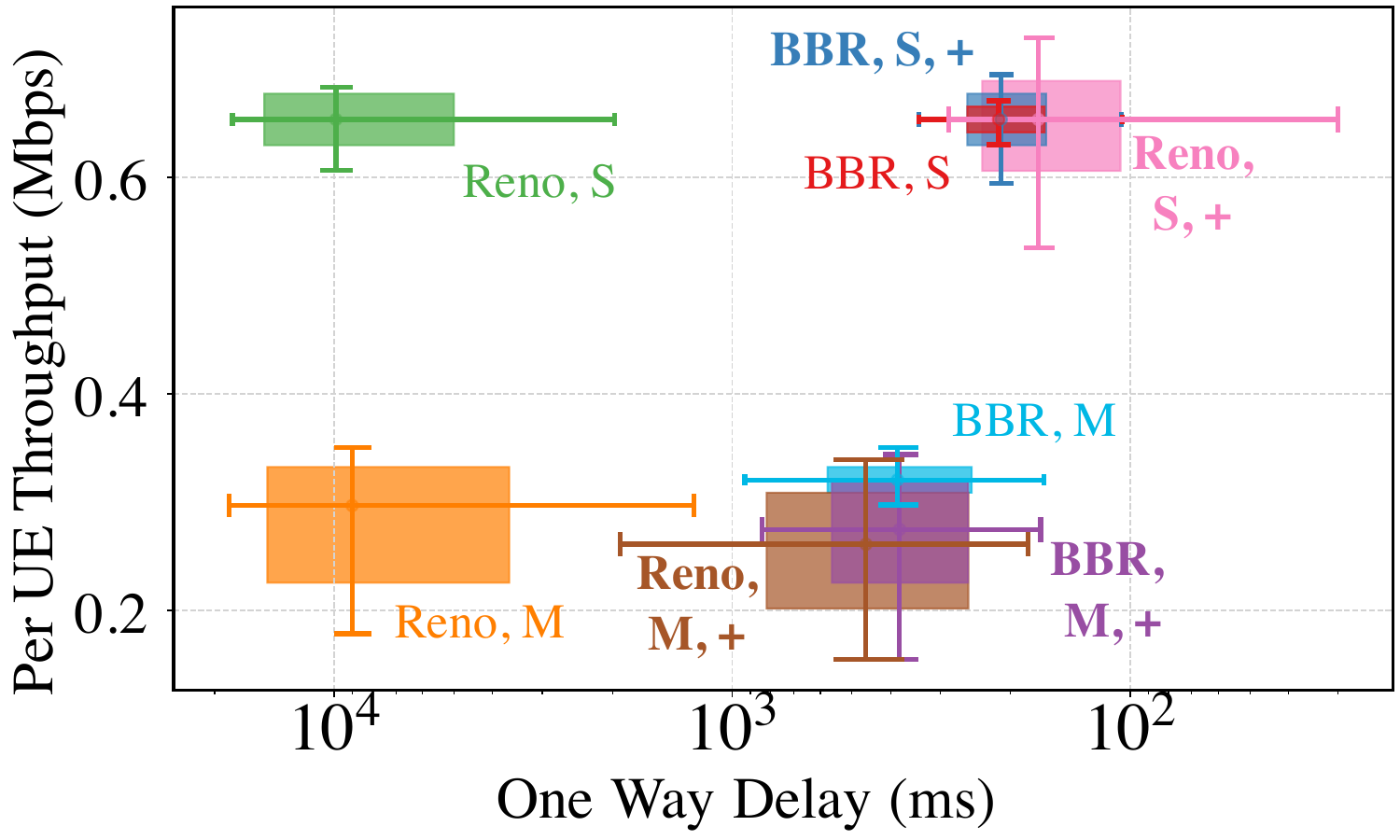}\label{fig:64ue_1000_reno}}

        \subfigure[16 UEs, RLC queue length 256, \revise{38 ms RTT}.]
        {\includegraphics[width=0.48\linewidth]{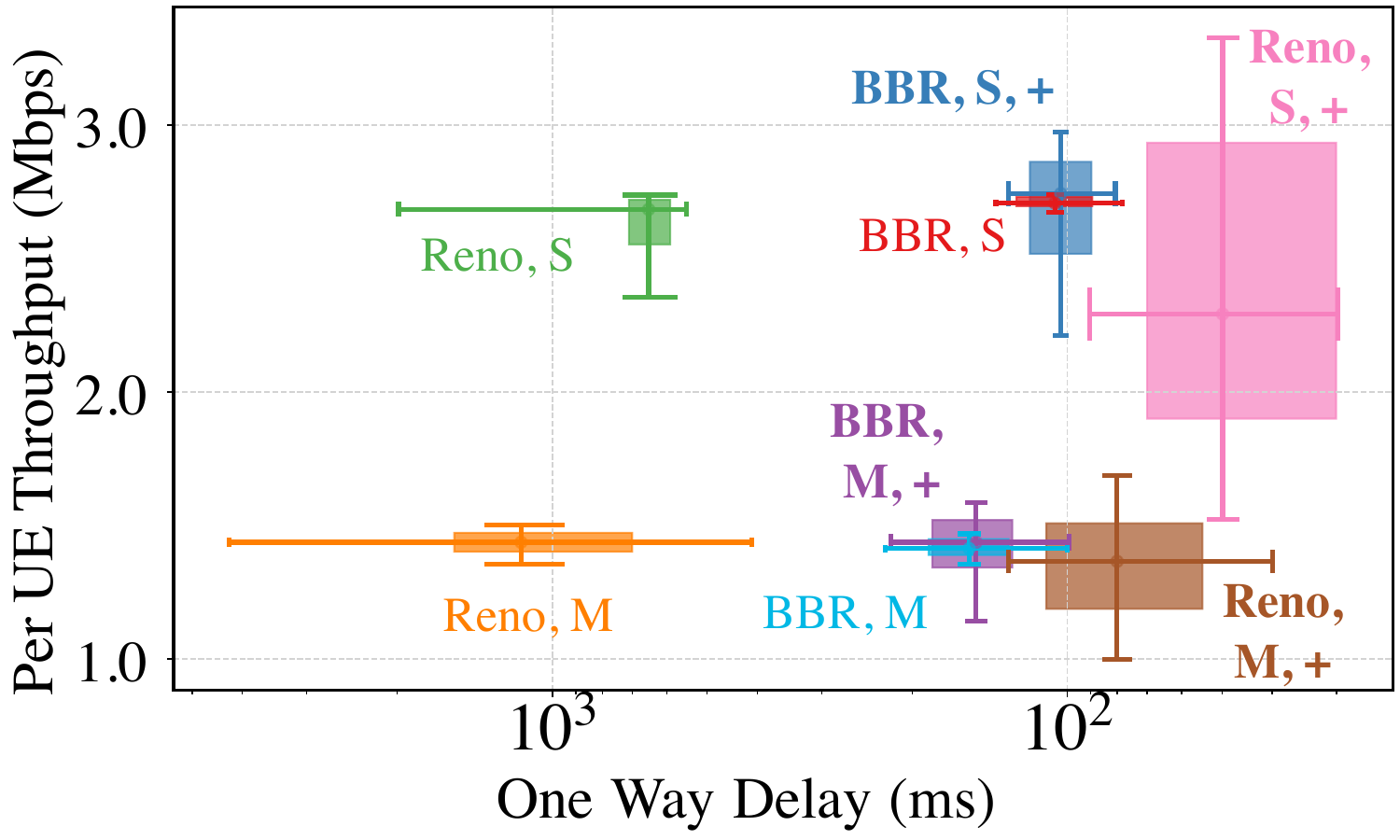}\label{fig:16ue_256_reno}}
        \hfill
        \subfigure[64 UEs, RLC queue length 256, \revise{38 ms RTT}.]
        {\includegraphics[width=0.48\linewidth]{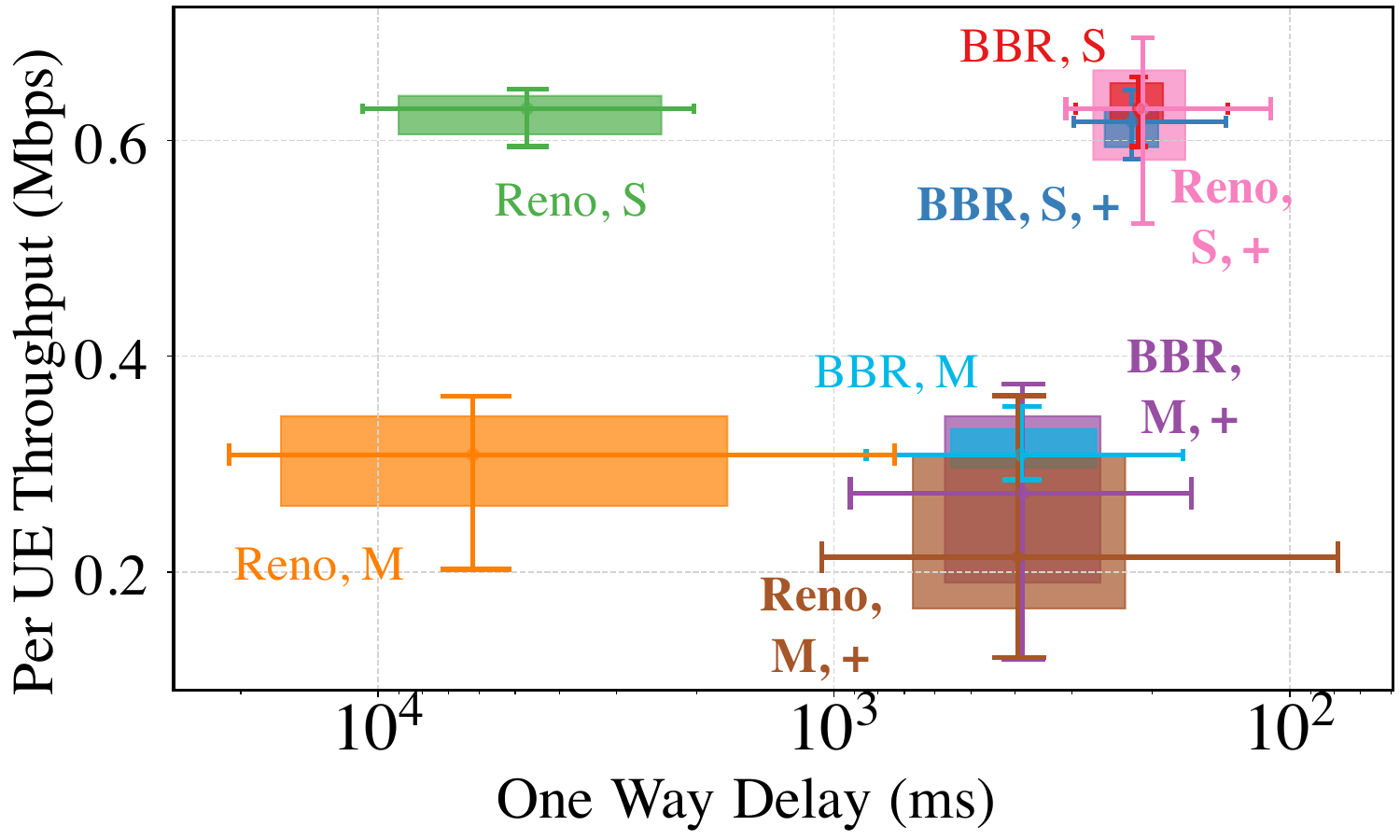}\label{fig:64ue_256_reno}}

        \subfigure[\revise{16 UEs, default RLC queue length, 106 ms RTT}.]
        {\includegraphics[width=0.48\linewidth]{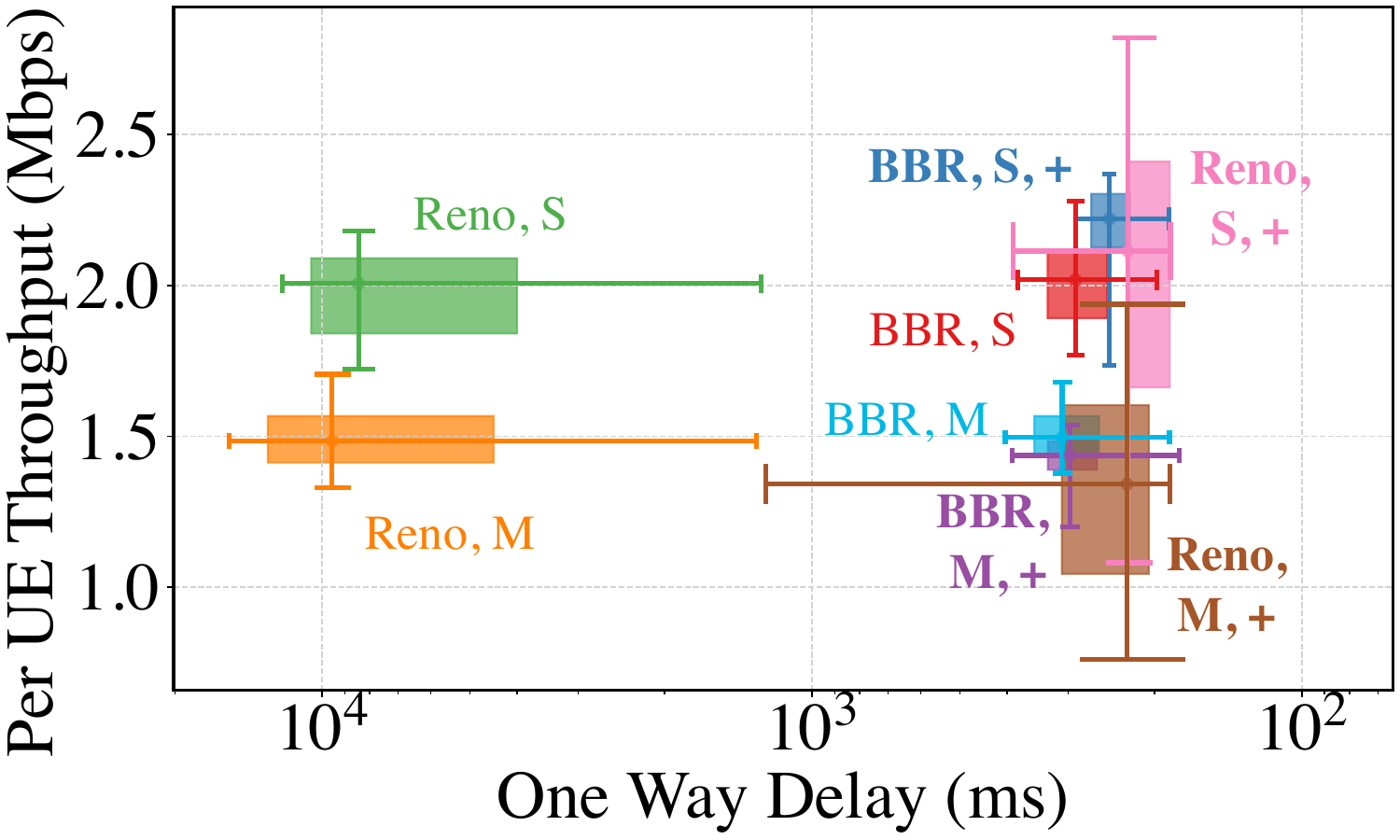}\label{fig:16ue_0916_reno}}
        \hfill
        \subfigure[\revise{64 UEs, default RLC queue length, 106 ms RTT}.]
        {\includegraphics[width=0.48\linewidth]{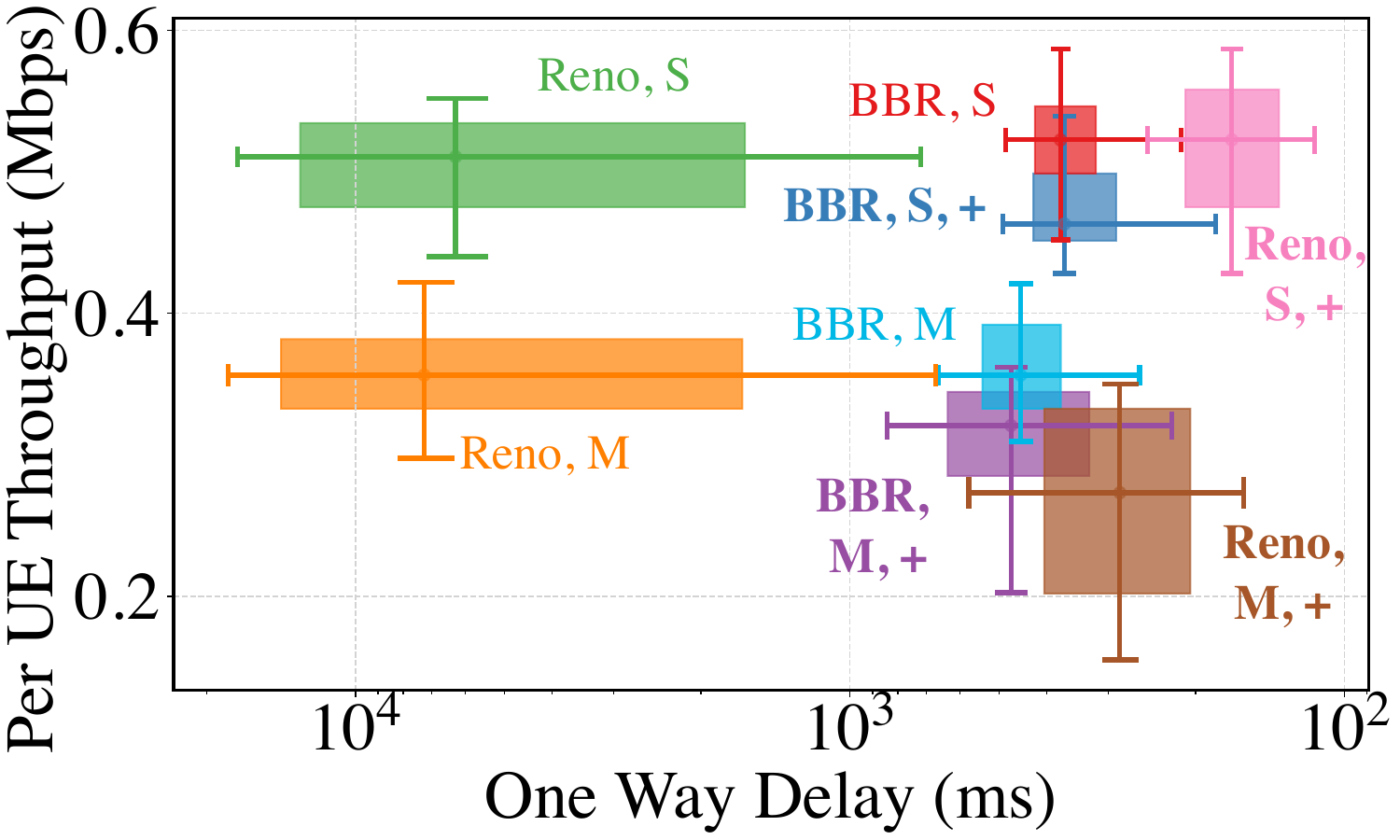}\label{fig:64ue_0916_reno}}

        \subfigure[16 UEs, RLC queue length 256, \revise{106 ms RTT}.]
        {\includegraphics[width=0.48\linewidth]{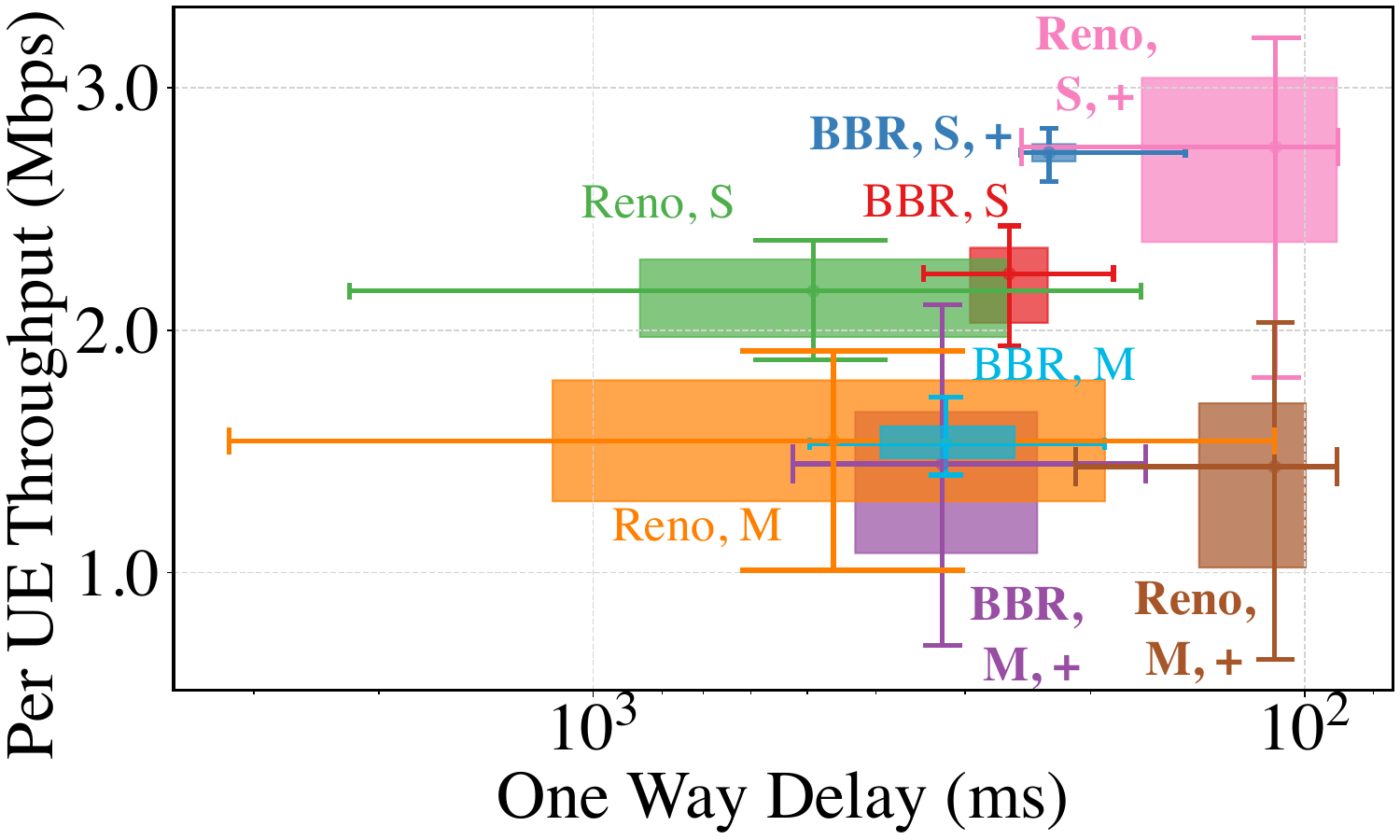}\label{fig:16ue_0916_256_reno}}
        \hfill
        \subfigure[\revise{64 UEs, RLC queue length 256, 106 ms RTT}.]
        {\includegraphics[width=0.48\linewidth]{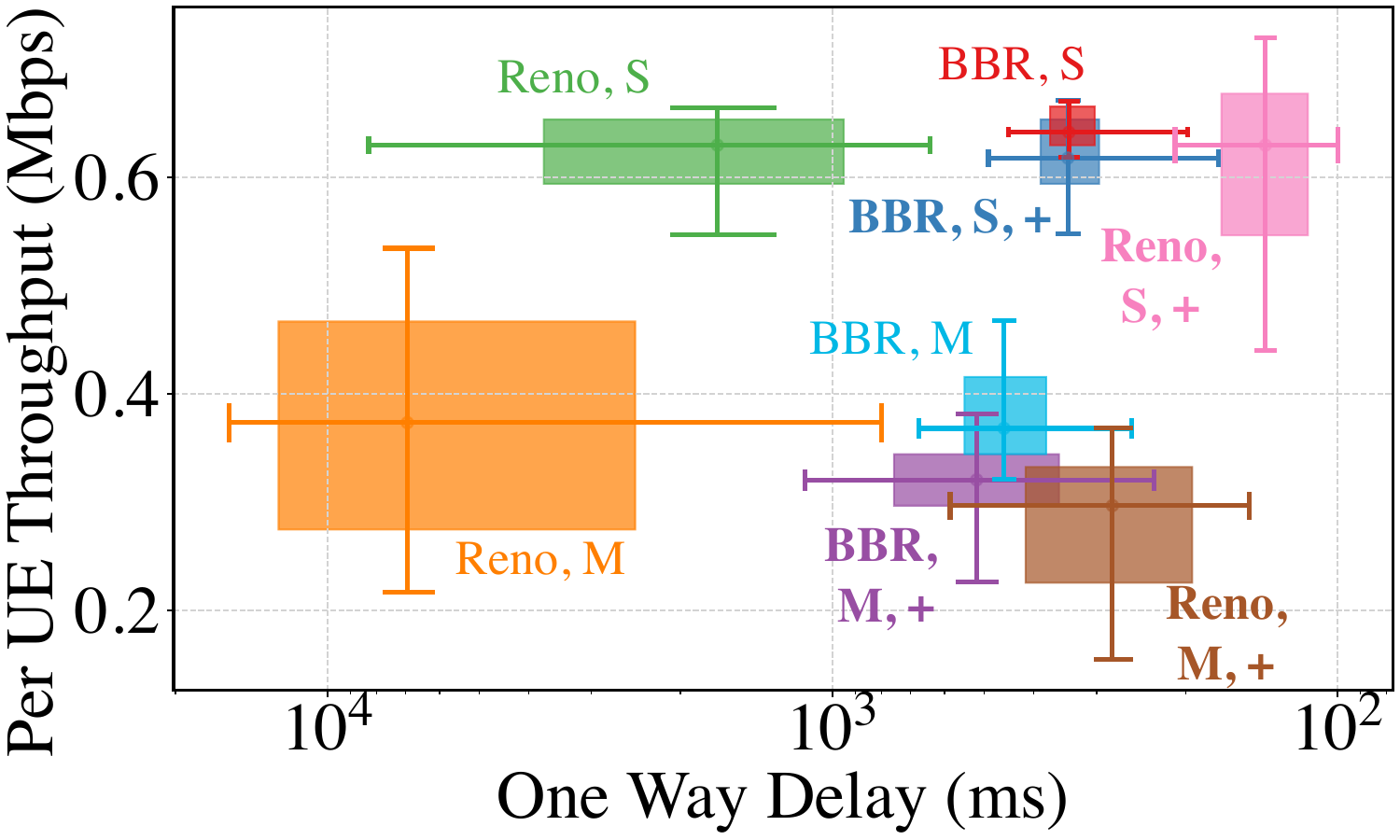}\label{fig:64ue_0916_256_reno}}
        
        \caption{\added{\sysname{} improves the performance of various congestion control algorithms in terms of reducing RTT while maintaining throughput, under severely congested RAN and different channel conditions (S: Static, M: Mobile). \textbf{Bold} font and adding symbol ($+$) indicate \sysname{} is deployed. \revise{Senders are Azure instances with uncongested RAN ping time marked in the caption.} Center point: median, box edges: 25th- and 75th-percentile, and whiskers: 10th- and 90th-percentile.}}
        \label{fig:l4s_tcp_end_to_end_reno}   
\end{figure*}

\end{document}